\documentclass[12pt,twoside,english]{article}
\usepackage[T1]{fontenc}
\usepackage{units}
\usepackage{mathtools}
\usepackage{amsmath}
\usepackage{amssymb}
\usepackage{graphicx}

\makeatletter

\providecommand{\tabularnewline}{\\}

\newcommand{\lyxaddress}[1]{
\par {\raggedright #1
\vspace{1.4em}
\noindent\par}
}

\usepackage{a4wide}
\usepackage{graphicx}
\usepackage{fancyhdr}
\usepackage{amsmath}
\usepackage{subfigure}
\usepackage{setspace}
\usepackage{verbatim} 
\numberwithin{equation}{section}

\newcommand{\vek}[1]{\mathchoice{\displaystyle\boldsymbol#1}
{\textstyle\boldsymbol#1}{\scriptstyle\boldsymbol#1}
{\scriptscriptstyle\boldsymbol#1}}
\newcommand{\mat}[1]{\mathchoice{\displaystyle\mathbf#1}
{\textstyle\mathbf#1}{\scriptstyle\mathbf#1}
{\scriptscriptstyle\mathbf#1}}

\makeatother

\usepackage{babel}
\begin{document}

\title{A Unified Finite Strain Theory for \\Membranes and Ropes}

\author{T.P. Fries, D. Sch{\"o}llhammer}

\maketitle

\lyxaddress{\begin{center}
Institute of Structural Analysis\\
Graz University of Technology\\
Lessingstr. 25/II, 8010 Graz, Austria\\
\texttt{www.ifb.tugraz.at}\\
\texttt{fries@tugraz.at}
\end{center}}
\begin{abstract}
The finite strain theory is reformulated in the frame of the Tangential
Differential Calculus (TDC) resulting in a unification in a threefold
sense. Firstly, ropes, membranes and three-dimensional continua are
treated with \emph{one} set of governing equations. Secondly, the
reformulated boundary value problem applies to parametrized \emph{and}
implicit geometries. Therefore, the formulation is more general than
classical ones as it does not rely on parametrizations implying curvilinear
coordinate systems and the concept of co- and contravariant base vectors.
This leads to the third unification: TDC-based models are applicable
to two fundamentally different numerical approaches. On the one hand,
one may use the classical Surface FEM where the geometry is discretized
by curved one-dimensional elements for ropes and two-dimensional surface
elements for membranes. On the other hand, it also applies to recent
Trace FEM approaches where the geometry is immersed in a higher-dimensional
background mesh. Then, the shape functions of the background mesh
are evaluated on the trace of the immersed geometry and used for the
approximation. As such, the Trace FEM is a fictitious domain method
for partial differential equations on manifolds. The numerical results
show that the proposed finite strain theory yields higher-order convergence
rates independent of the numerical methodology, the dimension of the
manifold, and the geometric representation type.

\end{abstract}
\newpage{}\tableofcontents{}\newpage{}

\section{Introduction\label{sec:Introduction}}

In structural analysis, there are many examples where a dimensional
reduction in the modeling is a key step to achieve simpler models
than considering the full (three-dimensional) continuum. The situation
is simple for flat domains such as straight beams and flat plates.
For curved structures with reduced dimensionality such as \emph{curved}
beams and shells, modeling becomes considerably more involved and
the governing equations are partial differential equations (PDEs)
on manifolds. The situation is even more complicated for structures
undergoing large displacements such as ropes (cables) and membranes.
These structures do not feature any bending resistance and often deform
largely such that one has to carefully distinguish between the undeformed
and deformed configurations. When equilibrium is to be fulfilled in
the deformed configuration, this is generally referred to as finite
strain theory, the geometrically non-linear situation, or large displacement
analysis.

Herein, we propose a new formulation of the mechanical models for
ropes and membranes in finite strain theory which employs the Tangential
Differential Calculus (TDC) for the definition of geometric and differential
quantities. The geometries of ropes and membranes may be seen as (curved)
manifolds embedded in a higher-dimensional space. Let the dimension
of the surrounding region be $d$ and the dimension of the manifold
be $q$ with $q=1$ for ropes and $q=2$ for membranes. The classical
modeling approach to ropes and membranes (including shells) may be
called parameter-based \cite{Chapelle_2011a}, because it relies on
a parametrization which maps from the $q$-dimensional to the $d$-dimensional
space. In this case, a curvilinear coordinate system is naturally
implied and co- and contra-variant base vectors are easily defined.
An alternative is to define the geometry implicitly, for example,
based on the level-set method \cite{Osher_2003a,Osher_2001a,Sethian_1999b}.
Then, the zero-level set of a scalar function implies the manifold.
In this case, no parametrization, hence, no curvilinear coordinates
exist on the manifold. This situation cannot be covered by the classical
models but the proposed TDC-based approach can.

When comparing the new TDC-based formulation with the classical parameter-based
formulation, we notice the following important aspects:
\begin{enumerate}
\item The first argument is \emph{geometrical} and is concerned with the
definition of the boundary value problem (BVP) of ropes and membranes.
It is desirable to have a mechanical model which applies to general
geometries no matter whether they are defined in parametric or implicit
form. The classical parameter-based approach does not apply to the
case where the geometry is, for example, defined by a zero-level set.
In that case, a parametrization only becomes available upon meshing
the zero-level set (resulting in an atlas of mappings implied by the
elements). However, then the BVP is rather defined with respect to
a discrete setting than a continuous one. Consequently, one advantage
of the TDC-based formulation is that it allows for a proper definition
of a continuous BVP which applies for both, parametrically and implicitly
defined geometries.
\item The second argument is \emph{mechanical}. The TDC-based formulation
treats ropes and membranes in a unified sense where all mechanical
quantities such as stress and strain tensors are based on differential
operators formulated with the TDC. These quantities are defined based
on the $d$-dimensional global coordinate system into which the manifold
is immersed. That is, these tensors are always $d\times d$ no matter
whether ropes or membranes are considered. The TDC-based formulation
may also be seen as a special case of a $d$-dimensional continuum
(non-manifold case) where the manifold-operators become classical
ones. In contrast, in the parameter-based formulation, where the $q$-dimensional
manifold results from mapping $q$ parameters to $d$ dimensions,
the related mechanical tensors are $q\times q$ and thus inherently
different for ropes, membranes and continua \cite{Chapelle_2011a}.
For the TDC-based situation, these tensors result from the same formulas,
however, based on different differential operators depending on the
mechanical situation. 
\item The third argument is \emph{numerical} and related to discretization
methods. The new formulation allows for two fundamentally different
numerical approaches. One may be seen as the classical approach which
relies on meshing the ropes and membranes by (curved) line or surface
meshes; this is called the Surface FEM herein. It is inherently linked
to the parameter-based formulation of the mechanical models. Due to
the fact that for implicit geometries, one may also provide surface
meshes, and then continue with the parameter-based formulation, there
was not really a strong need for a more general formulation of the
models until recently. However, there is an alternative approach to
solve ropes and membranes which uses $d$-dimensional, non-matching
background meshes to approximate the displacements instead of conforming,
curved $q$-dimensional surface meshes. This approach has been labeled
Trace FEM because for the integration of the weak form, the $d$-dimensional
shape functions are only evaluated on the trace of the manifold \cite{Olshanskii_2017a,Reusken_2015a,Grande_2018a}.
The Trace FEM is inherently linked to implicitly defined manifolds
and does not need any parametrization. The solution of BVPs based
on non-matching background methods is generally referred to as fictitious
(or embedded) domain methods (FDMs) with a large number of variants
existing today. Recently, the Cut FEM has emerged as a popular FDM
allowing for higher-order accuracy \cite{Burman_2014a,Burman_2010a,Burman_2012a}.
When using the Cut FEM for the solution of PDEs on manifolds as done
herein, the method becomes analogous to the Trace FEM.
\end{enumerate}
Consequently, the TDC-based formulation allows for a unification in
a geometrical (parametric and implicit geometries), mechanical (ropes,
membranes, and $d$-dimensional continua) and numerical (Surface and
Trace FEM) sense. For similar reasons, the authors have already used
the TDC-based approach to reformulate the mechanical models of linear
Kirchhoff-Love \cite{Schoellhammer_2019a,Schoellhammer_2018a,Schoellhammer_2018b}
and Reissner-Mindlin shells \cite{Schoellhammer_2019b}. Using the
TDC for the definition of BVPs on manifolds as discussed herein is
already well-accepted in the definition of transport phenomena on
manifolds where it has replaced the classical parameter-based definition
in many cases \cite{Dziuk_2013a,Fries_2018a,Jankuhn_2017a}. It is
thus coming timely and naturally to reformulate \emph{mechanical}
BVPs on manifolds based on the TDC in the same sense.

The \emph{classical} theory of large displacement membranes based
on curvilinear coordinates is described, e.g., in \cite{Bischoff_2017a,Calladine_1983a,Ciarlet_1997a,Chapelle_2011a}.
The general equivalence of models using parametrized or implicit manifolds
is outlined in \cite{Delfour_1997a}. Geometrically exact shell models
based on explicitly defined surfaces and locally using Cartesian coordinates
are defined in the linear setting in \cite{Simo_1989b} and for the
non-linear case in \cite{Ibrahimbegovic_1993a}. Local Cartesian coordinates
are also used in \cite{Pimenta_2009a}, where the initial configuration
is modeled with a stress-free deformation of a flat surface. Another
approach for explicitly defined geometries is the degenerated solid
approach \cite{Ahmad_1970a}. Concerning a TDC-based modeling of large
deformation membranes with the Surface FEM, we emphasize the work
in \cite{Hansbo_2015a}. A TDC-related approach in the field of composite
structures is presented in \cite{Monteiro_2011a} for embedded membranes
using complex material models and with focus on analytical solutions.
In contrast, herein, the novelty is in the unified treatment of ropes
\emph{and} membranes and the numerical treatment with Surface \emph{and}
Trace FEM. In this work, we also discuss in detail how ropes and membranes
(in two and three dimensions) are generally defined using the parametric
and the implicit approach, including all relevant geometric and differential
quantities. For example, implicitly defining a rope in three dimensions
requires \emph{two} level-set functions whereas one level-set function
is sufficient for membranes. Furthermore, the mechanical discussion
presented herein includes stress and strain tensors, so that the manifold
versions of the first and second Piola-Kirchhoff stress tensors, the
Cauchy stress tensor, the Euler-Almansi and Green-Lagrange strain
tensors are explicitly given.

Concerning the numerical approximation of ropes and membranes based
on the Surface and Trace FEM as discussed herein, it is mentioned
that the Trace FEM just as any other FDM simplifies the meshing of
geometries significantly. However, additional effort is required for
(i) the integration of the weak form, wherefore suitable integration
points on the manifolds have to be provided \cite{Fries_2015a,Omerovic_2016a,Fries_2016b,Fries_2017a},
(ii) the application of boundary conditions which is more involved
because the boundary is within the background elements and instead
of prescribing nodal values one may have to use Nitsche's method \cite{Burman_2012a,Burman_2012b,Hansbo_2002a,Schillinger_2016a,Ruess_2013a}
or other approaches for enforcing constraints, and (iii) stabilization
terms which are necessary to address the existence of shape functions
with small supports on the manifolds and to find unique solutions
with the background mesh although the BVP is only defined on the manifold
\cite{Olshanskii_2017a,Grande_2016a,Burman_2016b,Gross_2018a}. In
spite of the increased effort for implementing the Trace FEM, it may
have significant advantages over the Surface FEM. For example, when
ropes and/or membranes are reinforcing sub-structures embedded in
some three-dimensional continuum, there is no need to consider these
sub-structures in the meshing of the volume. This is the first work
where Trace FEM results for ropes and membranes are shown, enabling
the implicit analysis of these structures.

The paper is organized as follows: In Section \ref{X_Geometry}, the
parametric and implicit geometry definitions of manifolds are discussed
in detail for the situation of ropes and membranes. As usual in finite
strain theory, the undeformed and deformed situations are distinguished,
related by the sought displacements. In each configuration, normal
vectors, surface stretches, and differential surface operators are
defined, wherefore it is important to distinguish parametric from
implicit definitions. The mechanical modeling is outlined in Section
\ref{X_Mechanics} following the classical steps defining stress and
strain tensors and imposing equilibrium in the deformed configuration.
The strong and weak forms of the governing equations are given. It
is noteworthy that ropes and membranes are treated in the same manner
and only the geometry-dependent surface operators defined in Section
\ref{X_Geometry} differ. The numerical solution of the governing
equations is considered in Section \ref{X_Discretization} where the
discrete weak forms of the Surface and Trace FEM are given. The numerical
results in Section \ref{X_NumericalResults} confirm that both numerical
approaches achieve higher-order convergence rates. It is also confirmed
how ropes and membranes may easily be coupled with the presented formulation.
The paper ends in Section \ref{X_Conclusions} with a summary and
conclusions.

\section{Tangential differential calculus in finite strain theory\label{X_Geometry}}

The focus of this work is on ropes (also called cables) and membranes
undergoing large displacements which is covered by finite strain theory.
Only solids according to the Saint Venant--Kirchhoff material model
are considered herein which may be seen as the simplest extension
of a linear elastic material model to large displacements. Cables
may be modeled as one-dimensional lines in the two- or three-dimensional
space. Membranes are two-dimensional surfaces in the three-dimensional
space. Hence, we may state that cables \emph{and} membranes are curved
manifolds with a lower dimension $q$ than the surrounding space with
dimension $d$. In this section, we address the issue of how to define
such manifolds geometrically and formulate (differential) operators
needed later in the mechanical model, see Section \ref{X_Mechanics}.
For the geometry definition, we separately discuss the situation for
parametrized and implicit manifolds. Implicit manifolds are implied
by level-set(s) of scalar function(s) following the concept of the
level-set method \cite{Osher_2003a,Osher_2001a,Sethian_1999b}. For
the \emph{parametric} situation, the outline is related to the classical
setup using tensor notation rather than index notation and avoiding
any explicit reference to classical terms in the context of curvilinear
coordinate systems such as co- and contra-variant base vectors and
Christoffel symbols. For the \emph{implicit} situation, the presented
outline, systematically including all geometric and differential quantitites
for ropes and membranes in finite strain theory, is original.

\subsection{Deformed and undeformed configurations}

As usual in finite strain theory, we consider an undeformed material
configuration and a deformed spatial configuration. These are represented
by the $q$-dimensional manifolds $\Gamma_{\vek X}$ and $\Gamma_{\vek x}$,
respectively, which are immersed in a $d$-dimensional space $\mathbb{R}^{d}$,
herein, $d=\left\{ 2,3\right\} $. The difference $d-q$ is also called
codimension of the manifold. We follow the usual notation to relate
uppercase letters in variable and operator names with the undeformed
configuration and lowercase letters with the deformed one. The displacement
field $\vek u\left(\vek X\right)$ relates the two configurations
via 
\[
\vek x=\vek X+\vek u\left(\vek X\right)\quad\text{with}\;\vek X\in\Gamma_{\vek X}\subset\mathbb{R}^{d}\;\text{and}\;\vek x\in\Gamma_{\vek x}\subset\mathbb{R}^{d}.
\]

\subsection{Parametrized manifolds\label{XX_ParametrizedManifolds}}

For parametrized manifolds, there exists a map $\vek X\!\left(\vek r\right):\mathbb{R}^{q}\rightarrow\mathbb{R}^{d}$
from some lower-dimensional reference domain $\Omega_{\vek r}\subset\mathbb{R}^{q}$
to the undeformed configuration $\Gamma_{\vek X}\subset\mathbb{R}^{d}$.
An important consequence is that local curvilinear coordinate systems
result naturally on the manifolds. It is useful to describe the situation
separately for cables (one-dimensional manifolds) and membranes (two-dimensional
manifolds).

\begin{figure}
\centering

\subfigure[reference domain $\Omega_{\vek r}$]{\includegraphics[width=0.3\textwidth]{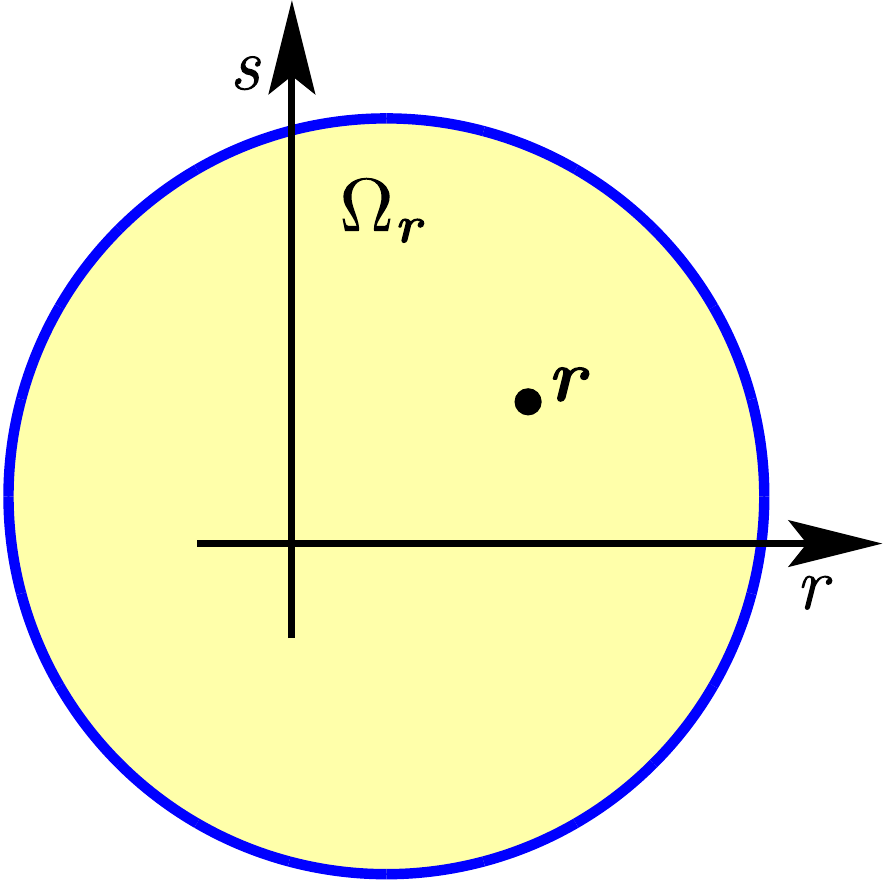}}\qquad\subfigure[def.~and undef.~configuration, $\Gamma_{\vek X}$ and $\Gamma_{\vek x}$]{\includegraphics[width=0.5\textwidth]{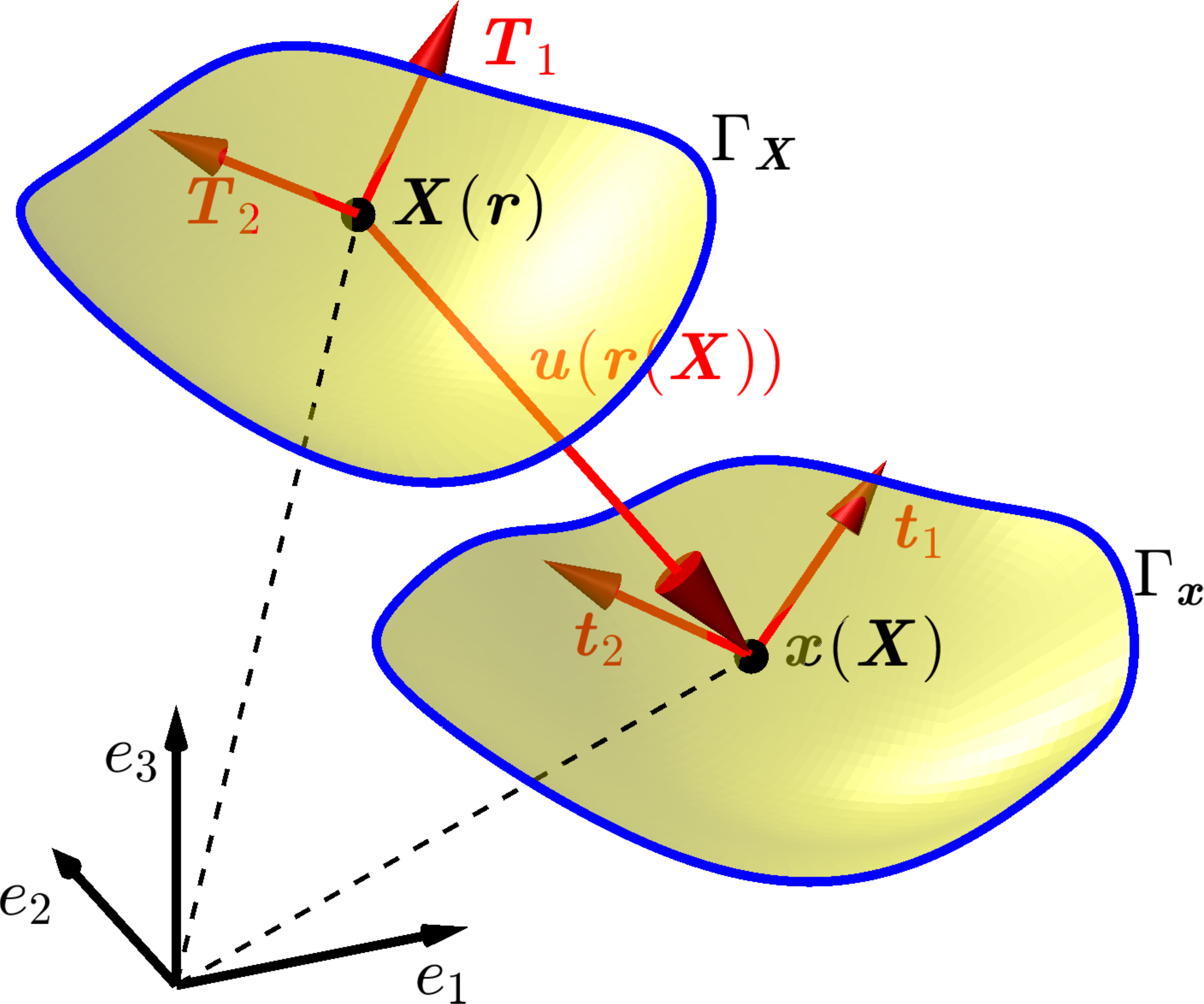}}

\caption{\label{fig:ParamManif2dTo3d}The situation for membranes given by
a parametrization: (a) The reference domain $\Omega_{\vek r}\subset\mathbb{R}^{2}$,
(b) the undeformed domain $\Gamma_{\vek X}$ resulting from a given
parametrization $\vek X\!\left(\vek r\right)$ and the deformed domain
$\Gamma_{\vek x}$ resulting from the deformation $\vek x=\vek X+\vek u$.}
\end{figure}

\subsubsection{Two-dimensional manifolds\label{XXX_ManifoldsIn2d}}

We start with two-dimensional manifolds ($q=2$) in the three-dimensional
space ($d=3$) which is relevant for membranes. Let there be a reference
domain $\Omega_{\vek r}\subset\mathbb{R}^{2}$ and a map $\vek X\!\left(\vek r\right):\Omega_{\vek r}\rightarrow\Gamma_{\vek X}\subset\mathbb{R}^{3}$,
see Fig.~\ref{fig:ParamManif2dTo3d}. We label the components $\vek r=\left[r,s\right]^{\mathrm{T}}$
and $\vek X=\left[X,Y,Z\right]^{\mathrm{T}}$. The Jacobi matrix 
\[
\mat J\left(\vek r\right)=\frac{\partial\vek X\!\left(\vek r\right)}{\partial\vek r}=\nabla_{\vek r}\vek X\!\left(\vek r\right)=\left[\begin{array}{cc}
\partial_{r}X & \partial_{s}X\\
\partial_{r}Y & \partial_{s}Y\\
\partial_{r}Z & \partial_{s}Z
\end{array}\right]
\]
has dimensions $\left(3\times2\right)$. One may easily obtain two
vectors $\vek T_{1}^{\star}=\frac{\partial\vek X}{\partial r}$ and
$\vek T_{2}^{\star}=\frac{\partial\vek X}{\partial s}$ from the columns
of $\mat J$ being tangential to $\Gamma_{\vek X}$ at a mapped point
$\vek X$. The Jacobi matrix is also used to compute the first fundamental
form $\mat G$ and the operator $\mat Q$, later needed for defining
surface gradients,
\begin{eqnarray}
\mat G=\mat J^{\mathrm{T}}\cdot\mat J & \;(q\times q)\text{-matrix},\qquad & \mat Q=\mat J\cdot\mat G^{-1}\;(d\times q)\text{-matrix}.\label{eq:OperatorsGandQ}
\end{eqnarray}

Next, we consider a displacement field $\vek u\left(\vek r\right)$
assuming that a point $\vek r$ is given which may also be seen as
a function $\vek u\left(\vek r\left(\vek X\right)\right)$ when a
point $\vek X\in\Gamma_{\vek X}$ is given (and back-projected to
the reference domain inverting the map $\vek X\!\left(\vek r\right)$).
We emphasize that in both cases, the displacement field only lives
on the manifold $\Gamma_{\vek X}$. Hence, no classical partial derivatives
with respect to $\vek X$ may be computed (unless $\vek u$ is smoothly
extended to some neighborhood of $\vek X$ which is not unique and
not considered here) so that the only useful gradient of $\vek u$
in this context is the \emph{surface gradient}. For some scalar function
$f\!\left(\vek r\right)$, e.g., each displacement component, the
surface gradient is
\[
\nabla_{\vek X}^{\Gamma}f\!\left(\vek r\right)=\mat Q\cdot\nabla_{\vek r}f\!\left(\vek r\right)\quad\Leftrightarrow\quad\left[\begin{array}{c}
\partial_{X}^{\Gamma}f\\
\partial_{Y}^{\Gamma}f\\
\partial_{Z}^{\Gamma}f
\end{array}\right]=\mat Q\cdot\left[\begin{array}{c}
\partial_{r}f\\
\partial_{s}f
\end{array}\right],
\]
For a vector function $\vek u\left(\vek r\right)=\left[u,v,w\right]^{\mathrm{T}}\in\mathbb{R}^{3}$,
we have the \emph{directional} surface gradient
\begin{equation}
\nabla_{\vek X}^{\Gamma,\mathrm{dir}}\vek u\left(\vek r\right)=\left[\begin{array}{c}
\left(\nabla_{\vek X}^{\Gamma}u\right)^{\mathrm{T}}\\
\left(\nabla_{\vek X}^{\Gamma}v\right)^{\mathrm{T}}\\
\left(\nabla_{\vek X}^{\Gamma}w\right)^{\mathrm{T}}
\end{array}\right]=\left[\begin{array}{ccc}
\partial_{X}^{\Gamma}u & \partial_{Y}^{\Gamma}u & \partial_{Z}^{\Gamma}u\\
\partial_{X}^{\Gamma}v & \partial_{Y}^{\Gamma}v & \partial_{Z}^{\Gamma}v\\
\partial_{X}^{\Gamma}w & \partial_{Y}^{\Gamma}w & \partial_{Z}^{\Gamma}w
\end{array}\right]=\nabla_{\vek r}\vek u\left(\vek r\right)\cdot\mat Q^{\mathrm{T}},\label{eq:SurfGradVectorParametric}
\end{equation}
which is to be distinguished from the \emph{covariant} surface gradient
of a vector field defined in Section \ref{XXX_CovariantSurfaceGradient}.

Let us next consider the map from the undeformed to the deformed configuration
$\vek x\left(\vek X\right)=\vek X+\vek u\left(\vek X\right)$ which
is $\Gamma_{\!\vek X}\rightarrow\mathbb{R}^{d}$. The related Jacobi
matrix is also called \emph{surface deformation gradient}, 
\begin{equation}
\mat F_{\Gamma}=\nabla_{\vek X}^{\Gamma,\mathrm{dir}}\vek x\left(\vek X\right)=\mat I+\nabla_{\vek X}^{\Gamma,\mathrm{dir}}\vek u\left(\vek X\right),\label{eq:SurfaceDeformationGradient}
\end{equation}
where $\mat I$ is a $(d\times d)$ identity matrix.

One may obtain all equivalent quantities in the \emph{deformed} configuration:
The Jacobi-matrix from the reference to the deformed configuration
$\mat j=\mat F_{\Gamma}\cdot\mat J$ and the tangent vectors $\vek t_{1}^{\star}$,
$\vek t_{2}^{\star}$ to the deformed configuration $\Gamma_{\vek x}$
at a mapped point $\vek x$ based on the Jacobi matrix $\mat j$.
Furthermore, the first fundamental form $\mat g=\mat j^{\mathrm{T}}\cdot\mat j$
and the operator $\mat q=\mat j\cdot\mat g^{-1}$ relating the classical
gradient of the reference configuration with the surface gradient
of the deformed configuration as $\nabla_{\vek x}^{\Gamma}f=\mat q\cdot\nabla_{\vek r}f$.

Based on the pairs of tangent vectors in the undeformed and deformed
configuration, one may compute unique normal vectors in each configuration,
$\vek N^{\star}=\vek T_{1}^{\star}\times\vek T_{2}^{\star}$ and $\vek n^{\star}=\vek t_{1}^{\star}\times\vek t_{2}^{\star}$.
Then, the projectors $\mat P\left(\vek X\right)$ and $\mat p\left(\vek x\right)$
are computed as
\begin{eqnarray}
\mat P & = & \mat I-\vek N\otimes\vek N\quad\text{with}\;\vek N=\frac{\vek N^{\star}}{\left\Vert \vek N^{\star}\right\Vert },\label{eq:ProjectorUndefNormal}\\
\mat p & = & \mat I-\vek n\otimes\vek n\quad\text{with}\;\vek n=\frac{\vek n^{\star}}{\left\Vert \vek n^{\star}\right\Vert }.\label{eq:ProjectorDefNormal}
\end{eqnarray}
The same result is obtained when computing a tangent vector $\vek T_{3}^{\star}$
in the tangent plane spanned by $\vek T_{1}^{\star}$ and $\vek T_{2}^{\star}$
being orthogonal to $\vek T_{1}^{\star}$ using Gram Schmidt orthogonalization,
$\vek T_{3}^{\star}=\vek T_{2}^{\star}-\frac{\vek T_{1}^{\star}\cdot\vek T_{2}^{\star}}{\vek T_{1}^{\star}\cdot\vek T_{1}^{\star}}\cdot\vek T_{1}^{\star}$,
then 
\[
\mat P=\vek T_{1}\otimes\vek T_{1}+\vek T_{3}\otimes\vek T_{3}\quad\text{with}\;\vek T_{1}=\frac{\vek T_{1}^{\star}}{\left\Vert \vek T_{1}^{\star}\right\Vert },\vek T_{3}=\frac{\vek T_{3}^{\star}}{\left\Vert \vek T_{3}^{\star}\right\Vert },
\]
analogously for $\mat p$. The projector $\mat P$ at some point $\vek X$
maps an arbitrary vector in $\mathbb{R}^{d}$ to the tangent space
at $\Gamma_{\vek X}$, hence, $\mat P\cdot\vek N=\vek0$. $\mat P$
is symmetric, $\mat P=\mat P^{\mathrm{T}}$, and idempotent, $\mat P\cdot\mat P=\mat P$,
which holds analogously for $\mat p$.

Next, we are interested in the stretch of a differential element of
the membrane when undergoing the deformation. This is interpreted
as an area stretch and defined as
\[
\Lambda=\frac{\sqrt{\det\mat g}}{\sqrt{\det\mat G}}=\frac{\left\Vert \vek n^{\star}\right\Vert }{\left\Vert \vek N^{\star}\right\Vert }=\frac{\left\Vert \vek t_{1}^{\star}\times\vek t_{2}^{\star}\right\Vert }{\left\Vert \vek T_{1}^{\star}\times\vek T_{2}^{\star}\right\Vert }.
\]

Finally, an operator $\mat W$ is introduced which relates surface
gradients of the undeformed and the deformed configuration as
\begin{equation}
\nabla_{\vek x}^{\Gamma}f=\mat W\cdot\nabla_{\vek X}^{\Gamma}f\quad\text{with}\;\mat W=\mat q\left(\mat Q^{\mathrm{T}}\cdot\mat Q\right)^{-1}\mat Q^{\mathrm{T}}.\label{eq:SurfGradUndefToDefParam}
\end{equation}
This result is obtained using $\nabla_{\vek X}^{\Gamma}f=\mat Q\cdot\nabla_{\vek r}f$
and $\nabla_{\vek x}^{\Gamma}f=\mat q\cdot\nabla_{\vek r}f$. Note
that $\mat Q$ and $\mat q$ are $\left(d\times q\right)$-matrices,
hence, not quadratic and the concept of generalized inverses (or pseudo
inverses) is needed to obtain Eq.~(\ref{eq:SurfGradUndefToDefParam}).

\begin{figure}
\centering

\subfigure[reference domain $\Omega_{r}$]{\includegraphics[width=0.4\textwidth]{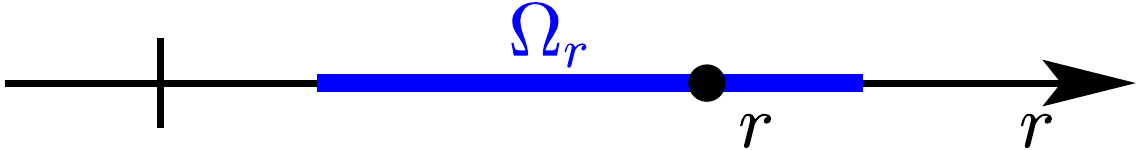}}\qquad\subfigure[def.~and undef.~config., $\Gamma_{\vek X}$ and $\Gamma_{\vek x}$]{\includegraphics[width=0.4\textwidth]{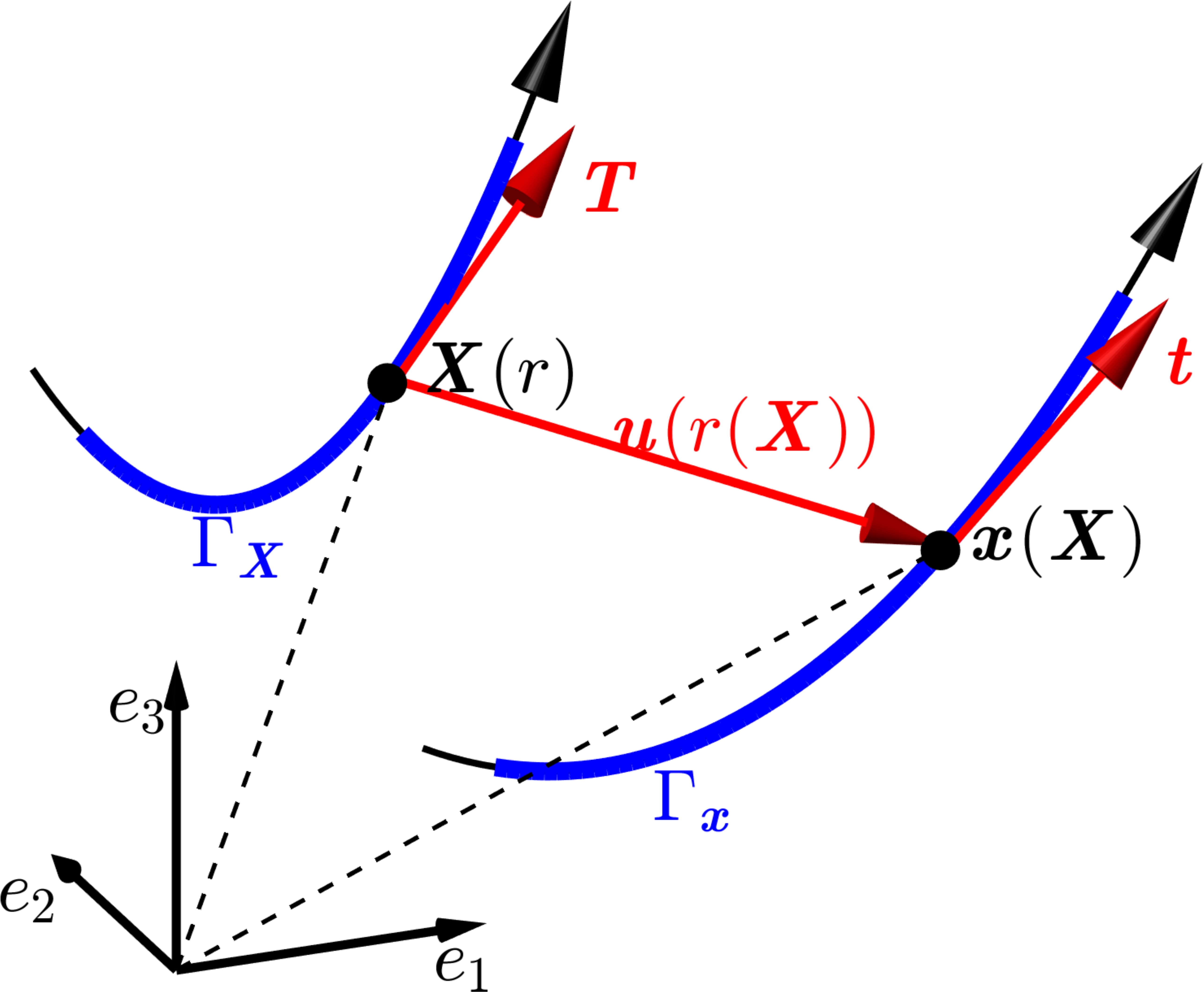}}

\caption{\label{fig:ParamManif1dTo3d}The situation for cables given by a parametrization:
(a) The reference domain $\Omega_{r}\subset\mathbb{R}$, (b) the undeformed
domain $\Gamma_{\vek X}$ resulting from a given parametrization $\vek X\!\left(r\right)$
and the deformed domain $\Gamma_{\vek x}$ resulting from the deformation
$\vek x=\vek X+\vek u$.}
\end{figure}

\subsubsection{One-dimensional manifolds}

Consider some one-dimensional reference domain $\Omega_{r}\subset\mathbb{R}$
and a map $\vek X\!\left(r\right):\Omega_{r}\rightarrow\Gamma_{\vek X}\subset\mathbb{R}^{d}$,
$d=\left\{ 2,3\right\} $, see Fig.~\ref{fig:ParamManif1dTo3d}.
This situation applies to cables in two and three dimensions. The
Jacobi matrix $\mat J\left(r\right)=\nabla_{r}\vek X\!\left(r\right)$
consists of one column only which implies a tangent vector $\vek T\in\mathbb{R}^{d}$
being tangential to the undeformed configuration at $\vek X$. For
the tangent vector in the deformed configuration follows $\vek t^{\star}=\mat F_{\Gamma}\cdot\vek T^{\star}$.
Most parts of the discussion from Section \ref{XXX_ManifoldsIn2d}
apply accordingly. However, the definition of the projectors changes
to 
\begin{eqnarray}
\mat P & = & \vek T\otimes\vek T\quad\text{with}\;\vek T=\frac{\vek T^{\star}}{\left\Vert \vek T^{\star}\right\Vert },\label{eq:ProjectorUndefTang}\\
\mat p & = & \vek t\otimes\vek t\quad\text{with}\;\vek t=\frac{\vek t^{\star}}{\left\Vert \vek t^{\star}\right\Vert }.\label{eq:ProjectorDefTang}
\end{eqnarray}
For the stretch of a differential element of the cable, which may
be seen as a line stretch, there follows
\[
\Lambda=\frac{\sqrt{\det\mat g}}{\sqrt{\det\mat G}}=\frac{\left\Vert \vek t^{\star}\right\Vert }{\left\Vert \vek T^{\star}\right\Vert }.
\]

\subsection{Implicit manifolds\label{XX_ImplicitManifolds}}

Implicit manifolds are implied by one or more level-set functions.
Generally speaking, the codimension determines the number of level-set
functions required to define a unique geometry of a manifold. For
the cases relevant in this work, this means that \emph{one} level-set
function $\phi\left(\vek X\right)$ is required for cables in $\mathbb{R}^{2}$
and membranes in $\mathbb{R}^{3}$ which have codimension $1$. For
cables in $\mathbb{R}^{3}$, on the other hand, \emph{two} level-set
functions $\phi_{1}\left(\vek X\right)$ and $\phi_{2}\left(\vek X\right)$
are needed. We split the discussion depending on the codimension.

\begin{figure}
\centering

\subfigure[]{\includegraphics[width=0.4\textwidth]{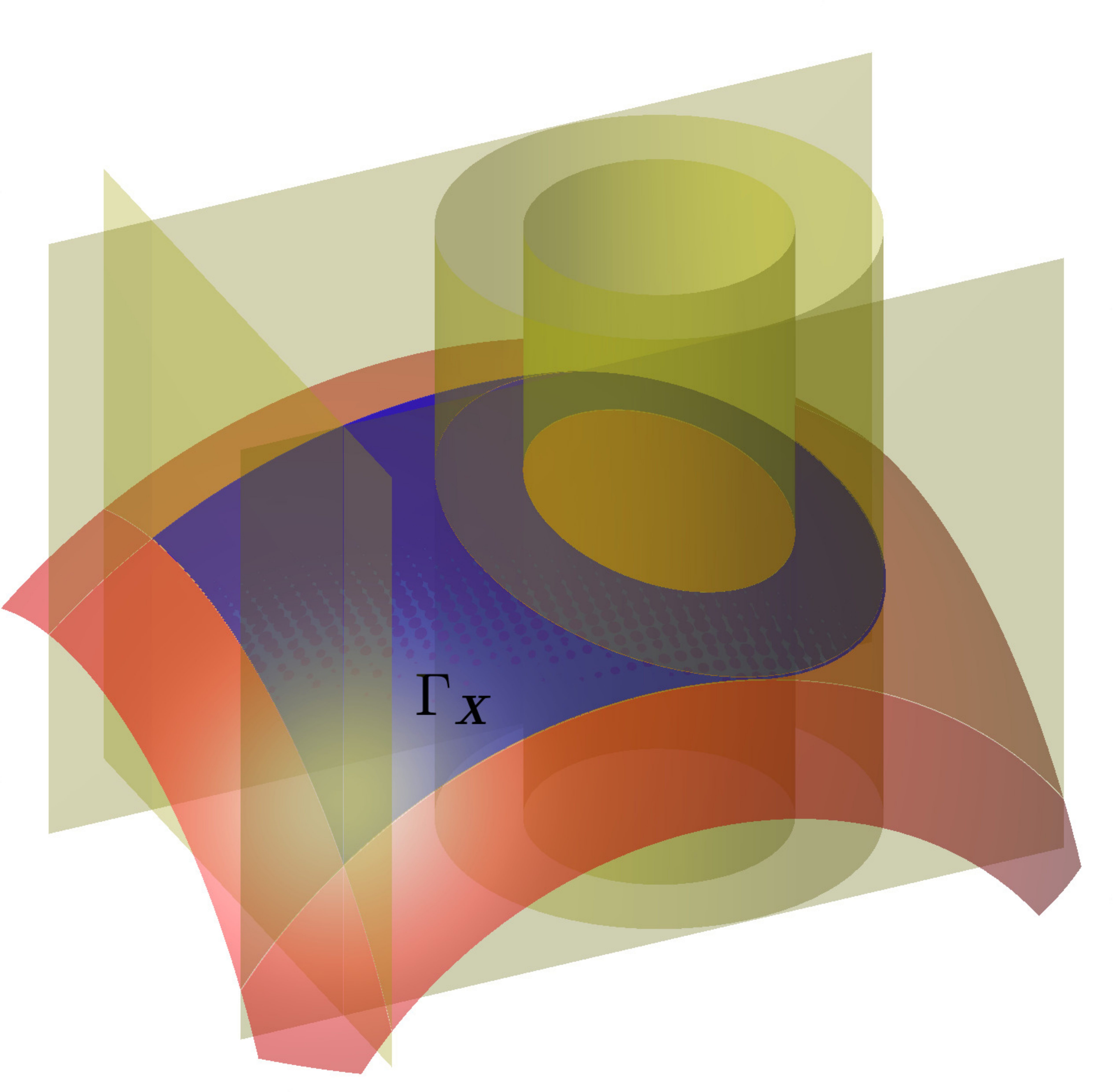}}\qquad\subfigure[]{\includegraphics[width=0.4\textwidth]{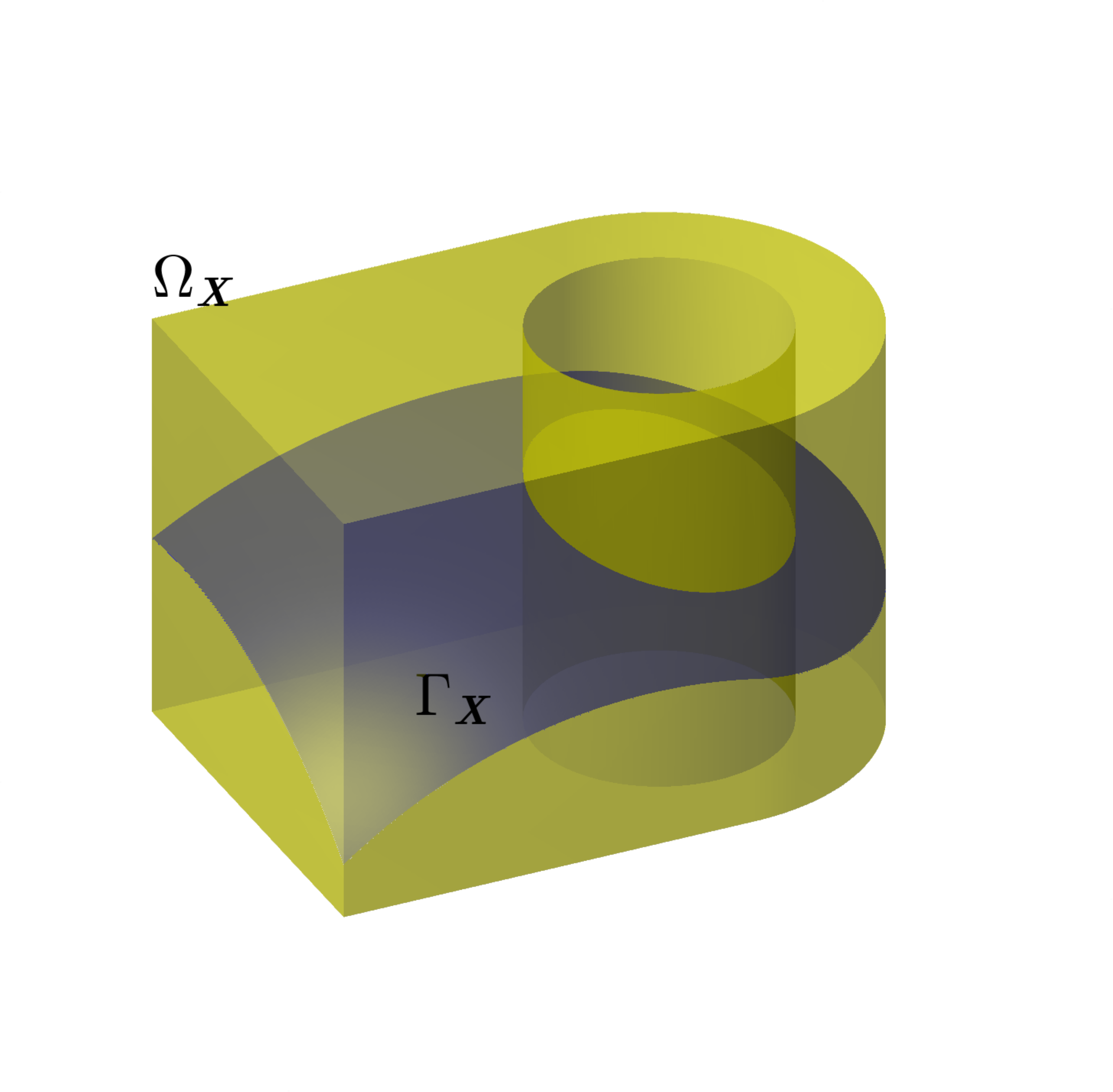}}

\caption{\label{fig:ManifoldDefinition}Two different implicit definitions
of the same manifold (blue surface) in $\mathbb{R}^{3}$ with codimension
$1$: (a) The manifold is defined by one (master) level-set $\phi\left(\vek X\right)$
(red surface) which is bounded by additional (slave) level-set functions
$\psi_{i}\left(\vek X\right)$ (yellow surfaces). (b) The manifold
is defined by \emph{one} level-set $\phi\left(\vek X\right)$ which
is evaluated in the domain of definition $\Omega_{\vek X}$ (yellow
body).}
\end{figure}

\subsubsection{Manifolds with codimension $1$}

Oriented manifolds with codimension $1$ may be defined by \emph{one}
level-set function $\phi\left(\vek X\right)$. Usually, the \emph{zero}-level
set of $\phi$ implies the manifold of interest and there are infinitely
many possible $\phi$ implying the same geometry. The signed distance
function is a particularly useful concrete example for $\phi\left(\vek X\right)$
and often used in practice. It is noteworthy that many level-sets
are \emph{unbounded} in $\mathbb{R}^{d}$ which is not desirable for
the definition of mechanical applications. Fortunately, it is easily
possible to define bounded manifolds by additional (slave) level-set
functions $\psi_{i}\left(\vek X\right)$, see \cite{Fries_2017c,Fries_2017d}.
Then, the undeformed configuration is, e.g., given by 
\begin{equation}
\Gamma_{\vek X}=\left\{ \vek X\in\mathbb{R}^{d}:\phi\left(\vek X\right)=0,\;\psi_{i}\left(\vek X\right)\geq0,\;i=1,\dots n\right\} ,\label{eq:ImplicitDef1}
\end{equation}
as shown in Fig.~\ref{fig:ManifoldDefinition}(a). Even simpler is
to associate a domain of definition $\Omega_{\vek X}\subset\mathbb{R}^{d}$
with $\phi$, then the bounded manifold results as
\begin{equation}
\Gamma_{\vek X}=\left\{ \vek X\in\Omega_{\vek X}:\phi\left(\vek X\right)=0\right\} .\label{eq:ImplicitDef2}
\end{equation}
In this case, the boundaries are the intersections of the zero-level
set with the boundary of the domain of definition, see Fig.~\ref{fig:ManifoldDefinition}(b).
The implicit definition according to Eq.~(\ref{eq:ImplicitDef2})
is used in the following unless noted otherwise.

\begin{figure}
\centering

\includegraphics[width=0.5\textwidth]{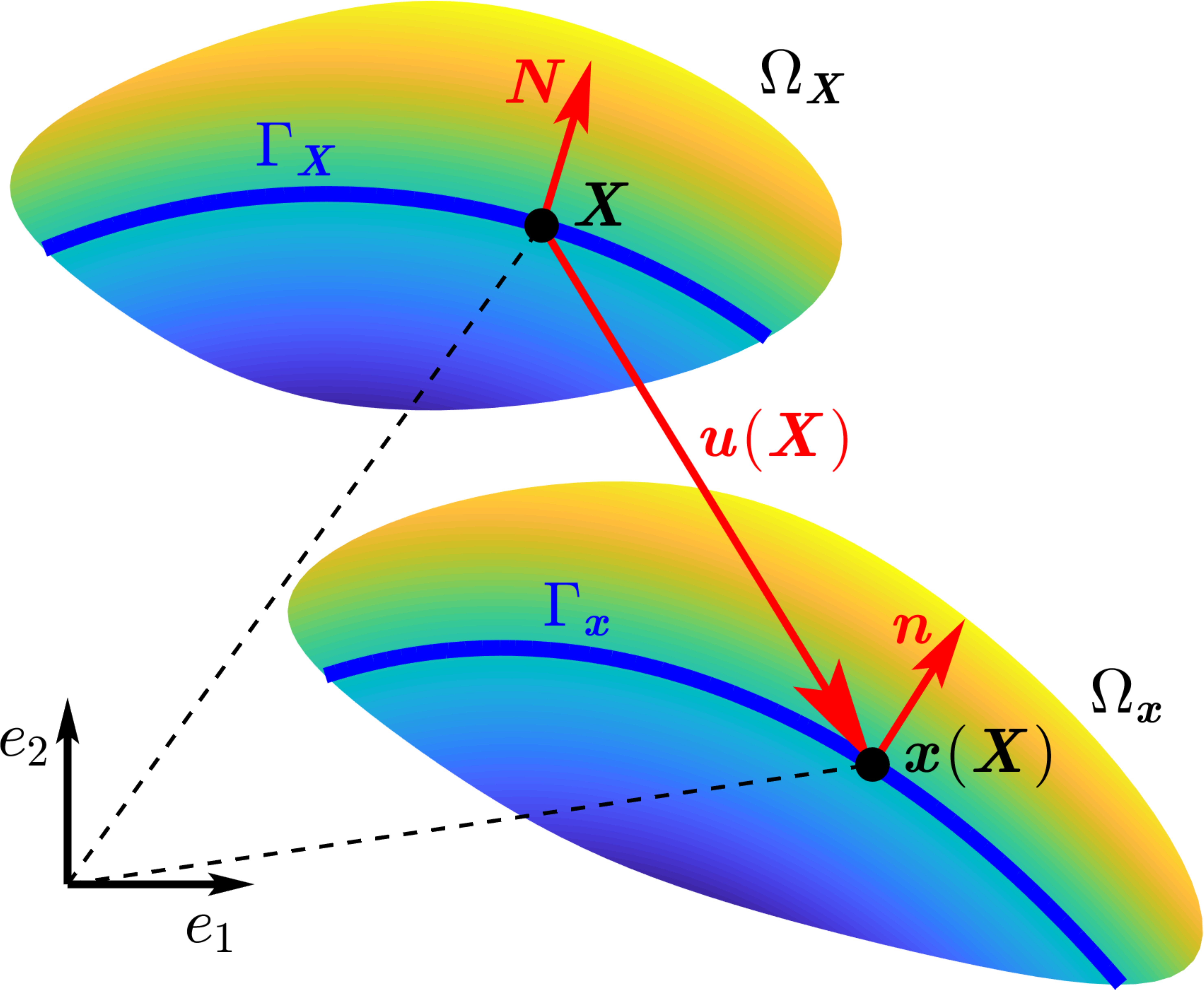}

\caption{\label{fig:ImplManifold1dTo2d}An implicitly defined manifold in $\mathbb{R}^{2}$
with codimension $1$, e.g., a cable in two dimensions: The undeformed
situation is implied by the zero-level set of $\phi\left(\vek X\right)$
(blue line), evaluated in the domain of definition $\Omega_{\vek X}$
(colored region). The deformed configuration results from the displacement
field $\vek u\left(\vek X\right)$ and $\vek x=\vek X+\vek u$.}
\end{figure}

Next, we focus on the situation in large displacement theory as shown
in Fig.~\ref{fig:ImplManifold1dTo2d} for this implicit setup. The
normal vector of the undeformed configuration is obtained by the gradient
of the level-set function,
\[
\vek N^{\star}\left(\vek X\right)=\nabla_{\vek X}\phi\left(\vek X\right)\quad\text{for}\;\vek X\in\Gamma_{\vek X}.
\]

Let there be a displacement field $\vek u\left(\vek X\right)$ which
lives in the full $d$-dimensional space (instead of only the manifold
itself as for parametric manifolds) so that the classical gradient
$\nabla_{\vek X}\vek u\left(\vek X\right)$ is available. The resulting
deformation gradient is
\begin{equation}
\mat F_{\Omega}=\nabla_{\vek X}\vek x\left(\vek X\right)=\mat I+\nabla_{\vek X}\vek u\left(\vek X\right)\label{eq:ClassicalDeformationGradient}
\end{equation}
which is different from the \emph{surface} deformation gradient $\mat F_{\Gamma}$
in Eq.~(\ref{eq:SurfaceDeformationGradient}). Based on this, one
may compute the normal vector of the deformed configuration at $\vek x=\vek X+\vek u\left(\vek X\right)$
as
\[
\vek n^{\star}\left(\vek x\right)=\nabla_{\vek x}\phi\left(\vek X\left(\vek x\right)\right)=\mat F_{\Omega}^{-\mathrm{T}}\cdot\vek N^{\star}\quad\text{for}\;\vek x\in\Gamma_{\vek x},
\]
which follows by the chain rule. Eqs.~(\ref{eq:ProjectorUndefNormal})
and (\ref{eq:ProjectorDefNormal}) are used to compute the projectors
$\mat P\left(\vek X\right)$ and $\mat p\left(\vek x\right)$, respectively.
The \emph{surface} gradient (with respect to the undeformed configuration)
of a scalar function $f\left(\vek X\right)$ with $\vek X\in\mathbb{R}^{d}$
results as
\begin{equation}
\nabla_{\vek X}^{\Gamma}f=\mat P\cdot\nabla_{\vek X}f.\label{eq:SurfGradScalarImplicit}
\end{equation}
As before, $\nabla_{\vek X}f$ is the classical gradient in the $d$-dimensional
space. The situation is analogous for each component $u_{i}$ of a
vector function $\vek u\left(\vek X\right)$, so that one obtains
for the \emph{directional} surface gradient
\begin{eqnarray}
\nabla_{\vek X}^{\Gamma,\mathrm{dir}}\vek u & = & \nabla_{\vek X}\vek u\cdot\mat P,\label{eq:SurfGradVectorImplicit}\\
\text{for }\vek u\in\mathbb{R}^{3}:\left[\begin{array}{ccc}
\partial_{X}^{\Gamma}u & \partial_{Y}^{\Gamma}u & \partial_{Z}^{\Gamma}u\\
\partial_{X}^{\Gamma}v & \partial_{Y}^{\Gamma}v & \partial_{Z}^{\Gamma}v\\
\partial_{X}^{\Gamma}w & \partial_{Y}^{\Gamma}w & \partial_{Z}^{\Gamma}w
\end{array}\right] & = & \left[\begin{array}{ccc}
\partial_{X}u & \partial_{Y}u & \partial_{Z}u\\
\partial_{X}v & \partial_{Y}v & \partial_{Z}v\\
\partial_{X}w & \partial_{Y}w & \partial_{Z}w
\end{array}\right]\cdot\left[\begin{array}{ccc}
P_{11} & P_{12} & P_{13}\\
P_{12} & P_{22} & P_{23}\\
P_{13} & P_{23} & P_{33}
\end{array}\right].\nonumber 
\end{eqnarray}
The \emph{surface} deformation gradient $\mat F_{\Gamma}$ follows
using Eq.~(\ref{eq:SurfaceDeformationGradient}). The stretch of
a differential element upon the deformation is 
\[
\Lambda=\frac{\left\Vert \nabla_{\vek x}\phi\right\Vert }{\left\Vert \nabla_{\vek X}\phi\right\Vert }\cdot\det\mat F_{\Omega}=\frac{\left\Vert \vek n^{\star}\right\Vert }{\left\Vert \vek N^{\star}\right\Vert }\cdot\det\mat F_{\Omega}.
\]
Finally, for the operator $\mat W$ relating surface gradients of
the undeformed and deformed configuration, one obtains
\begin{equation}
\nabla_{\vek x}^{\Gamma}f=\mat W\cdot\nabla_{\vek X}^{\Gamma}f\quad\text{with}\;\mat W=\mat p\cdot\mat F_{\Omega}^{-\mathrm{T}}.\label{eq:SurfGradUndefToDefImpl}
\end{equation}
Note that $\mat W\cdot\mat P=\mat W$, hence, when using the classical
derivatives one obtains $\nabla_{\vek x}^{\Gamma}f=\mat W\cdot\nabla_{\vek X}f$.

\subsubsection{Manifolds with codimension $2$}

\begin{figure}
\centering

\includegraphics[width=0.7\textwidth]{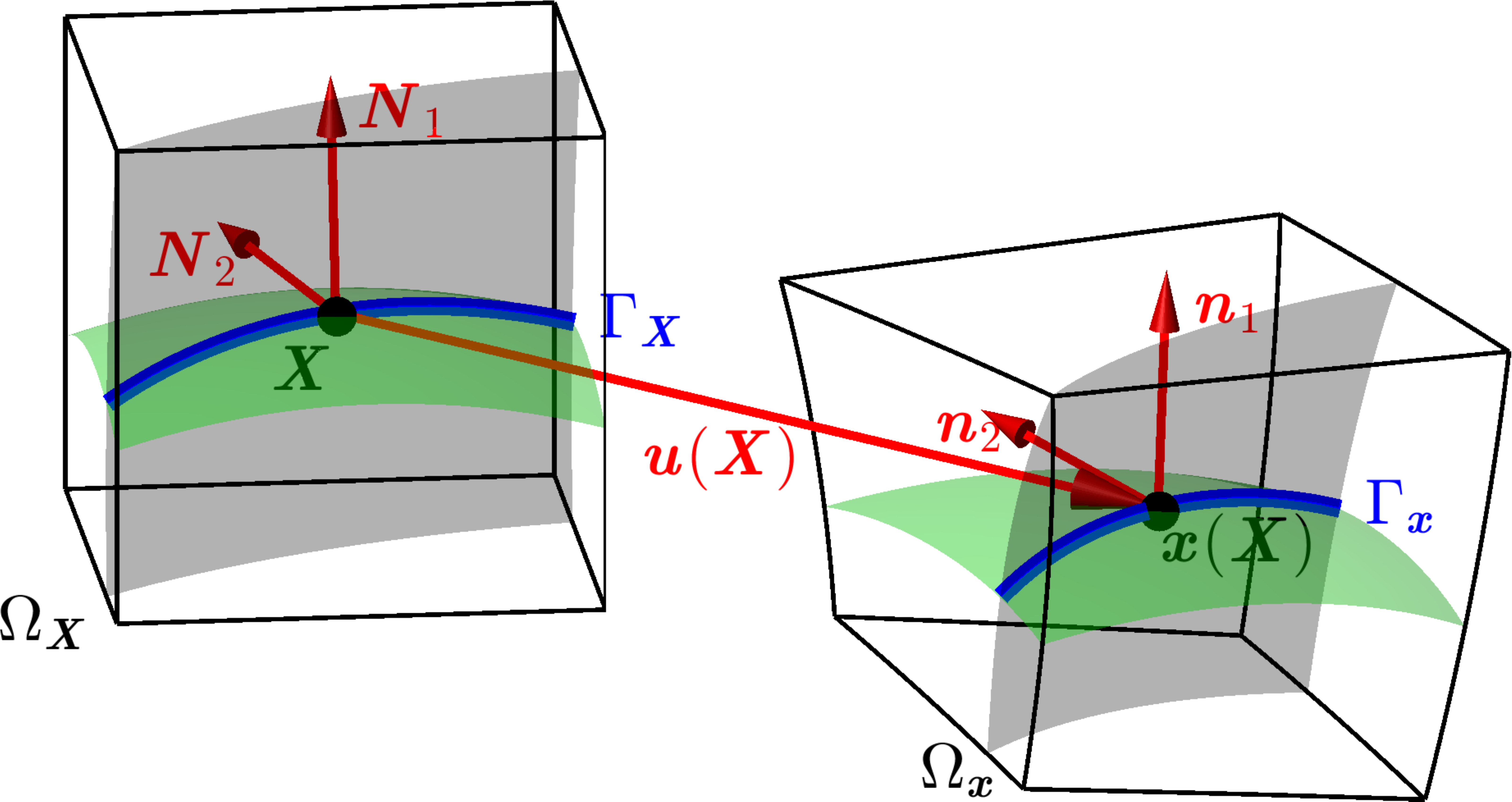}

\caption{\label{fig:ImplManifold1dTo3d}An implicitly defined manifold in $\mathbb{R}^{3}$
with codimension $2$, e.g., a cable in three dimensions: The undeformed
situation is implied by the zero-level sets of $\phi_{1}\left(\vek X\right)$
and $\phi_{2}\left(\vek X\right)$ (gray and yellow surfaces), evaluated
in the domain of definition $\Omega_{\vek X}$ (black box). The deformed
configuration results from the displacement field $\vek u\left(\vek X\right)$
and $\vek x=\vek X+\vek u$.}
\end{figure}

The focus is on one-dimensional manifolds in $\mathbb{R}^{3}$ such
as cables in the three-dimensional space. As mentioned before, two
level-set functions $\phi_{1}$ and $\phi_{2}$ are needed for the
geometry definition,
\[
\Gamma_{\vek X}=\left\{ \vek X\in\Omega_{\vek X}:\phi_{1}\left(\vek X\right)=0\text{ and }\phi_{2}\left(\vek X\right)=0\right\} ,
\]
see Fig.~\ref{fig:ImplManifold1dTo3d}. The $d$-dimensional displacement
field $\vek u$ is given as before with classical derivatives $\nabla_{\vek X}\vek u\left(\vek X\right)$
and the related deformation gradient $\mat F_{\Omega}$ as in Eq.~(\ref{eq:ClassicalDeformationGradient}).

The two normal vectors associated to the deformed and undeformed configuration
each, are given for $i=\left\{ 1,2\right\} $ as
\begin{eqnarray*}
\vek N_{i}^{\star} & = & \nabla_{\vek X}\phi_{i}\left(\vek X\right)\quad\text{for}\;\vek X\in\Gamma_{\vek X},\\
\vek n_{i}^{\star} & = & \nabla_{\vek x}\phi_{i}\left(\vek X\left(\vek x\right)\right)=\mat F_{\Omega}^{-\mathrm{T}}\cdot\vek N_{i}^{\star}\quad\text{for}\;\vek x\in\Gamma_{\vek x}.
\end{eqnarray*}
One may then compute the unique tangent vectors 
\begin{eqnarray*}
\vek T^{\star} & = & \vek N_{2}^{\star}\times\vek N_{1}^{\star},\\
\vek t^{\star} & = & \vek n_{2}^{\star}\times\vek n_{1}^{\star}.
\end{eqnarray*}
The projectors $\mat P$ and $\mat p$ follow using Eqs.~(\ref{eq:ProjectorUndefTang})
and (\ref{eq:ProjectorDefTang}), respectively. It is noted that the
same projector $\mat P$ is obtained when using $\vek N_{1}^{\star}$
and the orthogonalized normal vector $\vek N_{3}^{\star}=\vek N_{2}^{\star}-\frac{\vek N_{1}^{\star}\cdot\vek N_{2}^{\star}}{\vek N_{1}^{\star}\cdot\vek N_{1}^{\star}}\cdot\vek N_{1}^{\star}$
and
\[
\mat P=\mat I-\vek N_{1}\otimes\vek N_{1}-\vek N_{3}\otimes\vek N_{3}\quad\text{with}\;\vek N_{1}=\frac{\vek N_{1}^{\star}}{\left\Vert \vek N_{1}^{\star}\right\Vert },\vek N_{3}=\frac{\vek N_{3}^{\star}}{\left\Vert \vek N_{3}^{\star}\right\Vert },
\]
analogously for $\mat p$. As before, these projectors are used to
determine surface gradients of scalar and vector functions as in Eqs.~(\ref{eq:SurfGradScalarImplicit})
and (\ref{eq:SurfGradVectorImplicit}). The line stretch is given
as
\[
\Lambda=\frac{\left\Vert \vek t^{\star}\right\Vert }{\left\Vert \vek T^{\star}\right\Vert }\cdot\det\mat F_{\Omega}=\frac{\left\Vert \vek n_{2}^{\star}\times\vek n_{1}^{\star}\right\Vert }{\left\Vert \vek N_{2}^{\star}\times\vek N_{1}^{\star}\right\Vert }\cdot\det\mat F_{\Omega}.
\]

\subsection{Similarities and differences in the parametric and implicit descriptions}

In this section, purely based on geometric considerations related
to the undeformed and deformed situation, a number of useful quantities
and operators are given following the concept of the TDC. The situation
is summarized in Table \ref{tab:ParametricManifolds} for parametric
manifolds and in Table \ref{tab:ImplicitManifolds} for implicit ones.
For the mathematical equivalence of these two descriptions and more
details, see, e.g., \cite{Dziuk_2013a}. The order of the rows in
the tables is determined by the information available for parametric
and implicit manifolds. Thereby, it is made sure that the (geometric
and differential) quantities of interest may be computed in this order.

\begin{table}
\begin{tabular}{|c|c|c|}
\hline 
 & $\begin{array}{c}
\text{cables in }\mathbb{R}^{d}\text{ with }d=\left\{ 2,3\right\} \\
\text{(one-dimensional manifolds)}
\end{array}$  & $\begin{array}{c}
\text{membranes in }\mathbb{R}^{3}\\
\text{(two-dimensional manifolds)}
\end{array}$\tabularnewline
\hline 
\hline 
undeformed config.~$\Gamma_{\vek X}$ & $\vek X\!\left(r\right):\Omega_{r}\subset\mathbb{R}\rightarrow\Gamma_{\vek X}\subset\mathbb{R}^{d}$ & $\vek X\!\left(\vek r\right):\Omega_{\vek r}\subset\mathbb{R}^{2}\rightarrow\Gamma_{\vek X}\subset\mathbb{R}^{3}$\tabularnewline
\hline 
$\begin{array}{c}
\text{Jacobi matrix w.r.t. }\vek X\!\left(\vek r\right)\\
\text{and auxiliary operators}
\end{array}$ & \multicolumn{2}{c|}{$\mat J\left(\vek r\right)=\nabla_{\vek r}\vek X\!\left(\vek r\right),\;\mat G=\mat J^{\mathrm{T}}\cdot\mat J,\;\mat Q=\mat J\cdot\mat G^{-1}$}\tabularnewline
\hline 
$\begin{array}{c}
\text{surface gradients}\\
\text{w.r.t.\,}\Gamma_{\vek X}
\end{array}$ & \multicolumn{2}{c|}{$\nabla_{\vek X}^{\Gamma}f\!\left(\vek r\right)=\mat Q\cdot\nabla_{\vek r}f\!\left(\vek r\right),\;\nabla_{\vek X}^{\Gamma,\mathrm{dir}}\vek u=\nabla_{\vek r}\vek u\cdot\mat Q^{\mathrm{T}}$}\tabularnewline
\hline 
$\begin{array}{c}
\text{surface deformation}\\
\text{gradient\,}\mat F_{\Gamma}
\end{array}$ & \multicolumn{2}{c|}{$\mat F_{\Gamma}=\nabla_{\vek X}^{\Gamma,\mathrm{dir}}\vek x\left(\vek X\right)=\mat I+\nabla_{\vek X}^{\Gamma,\mathrm{dir}}\vek u$}\tabularnewline
\hline 
deformed config.~$\Gamma_{\vek x}$ & \multicolumn{2}{c|}{$\vek x\left(\vek X\right)=\vek X+\vek u\left(\vek X\right)\Leftrightarrow\vek x\left(\vek r\right)=\vek X\!\left(\vek r\right)+\vek u\left(\vek r\right)$}\tabularnewline
\hline 
$\begin{array}{c}
\text{Jacobi matrix w.r.t. }\vek x\left(\vek r\right)\\
\text{and help operators}
\end{array}$ & \multicolumn{2}{c|}{$\mat j\left(\vek r\right)=\nabla_{\vek r}\vek x\!\left(\vek r\right)=\mat F_{\Gamma}\cdot\mat J,\;\mat g=\mat j^{\mathrm{T}}\cdot\mat j,\;\mat q=\mat j\cdot\mat g^{-1}$}\tabularnewline
\hline 
$\begin{array}{c}
\text{surface gradients}\\
\text{w.r.t.\,}\Gamma_{\vek x}
\end{array}$ & \multicolumn{2}{c|}{$\nabla_{\vek x}^{\Gamma}f\!\left(\vek r\right)=\mat q\cdot\nabla_{\vek r}f\!\left(\vek r\right),\;\nabla_{\vek x}^{\Gamma,\mathrm{dir}}\vek u=\nabla_{\vek r}\vek u\cdot\mat q^{\mathrm{T}}$}\tabularnewline
\hline 
$\begin{array}{c}
\text{tangent vector(s)}\\
\text{in undef.\,config.}\,\Gamma_{\vek X}
\end{array}$ & $\vek T^{\star}=\frac{\partial\vek X}{\partial r},\;\vek T=\frac{\vek T^{\star}}{\left\Vert \vek T^{\star}\right\Vert }$ & $\begin{array}{c}
\vek T_{1}^{\star}=\frac{\partial\vek X}{\partial r},\;\vek T_{2}^{\star}=\frac{\partial\vek X}{\partial s}\\
\vek T_{3}^{\star}=\vek T_{2}^{\star}-\frac{\vek T_{1}^{\star}\cdot\vek T_{2}^{\star}}{\vek T_{1}^{\star}\cdot\vek T_{1}^{\star}}\cdot\vek T_{1}^{\star}\\
\vek T_{i}=\frac{\vek T_{i}^{\star}}{\left\Vert \vek T_{i}^{\star}\right\Vert }\text{ with }i=\left\{ 1,2,3\right\} 
\end{array}$\tabularnewline
\hline 
$\begin{array}{c}
\text{tangent vector(s)}\\
\text{in def.\,config.}\,\Gamma_{\vek x}
\end{array}$ & $\vek t^{\star}=\mat F_{\Gamma}\cdot\vek T^{\star},\;\vek t=\frac{\vek t^{\star}}{\left\Vert \vek t^{\star}\right\Vert }$ & $\begin{array}{c}
\vek t_{1}^{\star}=\mat F_{\Gamma}\cdot\vek T_{1}^{\star},\;\vek t_{2}^{\star}=\mat F_{\Gamma}\cdot\vek T_{2}^{\star}\\
\vek t_{3}^{\star}=\vek t_{2}^{\star}-\frac{\vek t_{1}^{\star}\cdot\vek t_{2}^{\star}}{\vek t_{1}^{\star}\cdot\vek t_{1}^{\star}}\cdot\vek t_{1}^{\star}\\
\vek t_{i}=\frac{\vek t_{i}^{\star}}{\left\Vert \vek t_{i}^{\star}\right\Vert }\text{ with }i=\left\{ 1,2,3\right\} 
\end{array}$\tabularnewline
\hline 
projectors & $\begin{array}{c}
\mat P=\vek T\otimes\vek T\\
\mat p=\vek t\otimes\vek t
\end{array}$ & $\begin{array}{c}
\mat P=\vek T_{1}\otimes\vek T_{1}+\vek T_{3}\otimes\vek T_{3}\\
\mat p=\vek t_{1}\otimes\vek t_{1}+\vek t_{3}\otimes\vek t_{3}
\end{array}$\tabularnewline
\hline 
line/area stretch & $\Lambda=\frac{\sqrt{\det\mat g}}{\sqrt{\det\mat G}}=\frac{\left\Vert \vek t^{\star}\right\Vert }{\left\Vert \vek T^{\star}\right\Vert }$ & $\Lambda=\frac{\sqrt{\det\mat g}}{\sqrt{\det\mat G}}=\frac{\left\Vert \vek t_{1}^{\star}\times\vek t_{2}^{\star}\right\Vert }{\left\Vert \vek T_{1}^{\star}\times\vek T_{2}^{\star}\right\Vert }$\tabularnewline
\hline 
relation $\nabla_{\vek x}^{\Gamma}f=\mat W\cdot\nabla_{\vek X}^{\Gamma}f$ & \multicolumn{2}{c|}{$\mat W=\mat q\left(\mat Q^{\mathrm{T}}\cdot\mat Q\right)^{-1}\mat Q^{\mathrm{T}}$}\tabularnewline
\hline 
\end{tabular}\caption{\label{tab:ParametricManifolds}Geometric quantities and differential
operators for parametric manifolds. Only tangent vectors are considered
here although normal vectors may be computed for manifolds with codimension
$1$. }

\end{table}

\begin{table}
\begin{tabular}{|c|c|c|}
\hline 
 & $\begin{array}{c}
\text{cables in }\mathbb{R}^{2}\text{, membr. in }\mathbb{R}^{3}\\
\text{(manifolds with codim. 1)}
\end{array}$ & $\begin{array}{c}
\text{cables in }\mathbb{R}^{3}\\
\text{(manifolds with codim. 2)}
\end{array}$\tabularnewline
\hline 
\hline 
undeformed config.~$\Gamma_{\vek X}$ & $\Gamma_{\vek X}=\left\{ \vek X\in\Omega_{\vek X}:\phi\left(\vek X\right)=0\right\} $ & $\begin{array}{c}
\Gamma_{\vek X}=\left\{ \vek X\in\Omega_{\vek X}:\phi_{1}\left(\vek X\right)=0\right)\\
\left.\text{ and }\phi_{2}\left(\vek X\right)=0\right\} 
\end{array}$\tabularnewline
\hline 
deformed config.~$\Gamma_{\vek x}$ & $\Gamma_{\vek x}=\left\{ \vek x\in\Omega_{\vek x}:\phi\left(\vek X\!\left(\vek x\right)\right)=0\right\} $ & $\begin{array}{c}
\Gamma_{\vek x}=\left\{ \vek x\in\Omega_{\vek x}:\phi_{1}\left(\vek X\!\left(\vek x\right)\right)=0\right)\\
\left.\text{ and }\phi_{2}\left(\vek X\!\left(\vek x\right)\right)=0\right\} 
\end{array}$\tabularnewline
\hline 
$\begin{array}{c}
\text{classical deformation}\\
\text{gradient\,}\mat F_{\Omega}
\end{array}$ & \multicolumn{2}{c|}{$\mat F_{\Omega}=\nabla_{\vek X}\vek x\left(\vek X\right)=\mat I+\nabla_{\vek X}\vek u\left(\vek X\right)$}\tabularnewline
\hline 
$\begin{array}{c}
\text{normal vector(s)}\\
\text{in undef.\,config.}\,\Gamma_{\vek X}
\end{array}$ & $\vek N^{\star}=\nabla_{\vek X}\phi,\;\vek N=\frac{\vek N^{\star}}{\left\Vert \vek N^{\star}\right\Vert }$ & $\begin{array}{c}
\vek N_{1}^{\star}=\nabla_{\vek X}\phi_{1},\;\vek N_{2}^{\star}=\nabla_{\vek X}\phi_{2}\\
\vek N_{3}^{\star}=\vek N_{2}^{\star}-\frac{\vek N_{1}^{\star}\cdot\vek N_{2}^{\star}}{\vek N_{1}^{\star}\cdot\vek N_{1}^{\star}}\cdot\vek N_{1}^{\star}\\
\vek N_{i}=\frac{\vek N_{i}^{\star}}{\left\Vert \vek N_{i}^{\star}\right\Vert }\text{ with }i=\left\{ 1,2,3\right\} 
\end{array}$\tabularnewline
\hline 
$\begin{array}{c}
\text{normal vector(s)}\\
\text{in def.\,config.}\,\Gamma_{\vek x}
\end{array}$ & $\vek n^{\star}=\mat F_{\Omega}^{-\mathrm{T}}\cdot\vek N^{\star},\;\vek n=\frac{\vek n^{\star}}{\left\Vert \vek n^{\star}\right\Vert }$ & $\begin{array}{c}
\vek n_{1}^{\star}=\mat F_{\Omega}^{-\mathrm{T}}\cdot\vek N_{1}^{\star},\;\vek n_{2}^{\star}=\mat F_{\Omega}^{-\mathrm{T}}\cdot\vek N_{2}^{\star}\\
\vek n_{3}^{\star}=\vek n_{2}^{\star}-\frac{\vek n_{1}^{\star}\cdot\vek n_{2}^{\star}}{\vek n_{1}^{\star}\cdot\vek n_{1}^{\star}}\cdot\vek n_{1}^{\star}\\
\vek n_{i}=\frac{\vek n_{i}^{\star}}{\left\Vert \vek n_{i}^{\star}\right\Vert }\text{ with }i=\left\{ 1,2,3\right\} 
\end{array}$\tabularnewline
\hline 
projectors & $\begin{array}{c}
\mat P=\mat I-\vek N\otimes\vek N\\
\mat p=\mat I-\vek n\otimes\vek n
\end{array}$ & $\begin{array}{c}
\mat P=\mat I-\vek N_{1}\otimes\vek N_{1}-\vek N_{3}\otimes\vek N_{3}\\
\mat p=\mat I-\vek n_{1}\otimes\vek n_{1}-\vek n_{3}\otimes\vek n_{3}
\end{array}$\tabularnewline
\hline 
line/area stretch & $\Lambda=\frac{\left\Vert \vek n^{\star}\right\Vert }{\left\Vert \vek N^{\star}\right\Vert }\cdot\det\mat F_{\Omega}$ & $\Lambda=\frac{\left\Vert \vek n_{2}^{\star}\times\vek n_{1}^{\star}\right\Vert }{\left\Vert \vek N_{2}^{\star}\times\vek N_{1}^{\star}\right\Vert }\cdot\det\mat F_{\Omega}$\tabularnewline
\hline 
$\begin{array}{c}
\text{surface gradients}\\
\text{w.r.t.\,}\Gamma_{\vek X}
\end{array}$ & \multicolumn{2}{c|}{$\begin{array}{c}
\nabla_{\vek X}^{\Gamma}f=\mat P\cdot\nabla_{\vek X}f\\
\nabla_{\vek X}^{\Gamma,\mathrm{dir}}\vek u=\nabla_{\vek X}\vek u\cdot\mat P
\end{array}$}\tabularnewline
\hline 
$\begin{array}{c}
\text{surface deformation}\\
\text{gradient\,}\mat F_{\Gamma}
\end{array}$ & \multicolumn{2}{c|}{$\mat F_{\Gamma}=\mat I+\nabla_{\vek X}^{\Gamma,\mathrm{dir}}\vek u$}\tabularnewline
\hline 
$\begin{array}{c}
\text{relation}\\
\nabla_{\vek x}^{\Gamma}f=\mat W\cdot\nabla_{\vek X}^{\Gamma}f
\end{array}$ & \multicolumn{2}{c|}{$\mat W=\mat p\cdot\mat F_{\Omega}^{-\mathrm{T}}$}\tabularnewline
\hline 
$\begin{array}{c}
\text{surface gradients}\\
\text{w.r.t.\,}\Gamma_{\vek x}
\end{array}$ & \multicolumn{2}{c|}{$\begin{array}{c}
\nabla_{\vek x}^{\Gamma}f=\mat p\cdot\nabla_{\vek x}f=\mat W\cdot\nabla_{\vek X}f\\
\nabla_{\vek x}^{\Gamma,\mathrm{dir}}\vek u=\nabla_{\vek x}\vek u\cdot\mat p=\nabla_{\vek X}\vek u\cdot\mat W^{\mathrm{T}}
\end{array}$}\tabularnewline
\hline 
\end{tabular}\caption{\label{tab:ImplicitManifolds}Geometric quantities and differential
operators for implicit manifolds. Only normal vectors are considered
here although tangent vectors may be computed for cables (one-dimensional
manifolds) as well.}
\end{table}

It was found above that for parametrized manifolds, tangent vectors
result naturally as primary quantities through the existence of Jacobi
matrices. For problems with codimension $1$, one may then obtain
normal vectors as secondary quantities (e.g., by a cross product of
tangent vectors), however, for higher codimensions, no unique normal
vector exists. For implicit manifolds, the situation is rather the
opposite: Here, normal vectors result naturally through the gradients
of level-set functions. A unique tangent vector is only applicable
for one-dimensional manifolds and may, for example, be computed by
a cross product of the normal vectors. This situation is an example
of the duality in the parametric and implicit description of manifolds.

\subsection{Further definitions}

Based on the previous definitions, some additional differential operators
are introduced which apply to parametric and implicit manifolds equivalently.

\subsubsection{Covariant surface gradient and divergence\label{XXX_CovariantSurfaceGradient}}

The \emph{covariant }surface gradient of a vector function $\vek u\left(\vek X\right):\Gamma_{\vek X}\to\mathbb{R}^{d}$
is based on the projection of the directional one (see Eqs.~(\ref{eq:SurfGradVectorParametric})
and (\ref{eq:SurfGradVectorImplicit}) for the parametric and implicit
situation, respectively) onto the tangent space. With respect to the
undeformed configuration, it is defined as 
\begin{equation}
\nabla_{\vek X}^{\Gamma,\mathrm{cov}}\vek u\left(\vek X\right)=\mat P\cdot\nabla_{\vek X}^{\Gamma,\mathrm{dir}}\vek u\left(\vek X\right).\label{eq:CovariantSurfaceGradient}
\end{equation}
Concerning the \emph{surface divergence }of vector functions $\vek u\left(\vek X\right)$
and tensor functions $\mat A\left(\vek X\right):\Gamma_{\vek X}\to\mathbb{R}^{d\times d}$,
there holds 
\begin{eqnarray}
\mathrm{Div}_{\Gamma}\,\vek u\left(\vek X\right) & = & \mathrm{tr}\left(\nabla_{\vek X}^{\Gamma,\mathrm{dir}}\vek u\right)=\mathrm{tr}\left(\nabla_{\vek X}^{\Gamma,\mathrm{cov}}\vek u\right)\eqqcolon\nabla_{\vek X}^{\Gamma}\cdot\vek u,\label{eq:DivergenceVector}\\
\mathrm{Div}_{\Gamma}\,\mat A\left(\vek X\right) & = & \left[\begin{array}{c}
\mathrm{Div}_{\Gamma}\left(A_{11},A_{12},A_{13}\right)\\
\mathrm{Div}_{\Gamma}\left(A_{21},A_{22},A_{23}\right)\\
\mathrm{Div}_{\Gamma}\left(A_{31},A_{32},A_{33}\right)
\end{array}\right]\eqqcolon\nabla_{\vek X}^{\Gamma}\cdot\mat A.\label{eq:DivergenceTensor}
\end{eqnarray}
The divergence operator with respect to the deformed configuration
follows accordingly as $\mathrm{div}_{\Gamma}\,\vek u\left(\vek X\right)=\nabla_{\vek x}^{\Gamma}\cdot\vek u$
and $\mathrm{div}_{\Gamma}\,\vek A\left(\vek X\right)=\nabla_{\vek x}^{\Gamma}\cdot\vek A$,
respectively.

\subsubsection{Conormal vector on the boundary}

Unit normal and tangent vectors on manifolds have already been used
before and exist with respect to the deformed and undeformed configuration,
respectively. For example, the unit normal vector on an undeformed
membrane is $\vek N\!\left(\vek X\right)$ for all $\vek X\in\Gamma_{\vek X}$
and, after the deformation, $\vek n\!\left(\vek x\right)$ for all
$\vek x\in\Gamma_{\vek x}$. It is important to note that, for physical
reasons, the manifolds used in this work are \emph{bounded}. The boundary
of the undeformed configuration is labeled $\partial\Gamma_{\vek X}$
and in the deformed situation $\partial\Gamma_{\vek x}$.

\begin{figure}
\centering

\subfigure[cable in $\mathbb{R}^2$]{\includegraphics[width=0.4\textwidth]{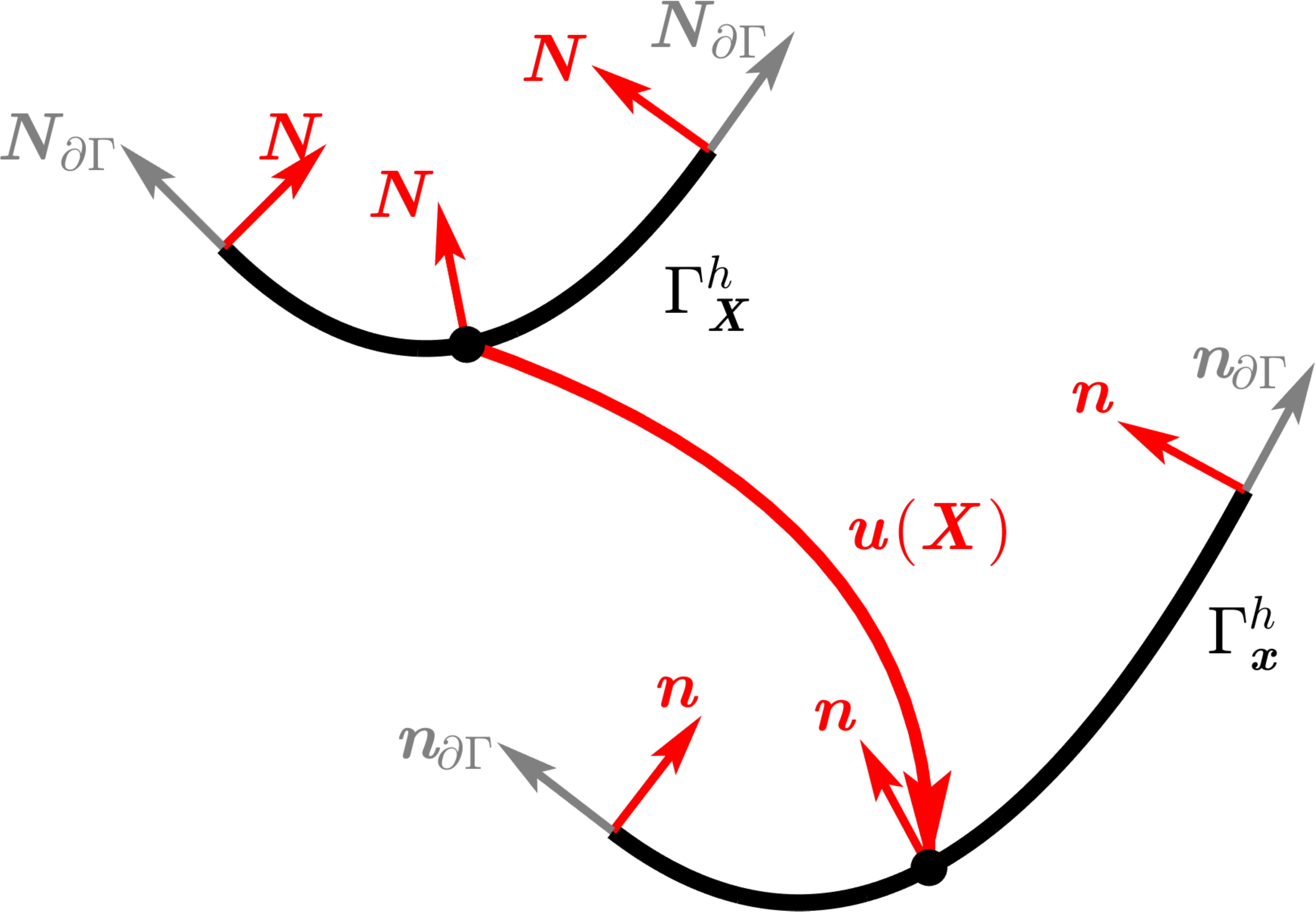}}\qquad\qquad\subfigure[membrane in $\mathbb{R}^3$]{\includegraphics[width=0.4\textwidth]{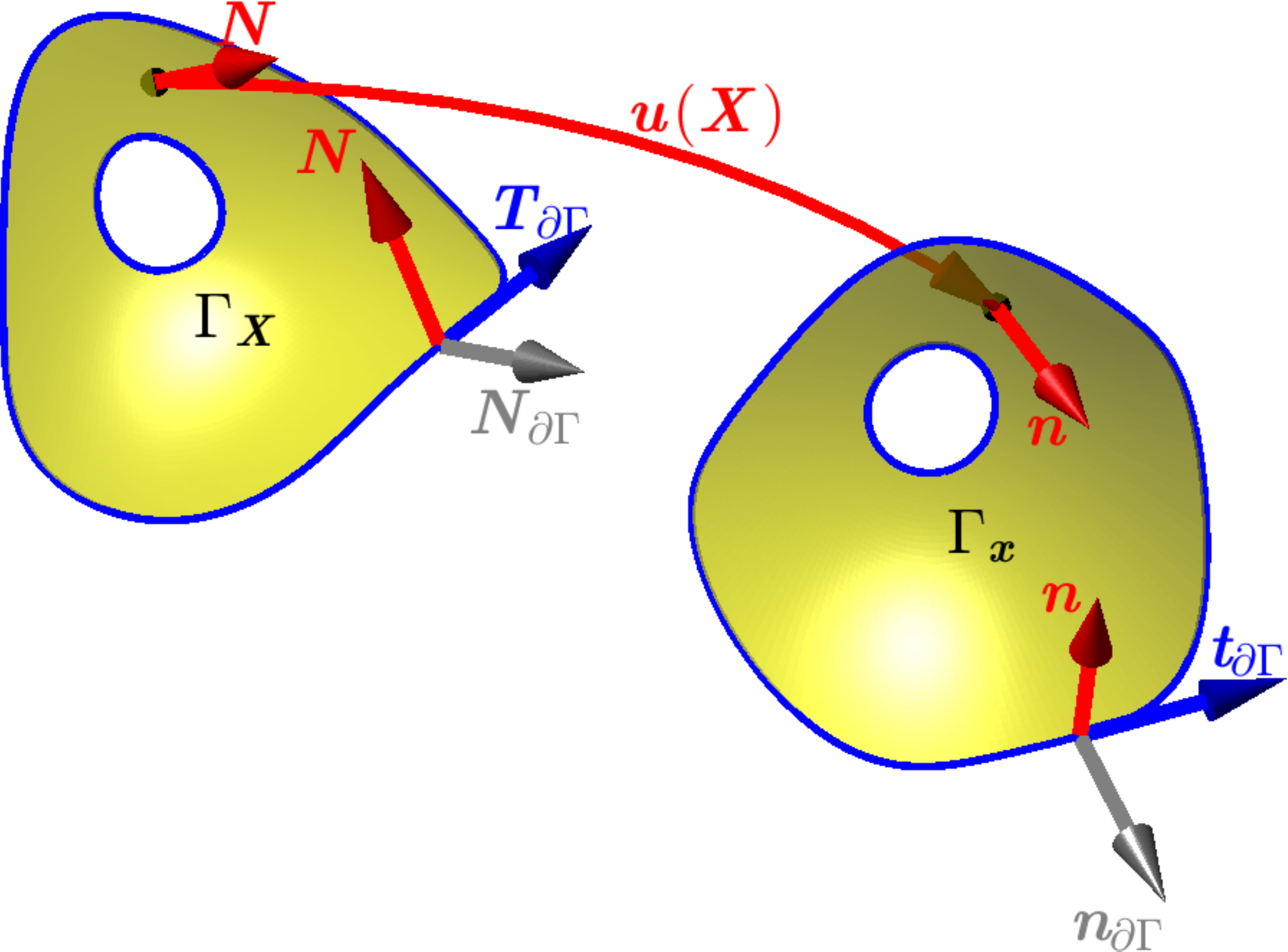}}

\caption{\label{fig:VisCoNormalVectors}Normal vectors, $\vek N$ and $\vek n$,
and conormal vectors, $\vek N_{\partial\Gamma}$ and $\vek n_{\partial\Gamma}$,
in undeformed and deformed manifolds. The vectors $\vek T_{\partial\Gamma}$
and $\vek t_{\partial\Gamma}$ in (b) point in tangential direction
along the boundary.}
\end{figure}

There exists a \emph{co}normal unit vector $\vek N_{\partial\Gamma}$
along the boundary $\partial\Gamma_{\vek X}$ which is in the tangent
plane of the manifold yet normal to $\partial\Gamma_{\vek X}$. This
vector points out of the manifold in the direction which naturally
extends the manifold, see Fig.~\ref{fig:VisCoNormalVectors}. The
computation of the conormal vectors is straightforward (often using
cross products) and depends on $q$ and $d$. In the deformed configuration,
the situation is similar for computing $\vek n_{\partial\Gamma}\left(\vek x\right)$
for all $\vek x\in\partial\Gamma_{\vek x}$. In the context of the
definition of boundary value problems on manifolds, the conormal vectors
play a crucial role for the consideration of boundary conditions as
shall be seen in Section \ref{XXX_BoundaryConditions}.

\subsubsection{Divergence theorem on manifolds}

To derive the weak form of the governing equations later on, the following
divergence theorem on manifolds is needed \cite{Delfour_1996a,Delfour_2011a},
\begin{equation}
\int_{\Gamma_{\vek X}}\vek u\cdot\mathrm{Div}_{\Gamma}\mat A\,\mathrm{d}\Gamma=-\int_{\Gamma_{\vek X}}\nabla_{\vek X}^{\Gamma,\mathrm{dir}}\vek u:\mat A\,\mathrm{d}\Gamma+\int_{\Gamma_{\vek X}}\varkappa\cdot\vek u\cdot\mat A\cdot\vek N\,\mathrm{d}\Gamma+\int_{\partial\Gamma_{\vek X}}\vek u\cdot\mat A\cdot\vek N_{\partial\Gamma}\,\mathrm{d}\partial\Gamma,\label{eq:DivTheorem}
\end{equation}
where $\nabla_{\vek X}^{\Gamma,\mathrm{dir}}\vek u:\mat A=\mathrm{tr}\left(\nabla_{\vek X}^{\Gamma,\mathrm{dir}}\vek u\cdot\mat A^{\mathrm{T}}\right)$
is a matrix scalar product. The mean curvature is $\varkappa=\mathrm{tr}\left(\mat H\right)$
with $\mat H=\nabla_{\vek X}^{\Gamma,\mathrm{dir}}\vek N=\nabla_{\vek X}^{\Gamma,\mathrm{cov}}\vek N$
being the second fundamental form. For \emph{in-plane} tensor functions
with $\mat A=\mat P\cdot\mat A\cdot\mat P$, the term involving the
curvature $\varkappa$ vanishes and one finds $\nabla_{\vek X}^{\Gamma,\mathrm{dir}}\vek u:\mat A=\nabla_{\vek X}^{\Gamma,\mathrm{cov}}\vek u:\mat A$.

\section{Mechanical model and governing equations\label{X_Mechanics}}

In Section \ref{X_Geometry}, a number of geometric quantities (such
as normal vectors, projectors, area/line stretches etc.) and differential
operators related to (surface) gradients are introduced. It was shown
how these quantities are obtained for parametrized and implicitly
defined manifolds. The focus is now turned to the mechanics and the
procedure follows the classical outline, however, it is based on the
TDC here. It is emphasized that \emph{all} tensors considered in the
following refer to the parametric as well as implicit situation. These
tensors have dimensions $d\times d$ (with $d$ being the dimension
into which the cable or membrane is immersed). A tensor $\mat A$
is called ``in-plane'' or ``tangential'' to the undeformed configuration
$\Gamma_{\vek X}$ if $\mat A=\mat P\cdot\mat A\cdot\mat P$ and to
the deformed configuration $\Gamma_{\vek x}$ if $\mat A=\mat p\cdot\mat A\cdot\mat p$.
An in-plane $\left(d\times d\right)$-tensor has $q$ non-zero eigenvalues
representing the principal mechanical quantity (with $q$ being the
dimension of the manifold: $q=1$ for cables, $q=2$ for membranes).

Starting point is the surface deformation gradient $\mat F_{\Gamma}\left(\vek X\right)$
at $\vek X\in\Gamma_{\vek X}$, specified previously in Eq.~(\ref{eq:SurfaceDeformationGradient}).
It may also be seen as a geometrical quantity mapping tangent vectors
from the undeformed to the deformed configuration. It is also noted
that the situation also applies to the ``volumetric'' case (where
$q=d=2$ are flat shells in two dimensions and $q=d=3$ are volumetric
continua in three dimensions). In this case, $\mat F_{\Gamma}=\mat F_{\Omega}$
as in Eq.~(\ref{eq:ClassicalDeformationGradient}), and for the projectors,
$\mat P=\mat p=\mat I$. In that sense, the presented mechanical outline
below applies to cables, membranes and continua in a unified sense.

\subsection{Governing equations in strong form\label{XX_StrongForm}}

\subsubsection{Strain tensors}

Based on the surface deformation gradient, the directional and tangential
Cauchy-Green strain tensors are defined as 
\begin{eqnarray}
\mat E_{\mathrm{dir}} & = & \nicefrac{1}{2}\cdot\left(\mat F_{\Gamma}^{\mathrm{T}}\cdot\mat F_{\Gamma}-\mat I\right),\label{eq:CauchyGreenDir}\\
\mat E_{\mathrm{tang}} & = & \mat P\cdot\mat E_{\mathrm{dir}}\cdot\mat P,\label{eq:CauchyGreenCov}
\end{eqnarray}
respectively. The Euler-Almansi strain tensors are
\begin{eqnarray}
\mat e_{\mathrm{dir}} & = & \nicefrac{1}{2}\cdot\left(\mat I-\left(\mat F_{\Gamma}\cdot\mat F_{\Gamma}^{\mathrm{T}}\right)^{-1}\right),\label{eq:EulerAlmansiDir}\\
\mat e_{\mathrm{tang}} & = & \mat p\cdot\mat e_{\mathrm{dir}}\cdot\mat p,\label{eq:EulerAlmansiCov}
\end{eqnarray}
where $\mat e_{\mathrm{tang}}$ is tangential to the deformed configuration
$\Gamma_{\vek x}$. As usual, there holds $\mat e_{\mathrm{dir}}=\mat F_{\Gamma}^{-\mathrm{T}}\cdot\mat E_{\mathrm{dir}}\cdot\mat F_{\Gamma}^{-1}$
(which is not true for the tangential versions of these strain tensors).

\subsubsection{Stress tensors}

Conjugated stress tensors are introduced next and only the tangential
versions are considered. Generally speaking, we assume some hyper-elastic
material with an elastic energy function $\Psi\left(\mat E_{\mathrm{tang}}\right)$
and obtain the second Piola-Kirchhoff stress tensor as $\mat S=\frac{\partial\Psi}{\partial\mat E_{\mathrm{tang}}}$.
For simplicity, only Saint Venant--Kirchhoff solids are considered
herein and there follows
\begin{eqnarray}
\mat S & = & \lambda\cdot\mathrm{trace}\left(\mat E_{\mathrm{tang}}\right)\cdot\mat P+2\mu\mat E_{\mathrm{tang}},\label{eq:2ndPiolaKirchhoff}\\
 & = & \mat P\cdot\left(\lambda\cdot\mathrm{trace}\left(\mat E_{\mathrm{dir}}\right)\cdot\mat I+2\mu\mat E_{\mathrm{dir}}\right)\cdot\mat P,\nonumber 
\end{eqnarray}
with $\mat S$ being tangential to $\Gamma_{\vek X}$. $\lambda$
and $\mu$ are the Lam\'e constants; for cables $\lambda$ becomes
$0$. On the other hand, the Cauchy stress tensor reads
\begin{equation}
\vek\sigma=\frac{1}{\Lambda}\cdot\mat F_{\Gamma}\cdot\mat S\cdot\mat F_{\Gamma}^{\mathrm{T}},\label{eq:CauchyStress}
\end{equation}
where $\Lambda$ is a line stretch for cables and an area stretch
for membranes when undergoing the displacement, see Section \ref{X_Geometry}.
For the volumetric case ($q=d$), $\Lambda=\det\mat F_{\Omega}$ is
the volumetric stretch. The Cauchy stress is tangential to the deformed
configuration $\Gamma_{\vek x}$ since $\mat F_{\Gamma}\cdot\mat P=\mat p\cdot\mat F_{\Gamma}\cdot\mat P$
and $\mat P\cdot\mat F_{\Gamma}^{\mathrm{T}}=\mat P\cdot\mat F_{\Gamma}^{\mathrm{T}}\cdot\mat p$,
hence $\vek\sigma=\mat p\cdot\vek\sigma\cdot\mat p.$ The first Piola-Kirchhoff
stress tensor is given by
\begin{equation}
\mat K=\mat F_{\Gamma}\cdot\mat S\label{eq:1stPiolaKirchhoff}
\end{equation}
and there holds $\mat K=\mat K\cdot\mat P=\mat p\cdot\mat K$.

\subsubsection{Relation of stress and strain tensors}

For every point $\vek X\in\Gamma_{\vek X}$ and its mapped counterpart
$\vek x\left(\vek X\right)\in\Gamma_{\vek x}$, we have the following
equality,
\begin{equation}
\mat S\left(\vek X\right):\mat E_{\mathrm{tang}}\left(\vek X\right)=\left(\vek\sigma\left(\vek x\right):\mat e_{\mathrm{tang}}\left(\vek x\right)\right)\cdot\Lambda\left(\vek X\right),\label{eq:EnergyRelation}
\end{equation}
where $:$ represents the matrix scalar product operator. In this
sense the two stress tensors $\mat S$ and $\vek\sigma$ are conjugated
to their related strain tensors $\mat E_{\mathrm{tang}}$ and $\mat e_{\mathrm{tang}}$,
respectively. It is noted that
\[
\mat S:\mat E_{\mathrm{tang}}=\mat S:\mat E_{\mathrm{dir}}\quad\text{and}\quad\vek\sigma:\mat e_{\mathrm{tang}}=\vek\sigma:\mat e_{\mathrm{dir}}
\]
which will be important later. Furthermore, the result of these matrix
scalar products may also be derived by the non-zero eigenvalues $S_{i}$,
$E_{\mathrm{tang},i}$, $\sigma_{i}$, $e_{\mathrm{tang},i}$, $i=1,\dots,q$,
of the tangential tensors $\mat S$, $\mat E_{\mathrm{tang}}$, $\vek\sigma$,
$\mat e_{\mathrm{tang}}$, respectively. Hence, we obtain
\[
\mat S:\mat E_{\mathrm{tang}}=\sum_{i=1}^{q}S_{i}\cdot E_{\mathrm{tang},i}\quad\text{and}\quad\vek\sigma:\mat e_{\mathrm{tang}}=\sum_{i=1}^{q}\sigma_{i}\cdot e_{\mathrm{tang},i}.
\]

\subsubsection{Equilibrium}

A crucial aspect of finite strain theory is that equilibrium is to
be fulfilled in the \emph{deformed} configuration which is expressed
in strong form as
\begin{equation}
\mathrm{div}_{\Gamma}\,\vek\sigma\!\left(\vek x\right)=-\vek f\!\left(\vek x\right)\quad\forall\vek x\in\Gamma_{\vek x},\label{eq:EquilibriumDefConfig}
\end{equation}
where $\vek f$ are body forces. Recall from (\ref{eq:DivergenceTensor})
that $\mathrm{div}_{\Gamma}\,\vek\sigma=\nabla_{\vek x}^{\Gamma,\mathrm{dir}}\cdot\vek\sigma=\nabla_{\vek x}^{\Gamma,\mathrm{cov}}\cdot\vek\sigma$
is the divergence of the Cauchy stress tensor with respect to $\Gamma_{\vek x}$.
Furthermore, we have the identity
\begin{equation}
\mathrm{Div}_{\Gamma}\,\mat K\left(\vek X\right)=\mathrm{div_{\Gamma}\,}\vek\sigma\!\left(\vek x\left(\vek X\right)\right)\cdot\Lambda\left(\vek X\right)\label{eq:EquilibriumDivRelation}
\end{equation}
with $\mathrm{Div}_{\Gamma}\,\mat K=\nabla_{\vek X}^{\Gamma,\mathrm{dir}}\cdot\mat K=\nabla_{\vek X}^{\Gamma,\mathrm{cov}}\cdot\mat K$
being the divergence of the first Piola-Kirchhoff stress tensor from
Eq.~(\ref{eq:1stPiolaKirchhoff}) with respect to $\Gamma_{\vek X}$.
In order to transform the derivatives in the divergence operators
from the undeformed to the deformed situation, use Eqs.~(\ref{eq:SurfGradUndefToDefParam})
and (\ref{eq:SurfGradUndefToDefImpl}) for parametric and implicit
manifolds, respectively. Due to $\vek F\!\left(\vek X\right)=\vek f\!\left(\vek x\left(\vek X\right)\right)\cdot\Lambda\left(\vek X\right)$,
the equilibrium in $\Gamma_{\vek x}$ can be stated equivalently to
Eq.~(\ref{eq:EquilibriumDefConfig}) based on quantities in the undeformed
configuration as
\begin{equation}
\mathrm{Div}_{\Gamma}\,\mat K\!\left(\vek X\right)=-\vek F\!\left(\vek X\right)\quad\forall\vek X\in\Gamma_{\vek X}.\label{eq:EquilibriumUndefConfig}
\end{equation}

\subsubsection{Boundary conditions\label{XXX_BoundaryConditions}}

The domain of interest is a bounded manifold where the boundary $\partial\Gamma$
falls into the two non-overlapping parts $\partial\Gamma_{\mathrm{D}}$
and $\partial\Gamma_{\mathrm{N}}$, which holds in the deformed and
undeformed configuration $\Gamma_{\vek X}$ and $\Gamma_{\vek x}$,
respectively. Hence, the boundary conditions in the deformed configuration
are 
\begin{align}
\vek u(\vek x) & =\hat{\vek g}(\vek x)\ \text{on}\ \partial\Gamma_{\vek x,\text{D}},\\
\vek\sigma\!\left(\vek x\right)\cdot\vek n_{\partial\Gamma}(\vek x) & =\hat{\vek h}(\vek x)\ \text{on}\ \partial\Gamma_{\vek x,\text{N}},
\end{align}
where $\hat{\vek g}$ are prescribed displacements and $\hat{\vek h}$
are tractions (force per area for $q=3$, force per length for $q=2$
or a single force for $q=1$). Note that for ropes and membranes,
$\hat{\vek{h}}$ must be in the tangent space of the deformed manifold
in order to satisfy the equilibrium due to the absence of bending
stresses or transverse shear stresses. The equivalent boundary conditions
formulated in the \emph{un}deformed configuration are 
\begin{align}
\vek u(\vek X) & =\hat{\vek G}(\vek X)\ \text{on}\ \partial\Gamma_{\vek X,\text{D}},\\
\mat K\!\left(\vek X\right)\cdot\vek N_{\partial\Gamma}(\vek X) & =\hat{\vek H}(\vek X)\ \text{on}\ \partial\Gamma_{\vek X,\text{N}}\ ,
\end{align}
where $\hat{\vek{G}}$ and $\hat{\vek{H}}$ have similar interpretations
as before. The relation between $\hat{\vek{h}}$ and $\hat{\vek{H}}$
is $\hat{\vek{H}}(\vek{X})=\bar{\Lambda}(\vek{X})\cdot\hat{\vek{h}}(\vek{x})$,
where 
\[
\bar{\Lambda}=\begin{cases}
1 & \text{for }q=1,\,d=\left\{ 2,3\right\} \text{ (cables)},\\
\text{line stretch along the boundary} & \text{for }q=2,\,d=2\mbox{\text{ (shells) or }}3\text{ (membranes)},\\
\text{area stretch of the face at the boundary} & \text{for }q=3,\,d=3\text{ (continuum)}.
\end{cases}
\]
Further information about boundary conditions for ropes and membranes
are given in \cite{Calladine_1983a}.

With the boundary conditions above, the complete second-order boundary
value problem (BVP) is defined in the deformed and undeformed configuration.
The obtained BVP in the frame of the TDC is valid for explicitly and
implicitly defined manifolds and does not rely on curvilinear coordinates
implied by a parametrization, which are typically used in classical
approaches, see, e.g., \cite{Calladine_1983a,Ciarlet_1997a}. Therefore,
the proposed formulation of ropes and membranes based on the TDC is
more general compared to the classical theory.

\subsection{Governing equations in weak form\label{XX_WeakForm}}

For stating the governing equations in weak form, the following test
and trial function spaces are introduced 
\begin{align}
\mathcal{S}_{\vek u} & =\left\lbrace \vek v\in\left[\mathcal{H}^{1}(\Gamma_{\vek X})\right]^{d}:\ \vek v=\hat{\vek G}\ \text{on}\ \partial\Gamma_{\vek X,\text{D}}\right\rbrace ,\label{eq:TrialFctSpaceCont}\\
\mathcal{V}_{\vek u} & =\left\lbrace \vek v\in\left[\mathcal{H}^{1}(\Gamma_{\vek X})\right]^{d}:\ \vek v=\vek0\ \text{on}\ \partial\Gamma_{\vek X,\text{D}}\right\rbrace ,\label{eq:TestFctSpaceCont}
\end{align}
where $\mathcal{H}^{1}$ is the Sobolev space of functions with square
integrable first derivatives. The task is to find $\vek u\in\mathcal{S}_{\vek u}$
such that for all $\vek w\in\mathcal{V}_{\vek u}$, there holds 
\begin{equation}
\eta\cdot\int_{\Gamma_{\vek X}}\nabla_{\vek X}^{\Gamma,\mathrm{dir}}\vek w:\mat K\left(\vek u\right)\,\mathrm{d}\Gamma=\eta\cdot\int_{\Gamma_{\vek X}}\vek w\cdot\vek F\,\mathrm{d}\Gamma+\int_{\partial\Gamma_{\vek X,\text{N}}}\vek w\cdot\hat{\vek H}\,\mathrm{d}\partial\Gamma.\label{eq:WeakFormCont}
\end{equation}
where 
\[
\eta=\begin{cases}
A & \text{for }q=1,\,d=\left\{ 2,3\right\} \text{ is the cross section of the cable},\\
t & \text{for }q=2,\,d=\left\{ 2,3\right\} \text{ is the thickness of the shell/membrane},\\
1 & \text{for }q=3,\,d=3\text{, i.e., a continuum}.
\end{cases}
\]
The integrals in Eq.~(\ref{eq:WeakFormCont}) are one-, two-, or
three-dimensional for cables, membranes and continua, respectively.
The multiplication with $\eta$ ensures that the units of the integration
are always consistent. Hence, it is possible to naturally consider
situations where cables, membranes and continua are coupled in one
setup by simply adding up the corresponding integrals as in Eq.~(\ref{eq:WeakFormCont})
for each structure. In order to obtain Eq.~(\ref{eq:WeakFormCont}),
we applied the usual procedure for converting the strong form to a
weak form: Multiply Eq.~(\ref{eq:EquilibriumUndefConfig}) with test
functions, integrate over the domain $\Gamma_{\vek X}$, and apply
the divergence theorem from Eq.~(\ref{eq:DivTheorem}). It is noteworthy
that the curvature term from Eq.~(\ref{eq:DivTheorem}) vanishes
also for cables and membranes due to $\mat K\cdot\vek N=\vek0$.

The weak form stated above is related to energy minimization in the
sense that
\[
\int_{\Gamma_{\vek X}}\nabla_{\vek X}^{\Gamma,\mathrm{dir}}\vek w:\mat K\left(\vek u\right)\,\mathrm{d}\Gamma=\int_{\Gamma_{\vek X}}\delta\mat E_{\mathrm{tang}}\left(\vek u\right):\mat S\left(\vek u\right)\,\mathrm{d}\Gamma,
\]
where $\delta$ is the variational operator.

\subsubsection{Energy relation\label{XXX_StoredElasticEnergy}}

An immediate consequence of Eq.~(\ref{eq:EnergyRelation}) is that
one may obtain the same stored potential energy of the deformed body
by integrating over the undeformed or deformed configuration as follows
\begin{eqnarray}
\mathfrak{e}\left(\vek u\right) & = & \frac{1}{2}\eta\cdot\int_{\Gamma_{\vek x}}\mat e_{\mathrm{tang}}\left(\vek u\right):\vek\sigma\left(\vek u\right)\;\mathrm{d}\Gamma,\label{eq:EnergyDef}\\
 & = & \frac{1}{2}\eta\cdot\int_{\Gamma_{\vek X}}\mat E_{\mathrm{tang}}\left(\vek u\right):\mat S\left(\vek u\right)\;\mathrm{d}\Gamma.\label{eq:EnergyUndef}
\end{eqnarray}

\section{Discretization and numerical methods\label{X_Discretization}}

In order to solve the boundary value problem, i.e., to approximate
the sought displacements, one may use two fundamentally different
approaches. Possibly the more intuitive one is to discretize the cable(s)
or membrane(s) by curved ($q$-dimensional) line or surface elements,
respectively. Then, each element results from an (often isoparametric)
map of some reference element so that this approach is naturally linked
to the parametric description of manifolds as discussed in Section
\ref{XX_ParametrizedManifolds}. This is the classical approach labeled
Surface FEM herein. An alternative is to use a ($d$-dimensional)
\emph{background} mesh for the approximation of the weak form. That
is, higher-dimensional shape functions (than the dimension of the
manifold) are used and evaluated on the trace of the manifold only.
This approach is called Trace FEM or Cut FEM. It is naturally related
to an implicit manifold description as discussed in Section \ref{XX_ImplicitManifolds}.

It is important to note that the classical definition of finite strain
theory based on curvilinear coordinates does not cover the latter
approach. This is another reason why the presented formulation based
on the TDC is more general as it supports both, the Surface and Trace
FEM. For the discussion below, we assume the manifold case, hence,
$q<d$. With $q=d$, the situation results in the standard FEM which
is not further outlined here.

\subsection{Surface FEM\label{XX_SurfaceFEM}}

\begin{figure}
\centering

\subfigure[cable in $\mathbb{R}^2$]{\includegraphics[width=0.3\textwidth]{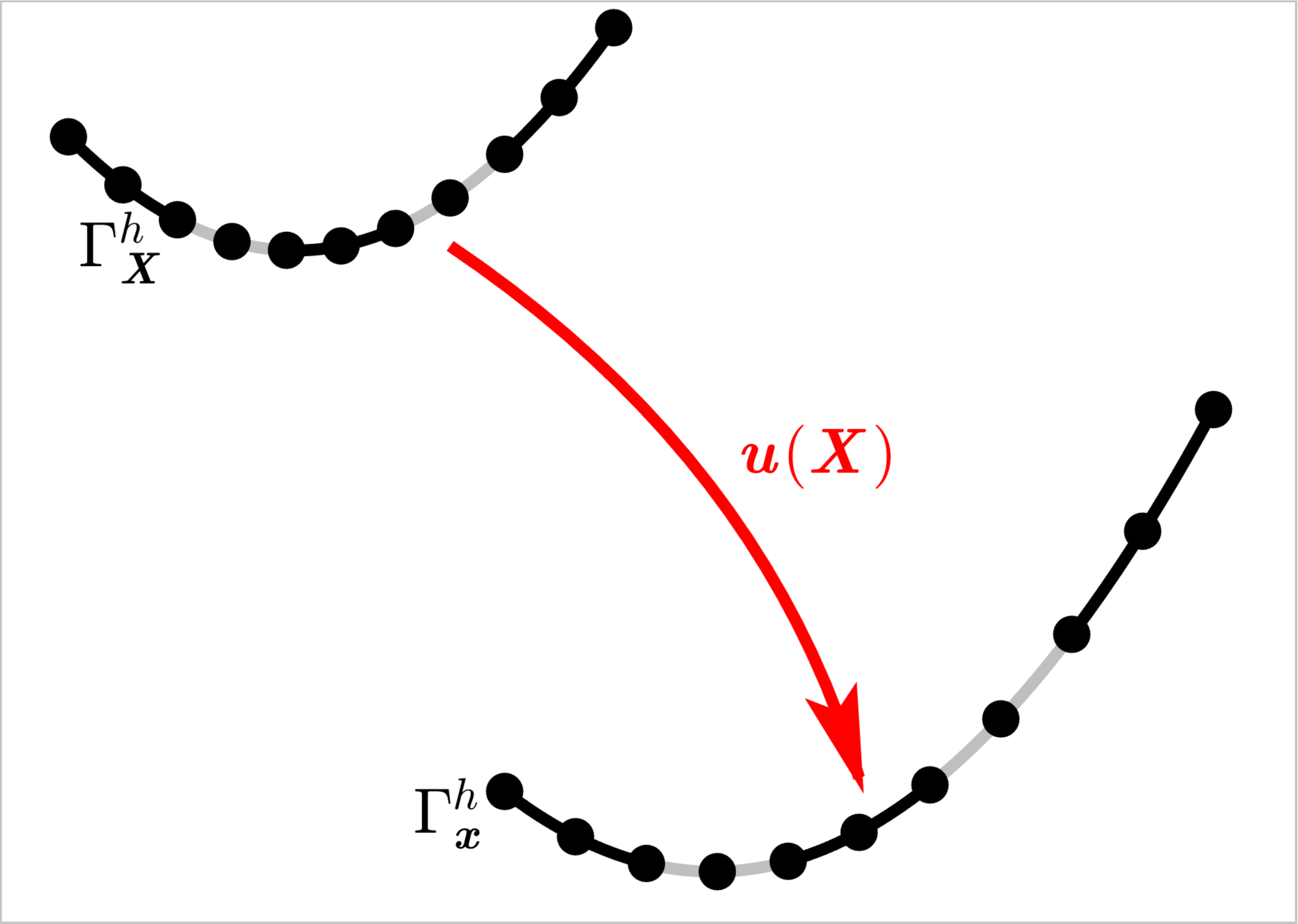}}\hfill\subfigure[cable in $\mathbb{R}^3$]{\includegraphics[width=0.3\textwidth]{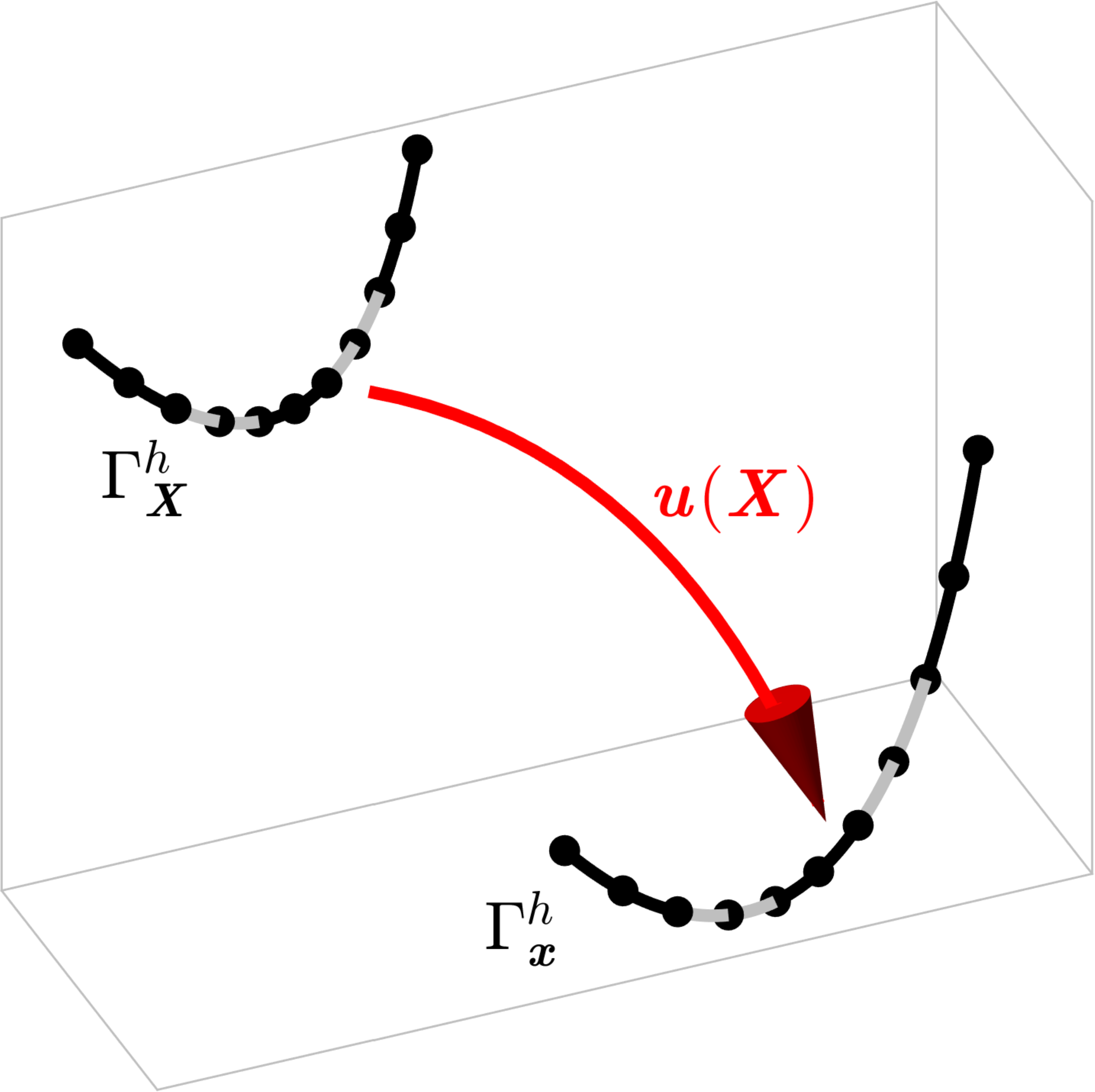}}\hfill\subfigure[membrane in $\mathbb{R}^3$]{\includegraphics[width=0.3\textwidth]{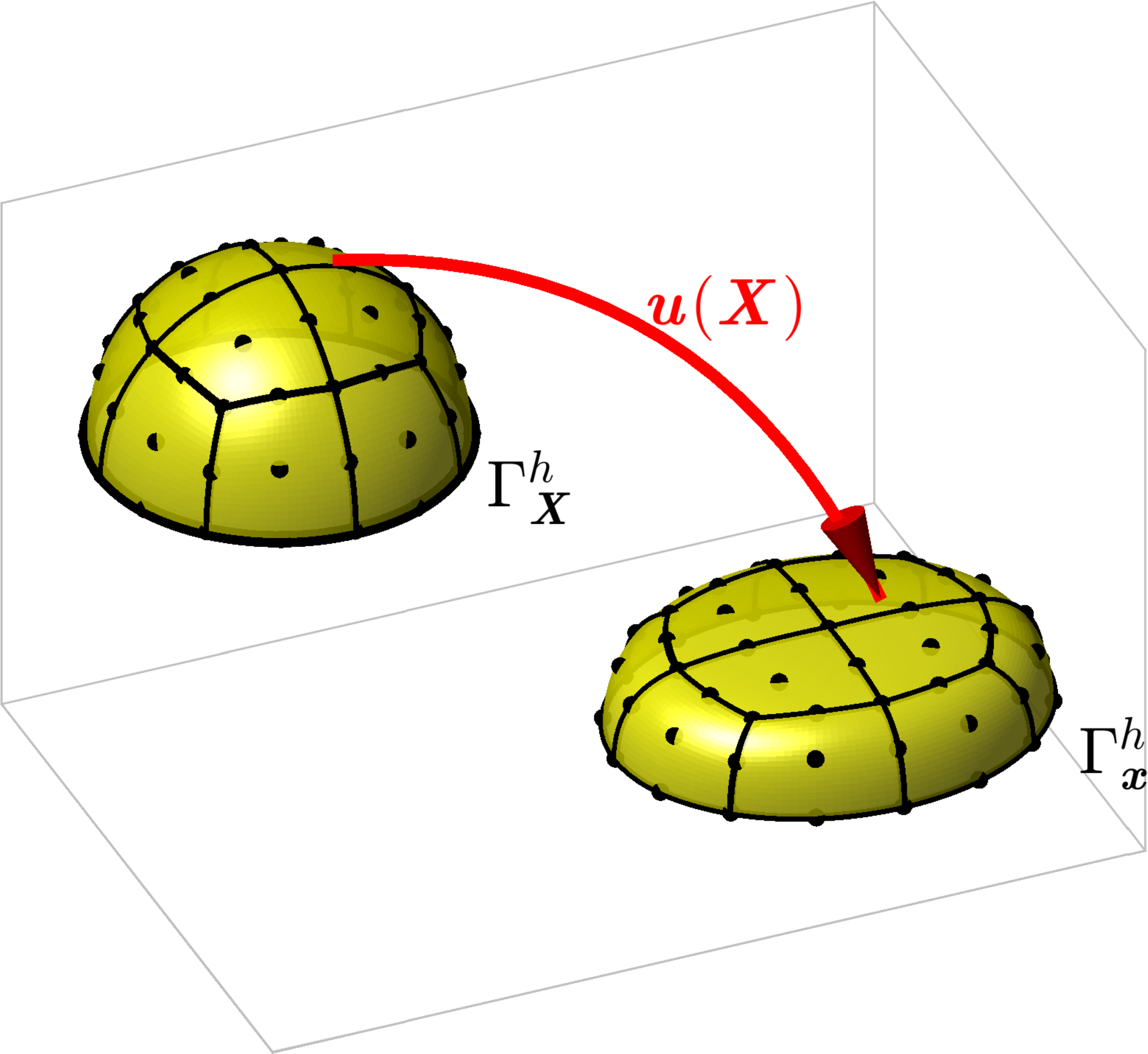}}

\caption{\label{fig:SurfaceFEM}The situation in the Surface FEM for (a) cables
in $\mathbb{R}^{2}$, (b) cables in $\mathbb{R}^{3}$, and (c) membranes
in $\mathbb{R}^{3}$. The discretized domains are shown for the undeformed
and deformed situations, $\Gamma_{\vek X}^{h}$ and $\Gamma_{\vek x}^{h}$,
respectively.}
\end{figure}

Starting point is the discretization $\Gamma_{\vek X}^{h}$ of the
undeformed cable ($q=1$) or membrane ($q=2$) by a line or surface
mesh, respectively. Herein, we use higher-order $q$-dimensional Lagrange
elements with equally spaced nodes in the reference element. The nodal
coordinates in the undeformed configuration are labeled $\vek X_{i}$
with $i=1,\dots,n_{q}$ and $n_{q}$ being the number of nodes in
the mesh, see Fig.~\ref{fig:SurfaceFEM}. The resulting shape functions
$M_{i}^{q}\!\left(\vek X\right)$ span a $C^{0}$-continuous finite
element space as 
\begin{align}
\mathcal{Q}_{\Gamma_{\vek X}}^{h}:=\left\lbrace v_{h}\in C^{0}(\Gamma_{\vek X}^{h}):\ v_{h}=\sum_{i=1}^{n_{q}}M_{i}^{q}(\vek X)\cdot\hat{v}_{i}\text{ with }\hat{v}_{i}\in\mathbb{R}\right\rbrace \subset\mathcal{H}^{1}(\Gamma_{\vek X}^{h})\ .\label{eq:SurfaceFEMFctSpace}
\end{align}
$M_{i}^{q}(\vek X)$ are obtained by isoparametric mappings from the
$q$-dimensional reference element to the physical elements in $d$
dimensions. Based on Eq.~(\ref{eq:SurfaceFEMFctSpace}), the following
discrete test and trial function spaces are introduced 
\begin{align}
\mathcal{S}_{\Gamma_{\vek X}}^{h} & =\left\lbrace \vek v_{h}\in\left[\mathcal{Q}_{\Gamma_{\vek X}}^{h}\right]^{d}:\ \vek v_{h}=\hat{\vek G}\ \text{on}\ \partial\Gamma_{\vek X,\text{D}}^{h}\right\rbrace ,\label{eq:TrialFctSpaceDiscr}\\
\mathcal{V}_{\Gamma_{\vek X}}^{h} & =\left\lbrace \vek v_{h}\in\left[\mathcal{Q}_{\Gamma_{\vek X}}^{h}\right]^{d}:\ \vek v_{h}=\vek0\ \text{on}\ \partial\Gamma_{\vek X,\text{D}}^{h}\right\rbrace .\label{eq:TestFctSpaceDiscr}
\end{align}
The discrete weak form of Eq.~(\ref{eq:WeakFormCont}) reads as follows:
Given Lam\'e constants $(\lambda,\mu)\in\mathbb{R}^{+}$, body forces
$\vek F\in\mathbb{R}^{d}$ on $\Gamma_{\vek X}^{h}$, tractions $\hat{\vek H}\in\mathbb{R}^{d}$
on $\partial\Gamma_{\vek X,\text{N}}^{h}$, find the displacement
field $\vek u_{h}\in\mathcal{S}_{\Gamma_{\vek X}}^{h}$ such that
for all test functions $\vek w_{h}\in\mathcal{V}_{\Gamma_{\vek X}}^{h}$
there holds in $\Gamma_{\vek X}^{h}$
\begin{equation}
\eta\cdot\int_{\Gamma_{\vek X}^{h}}\nabla_{\vek X}^{\Gamma,\mathrm{dir}}\vek w_{h}:\mat K\left(\vek u_{h}\right)\,\mathrm{d}\Gamma=\eta\cdot\int_{\Gamma_{\vek X}^{h}}\vek w_{h}\cdot\vek F\,\mathrm{d}\Gamma+\int_{\partial\Gamma_{\vek X,\text{N}}^{h}}\vek w_{h}\cdot\hat{\vek H}\,\mathrm{d}\partial\Gamma.\label{eq:WeakFormSurfaceFEM}
\end{equation}
The sought discrete displacement field $\vek u_{h}(\vek X)$ is obtained
solving a non-linear system of equations for the $n_{\text{DOF}}=d\cdot n_{q}$
nodal values (degrees of freedom) as usual in the context of finite
strain theory.

\subsection{Trace FEM\label{XX_TraceFEM}}

\begin{figure}
\centering

\subfigure[cable in $\mathbb{R}^2$]{\includegraphics[width=0.4\textwidth]{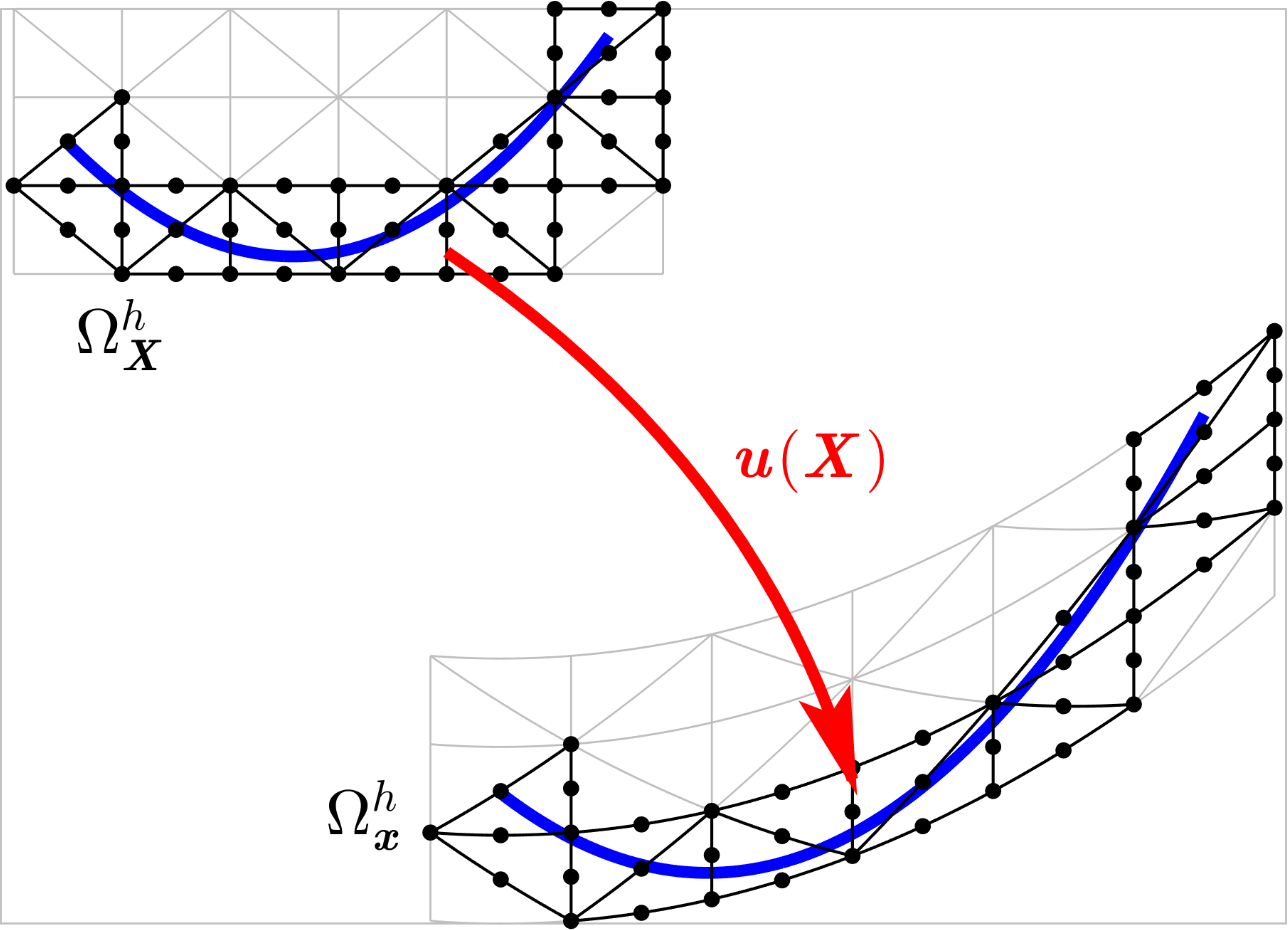}}\qquad\subfigure[cable in $\mathbb{R}^3$]{\includegraphics[width=0.35\textwidth]{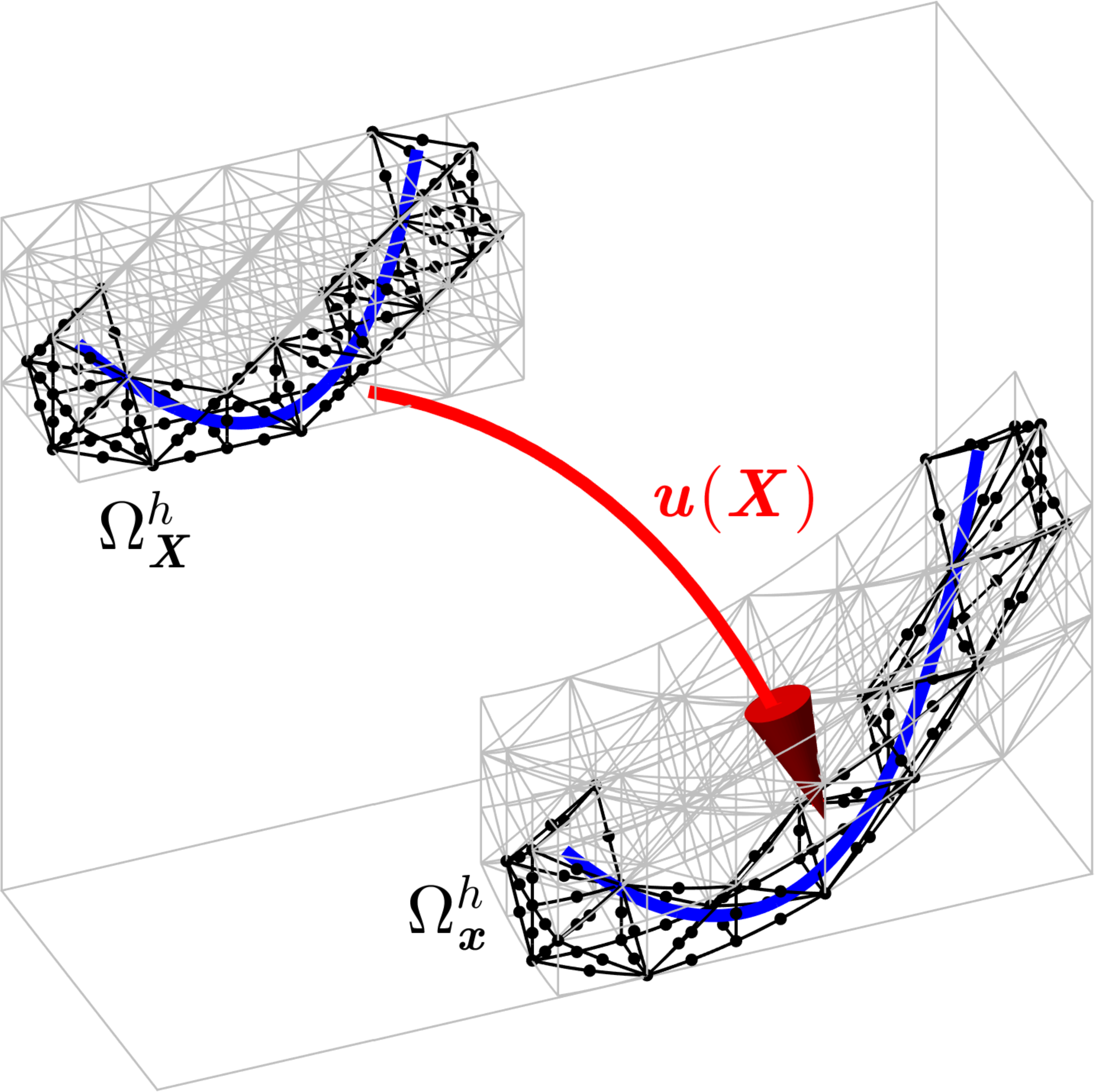}}

\subfigure[membrane in $\mathbb{R}^3$]{\includegraphics[width=0.4\textwidth]{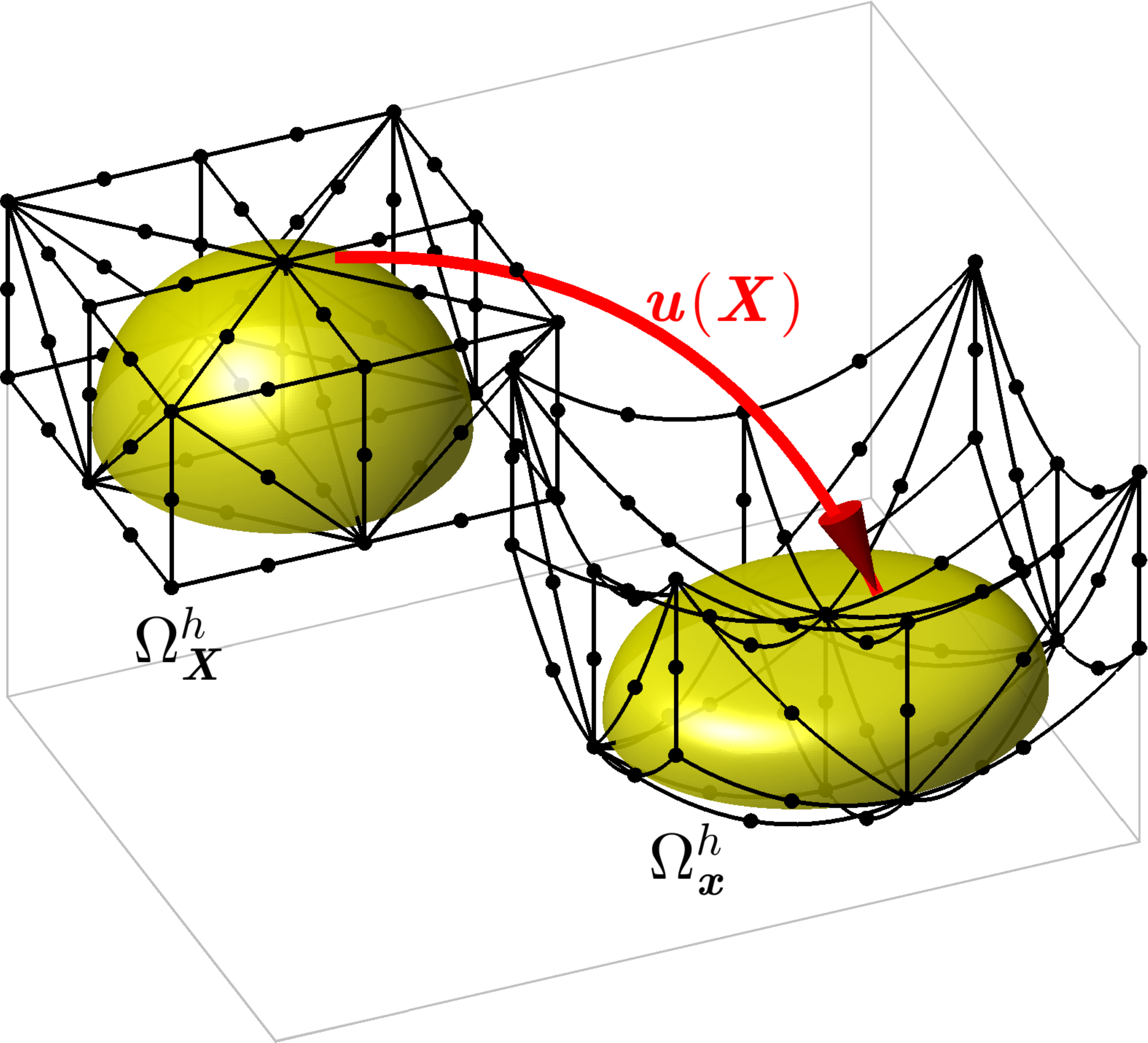}}

\caption{\label{fig:TraceFEM}The situation in the Trace FEM for (a) cables
in $\mathbb{R}^{2}$, (b) cables in $\mathbb{R}^{3}$, and (c) membranes
in $\mathbb{R}^{3}$. The discretized domains are shown for the undeformed
and deformed situations, $\Omega_{\vek X}^{h}$ and $\Omega_{\vek x}^{h}$,
respectively. Only the black background elements and their nodes are
\emph{active}. The undeformed and deformed manifolds coincide with
those shown in Fig.~\ref{fig:SurfaceFEM}.}
\end{figure}

\begin{figure}
\centering

\subfigure[active background elements]{\includegraphics[width=0.4\textwidth]{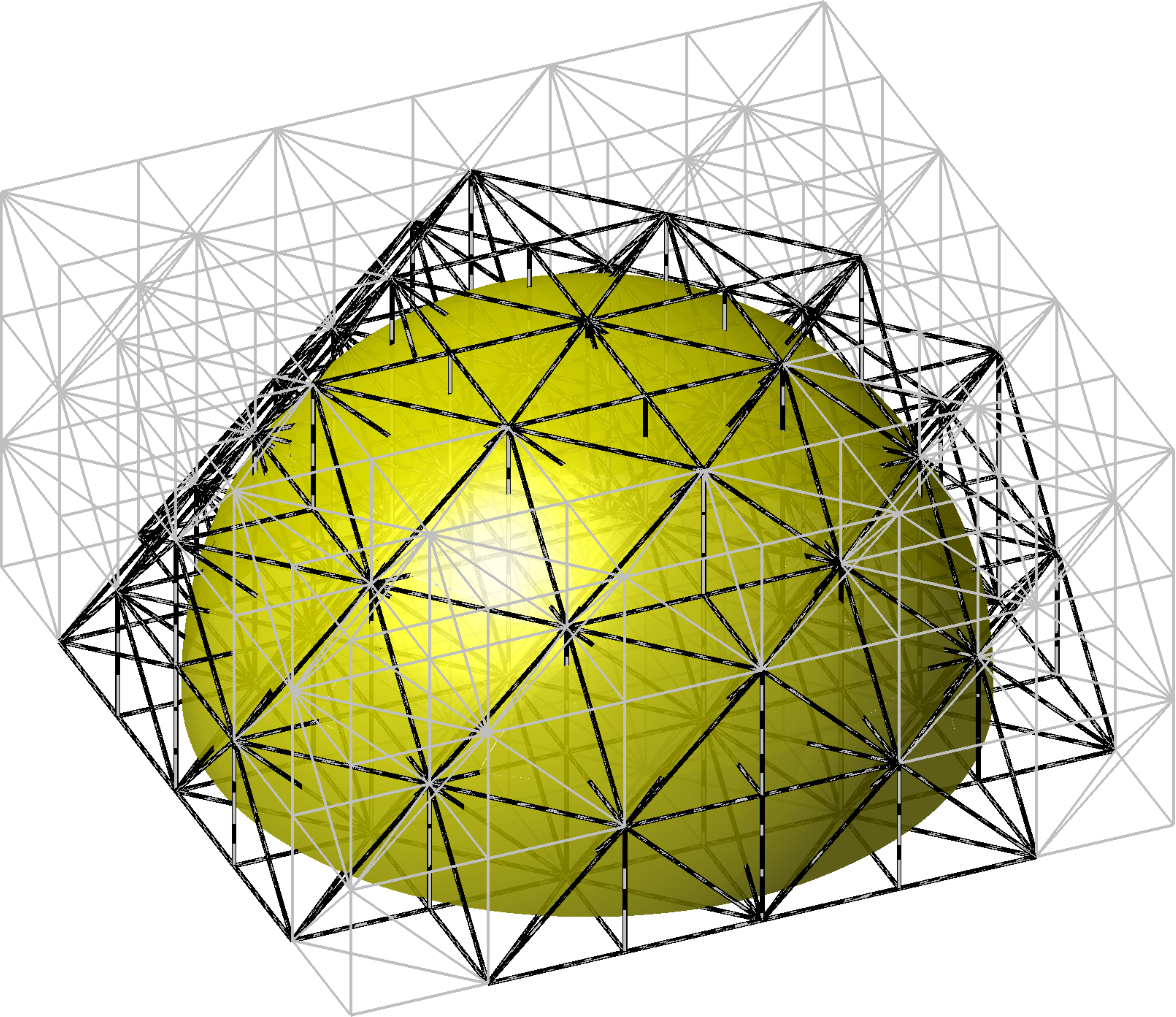}}\qquad\subfigure[integration cells]{\includegraphics[width=0.4\textwidth]{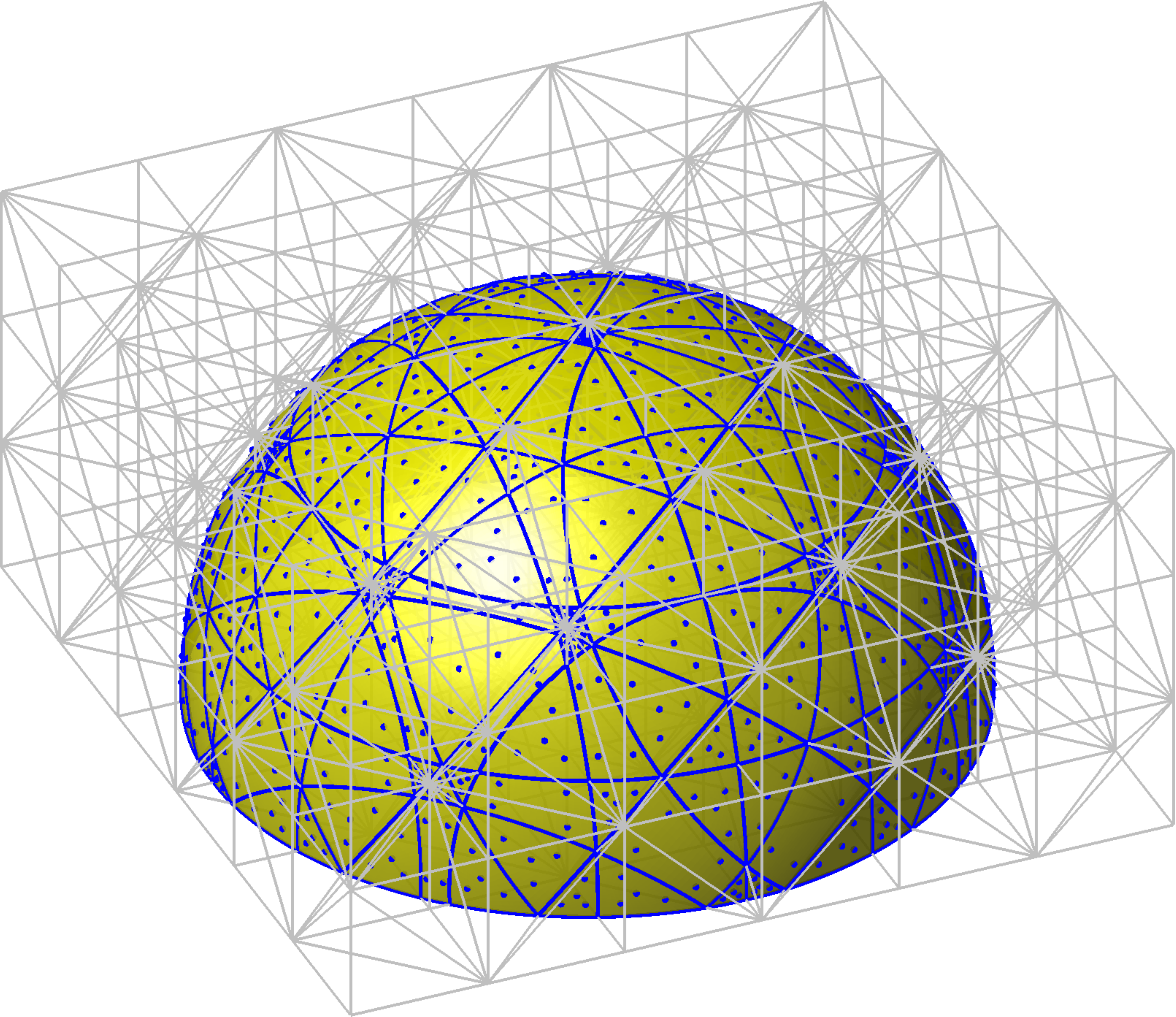}}

\caption{\label{fig:TraceFEMIntegration}(a) Active (black) elements in a background
mesh are those cut by the manifold. (b) Integration points have to
be identified within the active background elements. There, the shape
functions of the background elements are evaluated.}
\end{figure}

Let there be a $d$-dimensional background mesh into which the manifold
is completely immersed. Only those elements and corresponding nodes
are considered that are cut by the manifold, see Fig.~\ref{fig:TraceFEM}.
They may be labeled ``active'' elements and nodes, all others are
neglected. The shape functions of the active nodes are constructed
by (often isoparametric) mappings from a $d$-dimensional reference
element, but the shape functions are only evaluated on the $q$-dimensional
manifold.

The Trace FEM is a fictitious domain method (FDM) for PDEs on manifolds
\cite{Olshanskii_2009b,Olshanskii_2017a,Gross_2018a,Grande_2018a}.
As in any FDM, there is no boundary-conforming mesh but a background
mesh, herein further complicated by the fact that background mesh
and manifold have different dimensions. In general, the following
issues have to be properly addressed:
\begin{enumerate}
\item Integration points have to be defined for the integration of the weak
form of the governing equations---only at these points, shape functions
are evaluated. This requires the identification of the zero-level
set of some level-set function $\phi\left(\vek X\right)$ within each
active background element, see Fig.~\ref{fig:TraceFEMIntegration}.
The situation may be further complicated, when the boundary of the
manifold is within the background element, which may be defined by
additional slave level-set functions $\psi_{i}\left(\vek X\right)$
as mentioned in Section \ref{XX_ImplicitManifolds}. The placement
of integration points is an important and challenging task, in particular
with higher-order accuracy. For an overview, we refer to the references
\cite{Fries_2015a,Omerovic_2016a,Fries_2016b,Fries_2017a}. It is
useful to evaluate given level-set functions at the active nodes and
interpolate them in-between. The identification of the zero-level
sets and placement of integration points may then be achieved in the
$d$-dimensional reference element which simplifies the evaluation
of shape functions in the background elements. Of course, the \emph{interpolated}
zero-level set is only a (higher-order) approximation of the exact
manifolds $\Gamma_{\vek X}$ and $\Gamma_{\vek x}$ and labeled $\Gamma_{\vek X}^{h}$
and $\Gamma_{\vek x}^{h}$ , respectively. In the Trace FEM context,
$\Gamma_{\vek X}^{h}$ and $\Gamma_{\vek x}^{h}$ may be seen as integration
cells, see Fig.~\ref{fig:TraceFEMIntegration}(b), to define integration
points, they do not imply shape functions.
\item The treatment of boundary conditions is a challenging task in FDMs
as it is impossible to directly prescribe values of the nodes in the
active background elements. The additional constraints may, in principle,
be enforced using penalty methods, Lagrange multiplier methods, or
Nitsche's method. The latter has been developed to be a standard choice
in FDMs because the equations are formulated in a consistent way without
needing additional degrees of freedom \cite{Nitsche_1971a}. In the
standard (symmetric) form of the Nitsche's method, stabilization parameters
are required. In the simplest case, these parameters may be set to
a fixed user-defined number \cite{Burman_2012a,Hansbo_2002a}, however,
for background elements which are cut by a tiny fraction of the manifold,
resulting in small supports of the shape functions, this may lead
to unsatisfactory results. An alternative is to compute the stabilization
parameters based on global or element-wise generalized eigenvalue
problems \cite{Mendez_2004a,Ruess_2013a}. However, for the awkward
cut situations, this may result in unbounded values resulting in similar
problems known for penalty methods \cite{dePrenter_2018a}. Therefore,
we prefer the \emph{non}-symmetric Nitsche's method herein with the
main advantage that an additional stabilization is \emph{not} required
for imposing boundary conditions \cite{Burman_2012b,Schillinger_2016a}.
\item Stabilization is still necessary when applying FDMs in the context
of PDEs on manifolds to ensure the regularity of the resulting system
of equations. This may be traced back to two sources: One is found
in the shape functions with small supports and the other in the fact
that the approximated displacement field on the manifold may not necessarily
be represented by a \emph{unique} set of nodal values in the background
mesh. That is, the background shape functions restricted to the manifold
build a frame but not necessarily a basis \cite{Grande_2018a,Reusken_2014a,Olshanskii_2017a}.
Fortunately, different stabilization approaches exist to cure both
issues and we refer to the overview given in \cite{Olshanskii_2017a}
for the Trace FEM. Herein, we use the ``normal derivative volume
stabilization'', introduced for scalar-valued problems in \cite{Grande_2016a,Burman_2016b}
and in \cite{Gross_2018a} for vector-valued problems. This stabilization
technique enables higher-order accurate results, does not change the
sparsity pattern of the stiffness matrix, and only first derivatives
are needed.
\end{enumerate}
With these comments made, we are ready to define the discrete weak
form for the Trace FEM. Let $\Omega_{\vek X}^{h}$ be the background
mesh into which the undeformed configuration $\Gamma_{\vek X}$ is
immersed, only \emph{active} elements and nodes are present in $\Omega_{\vek X}^{h}$.
The resulting shape functions $M_{i}^{d}\!\left(\vek X\right)$ span
a $C_{0}$-continuous finite element space on the manifold as 
\begin{align}
\mathcal{Q}_{\Omega_{\vek X}}^{h}:=\left\lbrace v_{h}\in C^{0}(\Omega_{\vek X}^{h}):\ v_{h}=\sum_{i=1}^{n_{d}}M_{i}^{d}(\vek X)\cdot\hat{v}_{i}\text{ with }\hat{v}_{i}\in\mathbb{R}\right\rbrace \subset\mathcal{H}^{1}(\Omega_{\vek X}^{h})\ .\label{eq:TraceFEMFctSpace}
\end{align}
Based on Eq.~(\ref{eq:TraceFEMFctSpace}), the following discrete
\emph{trace} test and trial function spaces are introduced 
\begin{align}
\mathcal{T}_{\Gamma_{\vek X}}^{h} & =\left\lbrace \vek v_{h}\in\left[\mathcal{Q}_{\Omega_{\vek X}}^{h}\right]^{d}:\ \vek v_{h}\text{ only on }\Gamma_{\vek X}^{h}\right\rbrace ,\label{eq:TrialFctSpaceDiscrTraceFEM}\\
\mathcal{U}_{\Gamma_{\vek X}}^{h} & =\mathcal{T}_{\Gamma_{\vek X}}^{h}.\label{eq:TestFctSpaceDiscrTraceFEM}
\end{align}
For the Trace FEM, the discrete weak form of Eq.~(\ref{eq:WeakFormCont})
is: Given Lam\'e constants $(\lambda,\mu)\in\mathbb{R}^{+}$, body
forces $\vek F\in\mathbb{R}^{d}$ on $\Gamma_{\vek X}^{h}$, tractions
$\hat{\vek H}\in\mathbb{R}^{d}$ on $\partial\Gamma_{\vek X,\text{N}}^{h}$,
and stabilization parameter $\rho$, find the displacement field $\vek u_{h}\in\mathcal{T}_{\Gamma_{\vek X}}^{h}$
such that for all test functions $\vek w_{h}\in\mathcal{U}_{\Gamma{}_{\vek X}}^{h}$
there holds in $\Gamma_{\vek X}^{h}$
\begin{align}
\eta\cdot\int_{\Gamma_{\vek X}^{h}}\nabla_{\vek X}^{\Gamma,\mathrm{dir}}\vek w_{h}:\mat K\left(\vek u_{h}\right)\,\mathrm{d}\Gamma\underset{\text{boundary term due to \ensuremath{\vek w_{h}\neq\vek0} on }\partial\Gamma_{\vek X,\mathrm{D}}^{h}}{-\underbrace{\int_{\partial\Gamma_{\vek{X},\text{D}}^{h}}\vek w_{h}\cdot\left[\mat K(\vek u_{h})\cdot\vek N_{\partial\Gamma_{\vek X}}\right]\,\mathrm{d}\partial\Gamma}}+\label{eq:WeakFormTraceFEM-1}\\
\underset{\text{Nitsche term}}{\underbrace{\int_{\partial\Gamma_{\vek X,\text{D}}^{h}}\left(\vek u_{h}-\hat{\vek G}\right)\cdot\left[\mat K(\vek w_{h})\cdot\vek N_{\partial\Gamma_{\vek X}}\right]\,\mathrm{d}\partial\Gamma}}+\underset{\text{stabilization term}}{\underbrace{\rho\int_{\Omega_{\vek X}^{h}}(\vek N^{\mathrm{e}}\cdot\nabla_{\!\vek X}\vek u_{h})\cdot(\vek N^{\mathrm{e}}\cdot\nabla_{\!\vek X}\vek w_{h})\,\mathrm{d}\Omega}}\nonumber \\
=\eta\cdot\int_{\Gamma_{\vek X}^{h}}\vek w_{h}\cdot\vek F\,\mathrm{d}\Gamma+\int_{\partial\Gamma_{\vek X,\text{N}}^{h}}\vek w_{h}\cdot\hat{\vek H}\,\mathrm{d}\partial\Gamma\ .\nonumber 
\end{align}
In comparison to Eq.~(\ref{eq:WeakFormSurfaceFEM}), additional terms
occur in the discrete weak form due to the weak enforcement of essential
boundary conditions with Nitsche's method and the stabilization. The
sought discrete displacement field $\vek u_{h}(\vek X)$ is obtained
solving a non-linear system of equations for the $n_{\text{DOF}}=d\cdot n_{d}$
nodal values (degrees of freedom). 

Note that the stabilization term is the only term which is not evaluated
on the manifold $\Gamma_{\vek X}^{h}$ but in the volumetric background
mesh $\Omega_{\vek X}^{h}$ (using standard Gauss integration). Therefore,
one has to extend the normal vector $\vek N\!\left(\vek X\right)$
from the undeformed manifold $\Gamma_{\vek X}^{h}$ to the neighborhood,
resulting in $\vek N^{\mathrm{e}}\!\left(\vek X\right)$ for all $\vek X\in\Omega_{\vek X}^{h}$.
This is particularly simple for implicitly defined manifolds, e.g.,
using $\vek N^{\mathrm{e}}(\vek X)=\frac{\nabla_{\!\vek X}\phi^{h}(\vek X)}{\Vert\nabla_{\!\vek X}\phi^{h}(\vek X)\Vert}$
for all $\vek X\in\Omega_{\vek X}^{h}$. Furthermore, in the stabilization
term, the \emph{classical} gradient operator $\nabla_{\!\vek X}$
is used instead of the \emph{surface} operators used in the other
terms. In \cite{Grande_2016a}, it is recommended that the stabilization
parameter should be chosen in the range $O\left(h\right)\lesssim\rho\lesssim O\left(h^{-1}\right)$,
where $h$ is the element size in the active background mesh.

A remark is added concerning \emph{slip supports} because the above
mentioned weak form rather expects that all displacement components
are prescribed through $\hat{\vek G}$ along the Dirichlet boundary
$\partial\Gamma_{\vek X,\mathrm{D}}^{h}$. In the Nitsche's method,
displacement constraints in selected, arbitrary unit directions $\vek v_{d}$
with magnitude $\hat{G}$ may be prescribed by replacing the corresponding
terms in Eq.~(\ref{eq:WeakFormTraceFEM-1}) with
\begin{align}
\begin{split}-\int_{\partial\Gamma_{\vek X,\text{D}}^{h}}\left(\vek w_{h}\cdot\vek v_{d}\right)\,\left[\mat K(\vek u_{h})\cdot\vek N_{\partial\Gamma_{\vek X}}\right]\cdot\vek v_{d}\,\mathrm{d}\partial\Gamma\\
+\int_{\partial\Gamma_{\vek X,\text{D}}^{h}}\left(\vek u_{h}\cdot\vek v_{d}-\hat{G}\right)\,\left[\mat K(\vek w_{h})\cdot\vek N_{\partial\Gamma_{\vek X}}\right]\cdot\vek v_{d}\,\mathrm{d}\partial\Gamma\ .
\end{split}
\end{align}

\section{Numerical results\label{X_NumericalResults}}

A number of test cases for ropes and membranes in two and three dimensions
are considered in this section. The numerical results focus on the
convergence rates for two different types of errors. The ``energy
error'' $\varepsilon_{\mathfrak{e}}$ compares the approximated stored
elastic energy with the analytical one,
\begin{equation}
\varepsilon_{\mathfrak{e}}=\left|\mathfrak{e}\left(\vek u\right)-\mathfrak{e}\left(\vek u_{h}\right)\right|,\label{eq:EnergyError}
\end{equation}
with $\mathfrak{e}$ computed based on Eq.~(\ref{eq:EnergyDef}).
The analytical energy $\mathfrak{e}\left(\vek u\right)$ may also
be computed by an overkill approximation, i.e., based on an extremely
fine mesh with higher-order elements. Provided that geometry and boundary
conditions allow for sufficiently smooth solutions, the expected convergence
rates in this error norm are $p+1$ with $p$ being the order of the
elements.

The ``residual error'' $\varepsilon_{\mathrm{res}}$ integrates
the error in the equilibrium as stated in Eq.~(\ref{eq:EquilibriumDefConfig}),
that is,
\begin{equation}
\varepsilon_{\mathrm{res}}=\sqrt{\sum_{\mathrm{e}=1}^{n_{\mathrm{el}}}\int_{\Gamma_{\vek X}^{h,\mathrm{e}}}\mathfrak{r}\left(\vek u_{h}\right)\cdot\mathfrak{r}\left(\vek u_{h}\right)\,\mathrm{d}\Gamma}\quad\text{with }\mathfrak{r}\left(\vek u_{h}\right)=\mathrm{div}_{\Gamma}\,\vek\sigma\!\left(\vek u_{h}\right)+\vek f\!\left(\vek x\right)\label{eq:ResidualError}
\end{equation}
This error obviously vanishes for the analytical solution. It is important
to note that the integrand in (\ref{eq:ResidualError}) involves second-order
derivatives. Therefore, the integral must not be carried out over
the whole (discretized) domain $\Gamma_{\vek X}^{h}$ but integrated
element by element as indicated by the summation. That is, element
boundaries, where already the first derivatives of the $C^{0}$-continuous
shape functions feature jumps, are neglected in computing $\varepsilon_{\mathrm{res}}$.
Due to the presence of second-order derivatives, the expected convergence
rates are $p-1$ which also indicates that higher-order elements are
crucial for convergence in $\varepsilon_{\mathrm{res}}$.

\subsection{Membrane with given deformation}

In the first test case, a membrane in the shape of a half sphere with
radius $r=1.0$ undergoes a prescribed displacement and the stored
elastic energy is computed from the viewpoint of the Surface FEM and
the Trace FEM. That is, in the Surface FEM, \emph{surface} meshes
with different resolutions and element orders are generated. See Fig.~\ref{fig:SurfaceFEM}(c)
for some example mesh composed by quadratic elements. The displacements
\begin{equation}
\vek u\left(\vek X\right)=\left[\begin{array}{c}
\nicefrac{5}{2}+\nicefrac{1}{5}\left(X+1\right)\\
-\nicefrac{3}{2}\\
-\nicefrac{1}{2}\left[1-\left(X^{2}+Y^{2}\right)\right]-\nicefrac{3}{2}
\end{array}\right]\label{eq:TC1_DisplField}
\end{equation}
are evaluated at the nodes and interpolated based on the shape functions
implied by the \emph{surface} meshes, yielding $\vek u_{h}\!\left(\vek X\right)$;
see Fig.~\ref{fig:SurfaceFEM}(c) for the resulting deformed membrane.
Then, the elastic energy of the deformed configuration $\mathfrak{e}\left(\vek u_{h}\right)$
is computed with Eq.~(\ref{eq:EnergyDef}).

For the Trace FEM viewpoint, \emph{background} meshes of different
resolutions and orders are generated in $\Omega_{\vek X}=\left[-1,1\right]\times\left[-1,1\right]\times\left[0,1\right]$
and the geometry is defined based on the level-set function $\phi\left(\vek X\right)=\left\Vert \vek X\right\Vert -r$.
See Fig.~\ref{fig:TraceFEM}(c) for a sketch of the situation using
quadratic background elements. Then, the (active) nodes of the background
mesh are deformed by the given displacement field (\ref{eq:TC1_DisplField}),
yielding $\vek u_{h}\!\left(\vek X\right)$ based on the shape functions
implied by the \emph{background} meshes. This displacement field living
in the whole background mesh is only evaluated on the membrane surface
in order to compute the stored energy $\mathfrak{e}\left(\vek u_{h}\right)$
according to the Trace FEM.

\begin{figure}
\centering

\subfigure[Surface FEM, $\varepsilon_{\mathfrak{e}}$]{\includegraphics[width=0.45\textwidth]{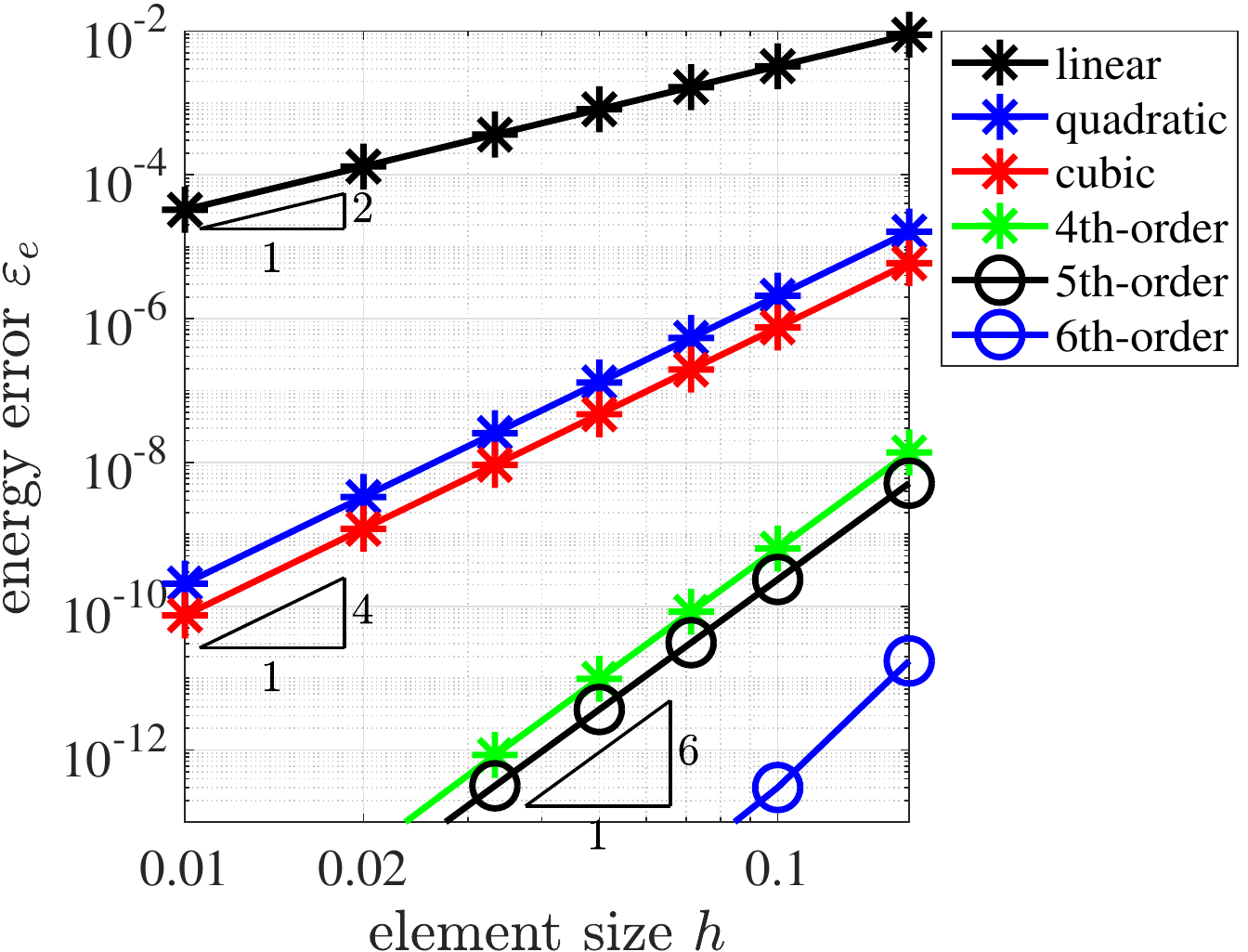}}\hfill\subfigure[Trace FEM, $\varepsilon_{\mathfrak{e}}$]{\includegraphics[width=0.45\textwidth]{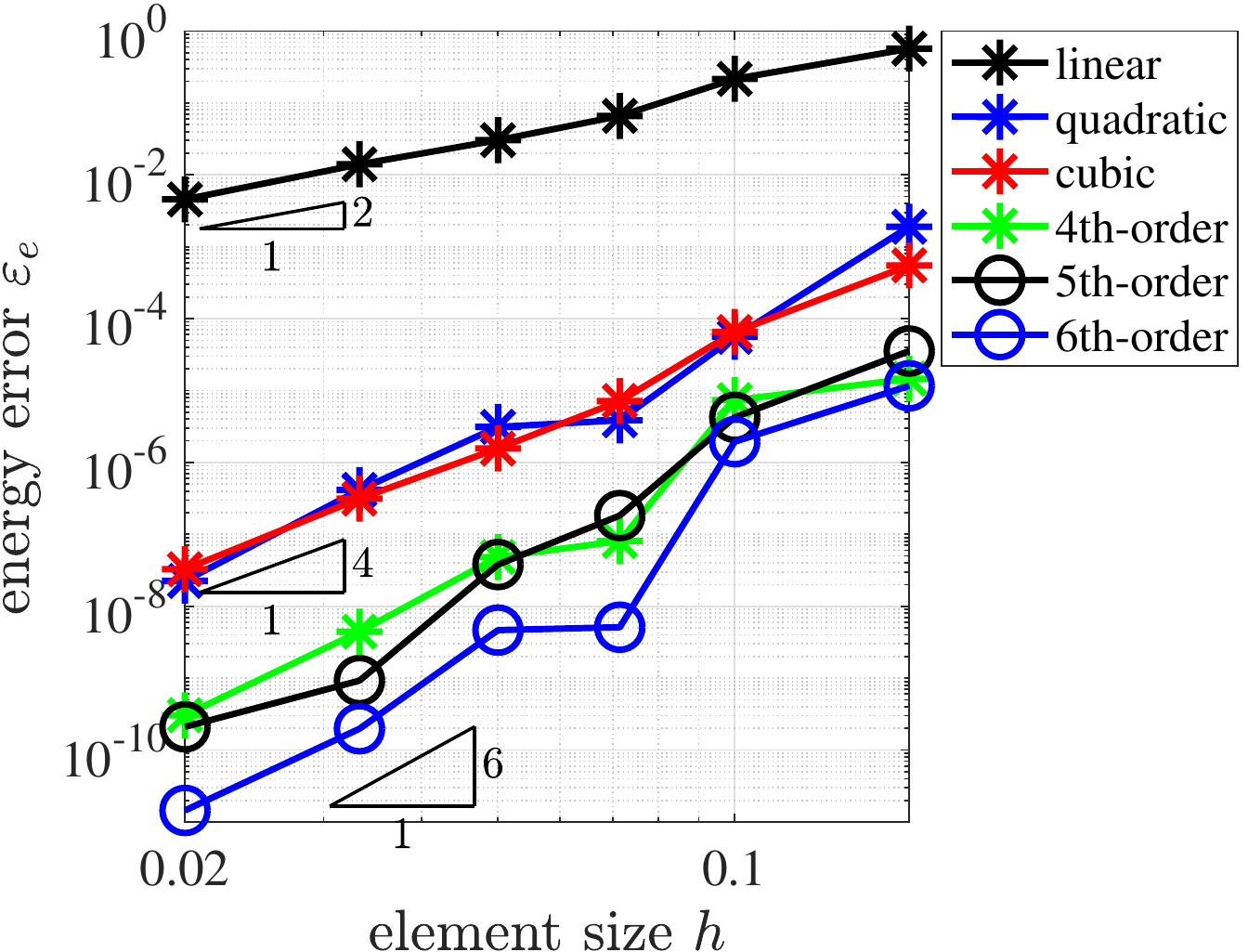}}

\caption{\label{fig:TC1_Results}Convergence results for test case 1: The energy
error $\varepsilon_{\mathfrak{e}}$ for the (a) Surface FEM and (b)
Trace FEM.}
\end{figure}

We set the Lam\'e constants to $\lambda=3$ and $\mu=2$. The resulting
energy is given by the value $\mathfrak{e}\left(\vek u\right)=1.642871443585262$.
In Fig.~\ref{fig:TC1_Results}, the convergence results for the various
meshes are shown for the Surface and the Trace FEM. It is seen that
in both cases \emph{optimal} convergence results are achieved. The
energy error converges one order higher than expected for even element
orders. In the Trace FEM, the convergence curves are less smooth than
in the Surface FEM because the approximation spaces are not nested
upon refinement; this is well-known for results obtained with FDMs
in general.

It is thus seen that the Surface FEM as well as the Trace FEM have
the potential to achieve optimal results. For all other test cases
below, we shall now obtain the discrete displacement fields based
on solving the non-linear systems of equations resulting from the
weak forms given in Section \ref{X_Discretization}.

\subsection{Rope with Surface and Trace FEM}

\begin{figure}
\centering

\subfigure[overview]{\includegraphics[height=0.28\textwidth]{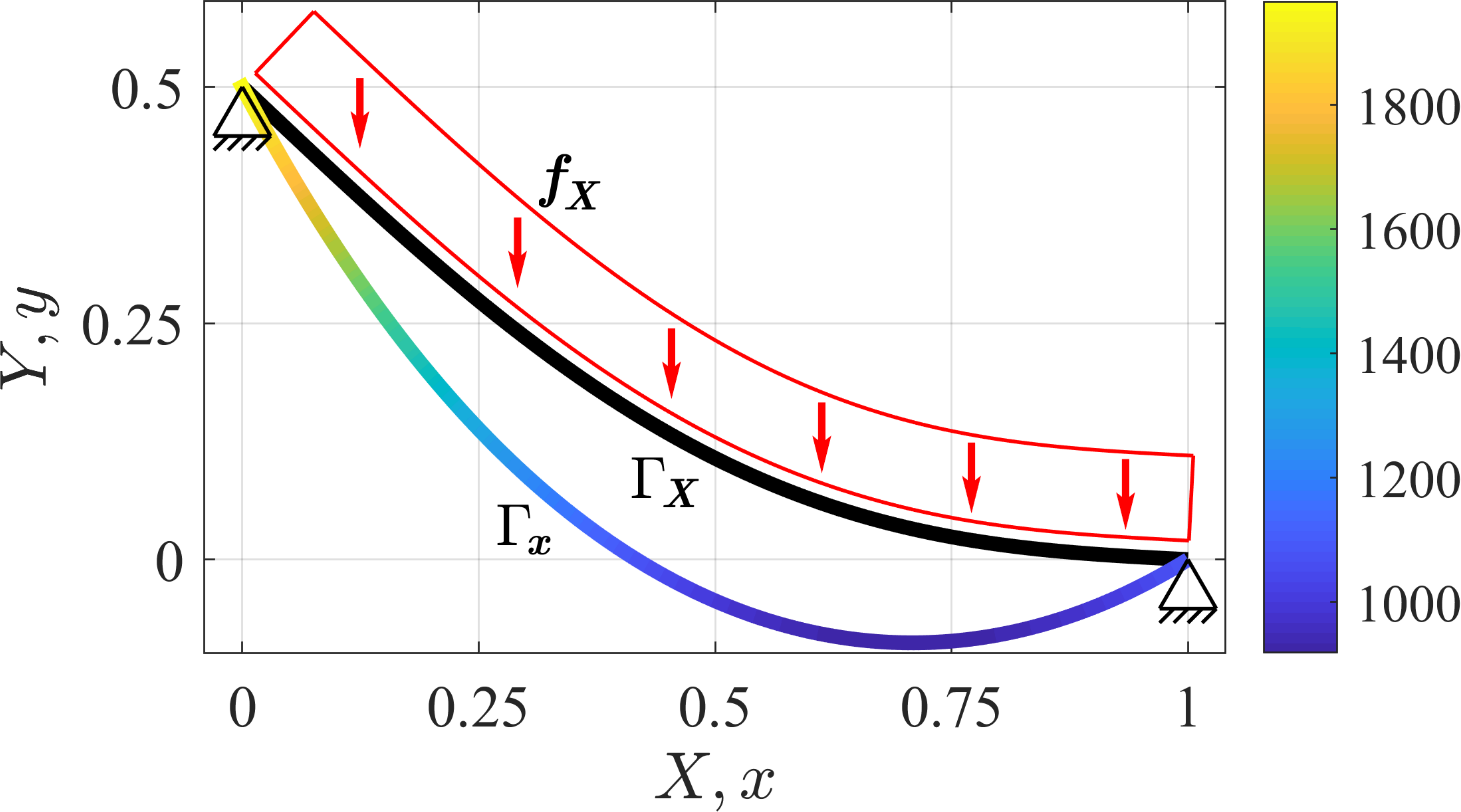}}\qquad\subfigure[Trace FEM]{\includegraphics[height=0.28\textwidth]{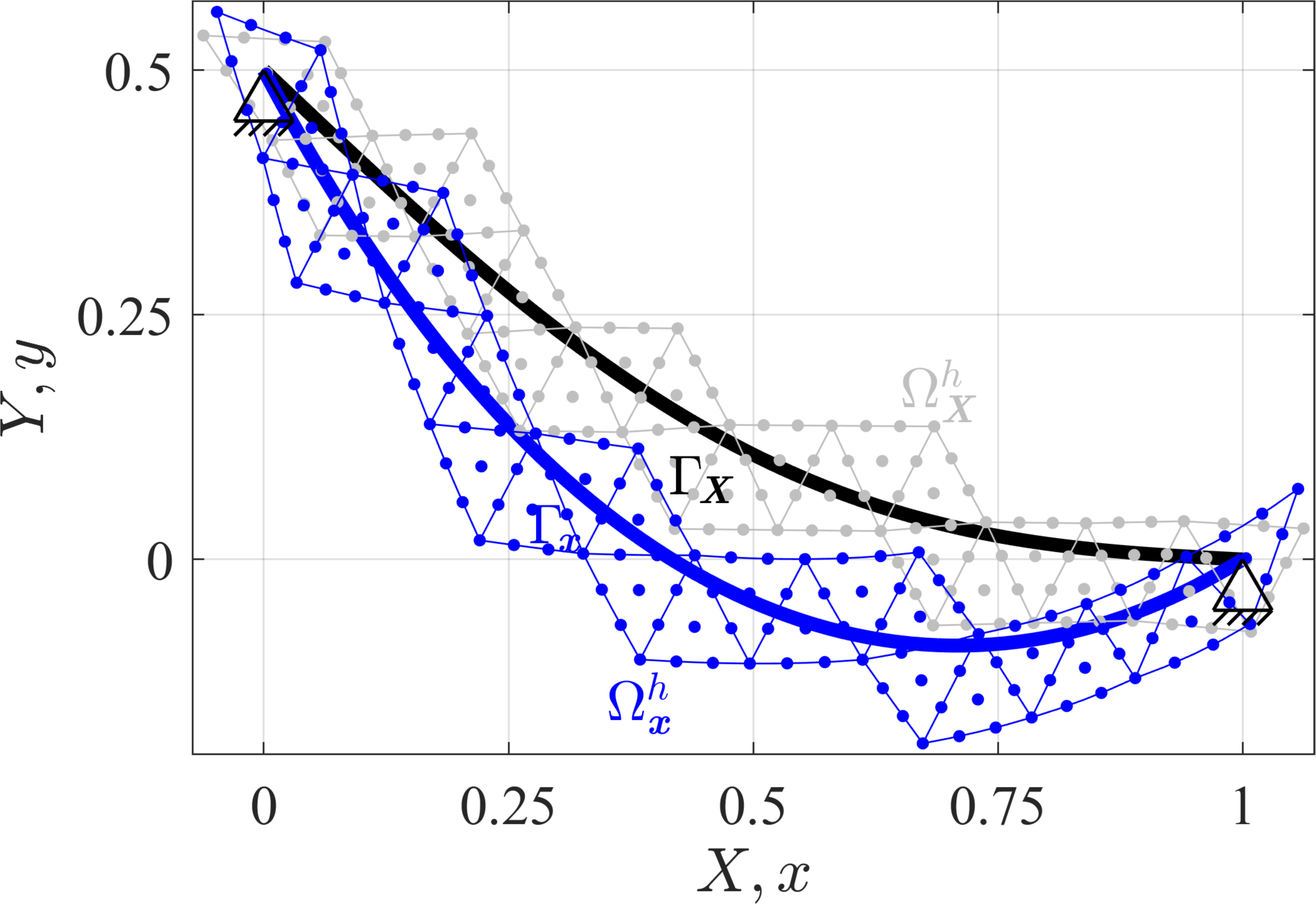}}

\caption{\label{fig:TC2_Situation}(a) Sketch of test case 2 including the
deformed and undeformed configuration. The colored deformed configuration
shows the principal Cauchy stress. (b) The situation in the Trace
FEM using two-dimensional cubic background elements.}
\end{figure}

The second test case considers a rope in two dimensions as shown in
Fig.~\ref{fig:TC2_Situation}. The cross section of the rope is $A=0.01$,
Young's modulus is $E=10\,000$, and the Lam\'e constants are $\mu=\nicefrac{1}{2}E$
and $\lambda=0$. The left support is located at $\left(0,\nicefrac{1}{2}\right)$
and the right at $\left(1,0\right)$. The geometry may be given in
parametric form for $r\in\left(0,1\right)$ as 
\[
\vek X\!\left(r\right)=\left[\begin{array}{c}
X\!\left(r\right)\\
Y\!\left(r\right)
\end{array}\right]=\left[\begin{array}{c}
r\\
\nicefrac{1}{2}\left(1-r\right)-\nicefrac{1}{7}\sin\left(\pi\cdot r\right)
\end{array}\right].
\]
Alternatively, the same geometry may be implied by the level-set function
\[
\phi\left(\vek X\right)=Y-\left[\nicefrac{1}{2}\left(1-X\right)-\nicefrac{1}{7}\sin\left(\pi\cdot X\right)\right]\quad\forall X\in\left(0,1\right).
\]
The structure is loaded by its own weight with $\vek F\!\left(\vek X\right)=\left[0,-2\,000\cdot A\right]^{\mathrm{T}}$
for all $\vek X\in\Gamma_{\!\vek X}$. The deformed rope is illustrated
in Fig.~\ref{fig:TC2_Situation}(a) where the color information indicates
the principal Cauchy stress in the rope. The stored elastic energy
is $\mathfrak{e}=0.7528302283000$ and the length of the rope is increased
by a factor of $1.1053648264108$ due to the deformation.

\begin{figure}
\centering

\subfigure[Surface FEM, $\varepsilon_{\mathfrak{e}}$]{\includegraphics[width=0.45\textwidth]{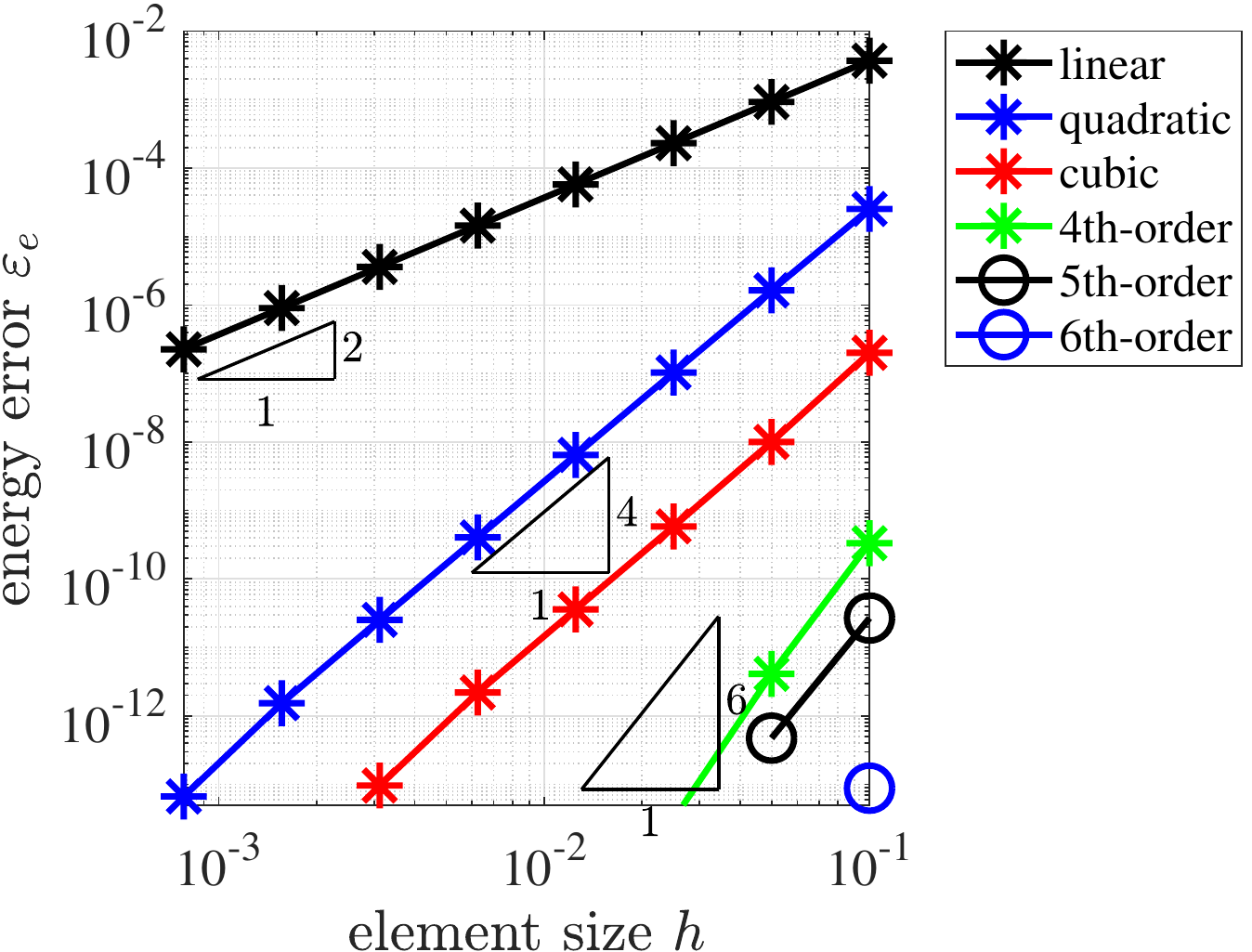}}\hfill\subfigure[Trace FEM, $\varepsilon_{\mathfrak{e}}$]{\includegraphics[width=0.45\textwidth]{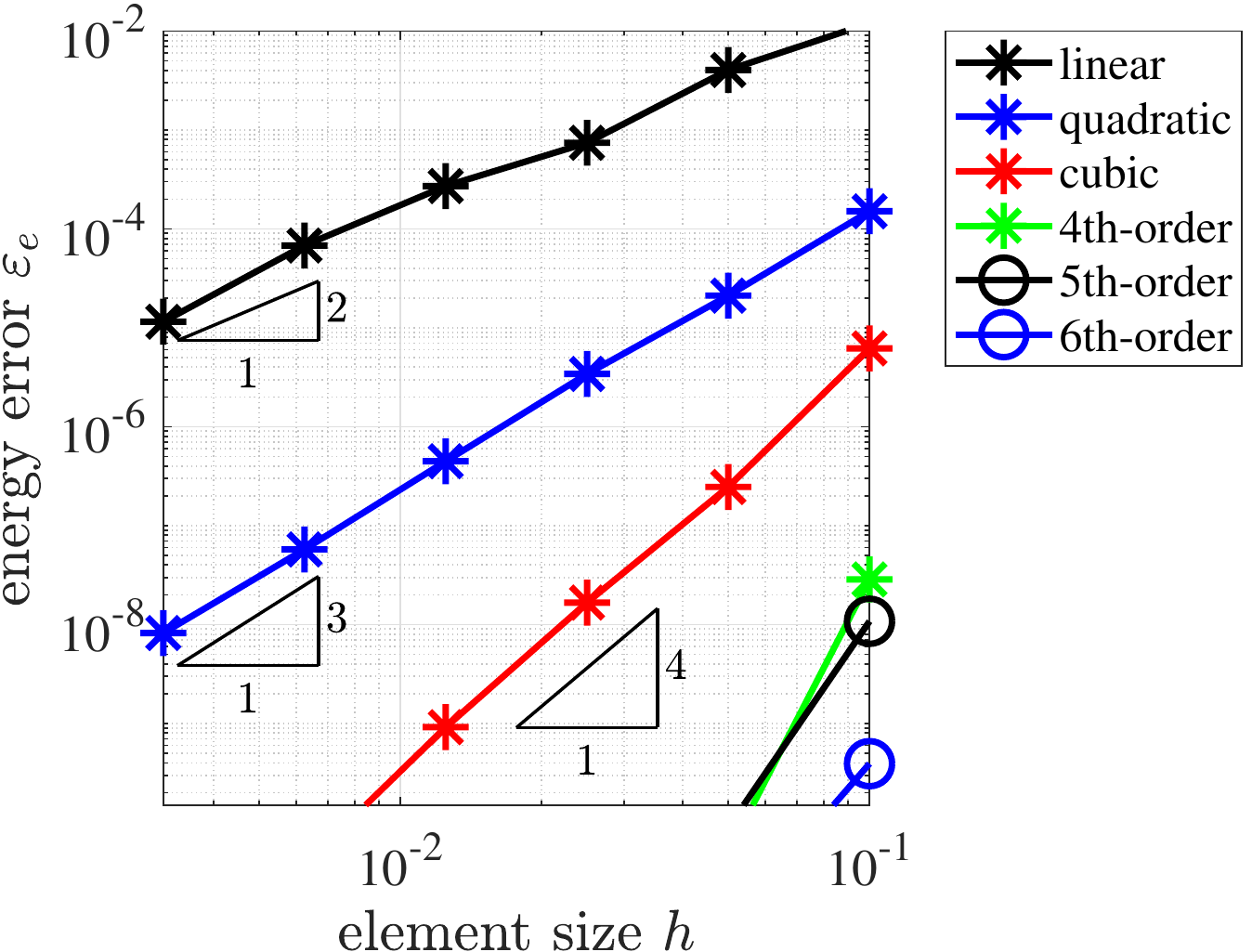}}

\subfigure[Surface FEM, $\varepsilon_{\mathrm{res}}$]{\includegraphics[width=0.45\textwidth]{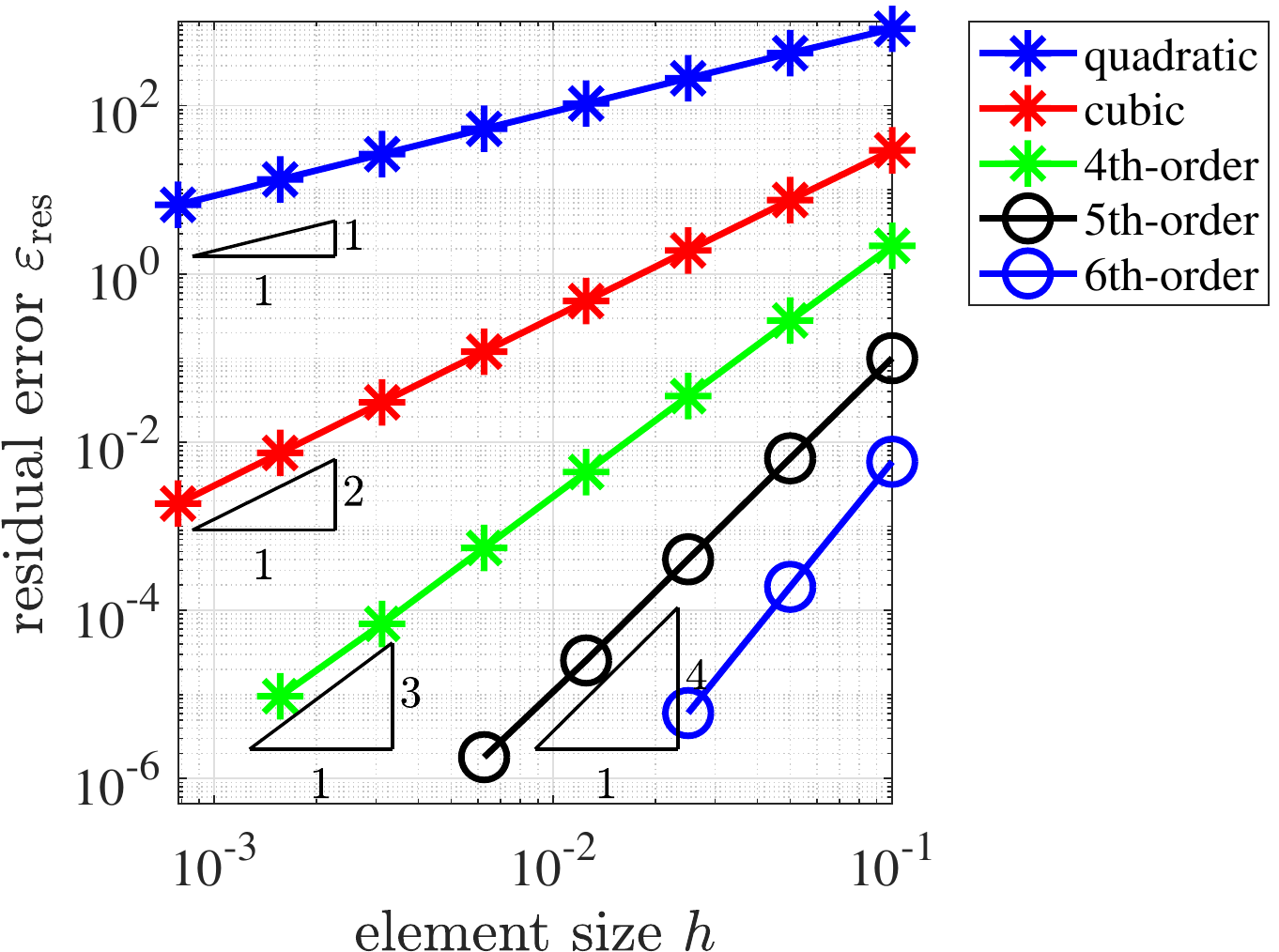}}\hfill\subfigure[Trace FEM, $\varepsilon_{\mathrm{res}}$]{\includegraphics[width=0.45\textwidth]{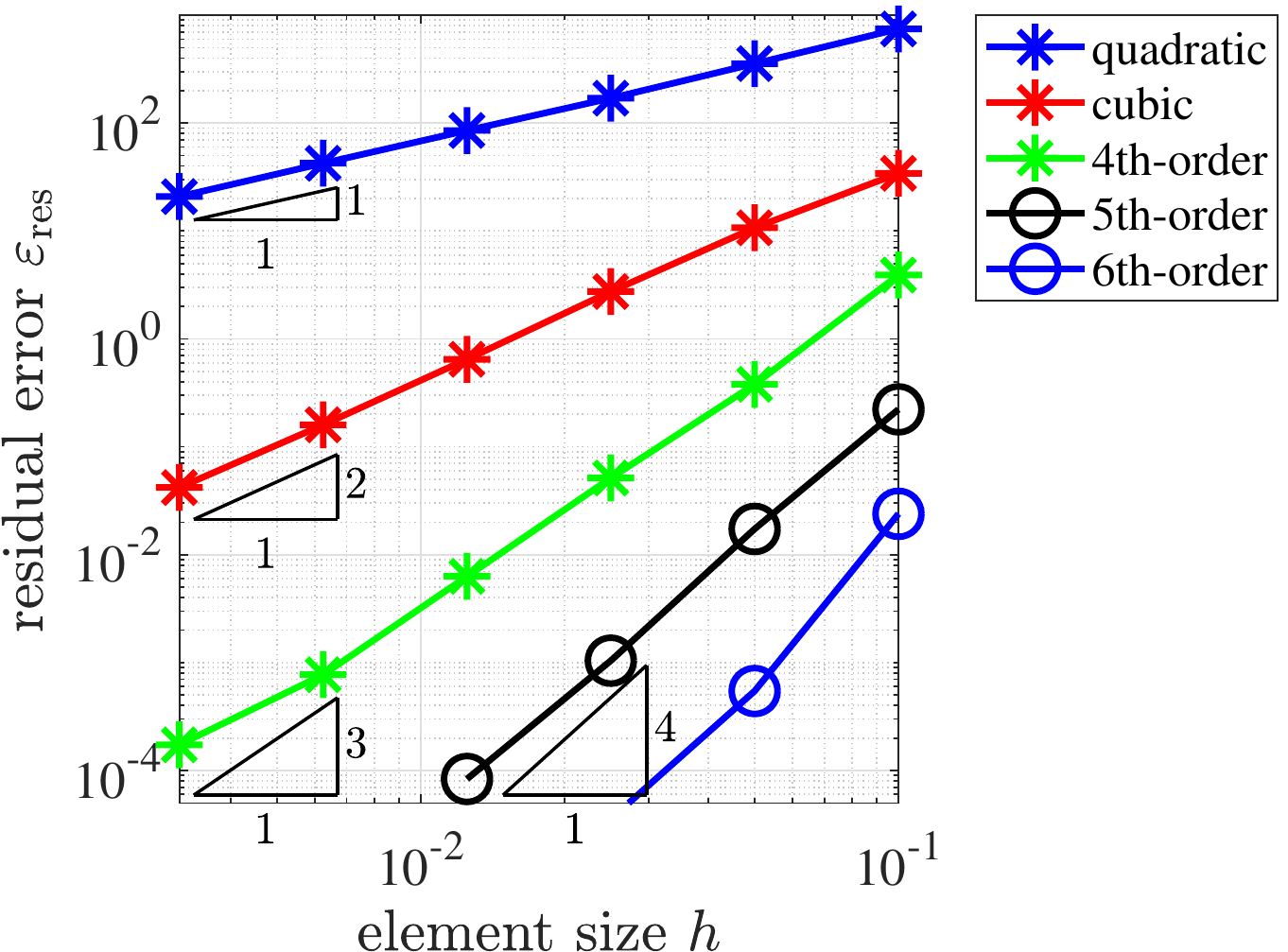}}

\caption{\label{fig:TC2_Results}Convergence results for test case 2: (a) and
(b) show the energy error $\varepsilon_{\mathfrak{e}}$, (c) and (d)
the residual error $\varepsilon_{\mathrm{res}}$ for the Surface and
Trace FEM, respectively.}
\end{figure}

The displacements are computed with the Surface and Trace FEM, respectively.
For the Trace FEM, the stabilization parameter in Eq.~(\ref{eq:WeakFormTraceFEM-1})
is set to $\rho=\nicefrac{1000}{h}$. The mesh resolutions and orders
are systematically varied and convergence results are seen in Fig.~\ref{fig:TC2_Results}.
Figs.~\ref{fig:TC2_Results}(a) and (b) show the error in the elastic
energy $\varepsilon_{\mathfrak{e}}$ and Figs.~\ref{fig:TC2_Results}(c)
and (d) the residual error $\varepsilon_{\mathrm{res}}$. Of course,
for some given element length $h$, the use of (curved) line meshes
in the Surface FEM results in considerably less degrees of freedom
than using background meshes in the Trace FEM. Therefore, the convergence
studies for the Surface FEM are realized for up to $1024$ elements
whereas the background meshes in the Trace FEM feature up to $320\times160$
elements (of which only those cut by the rope are active and, hence,
taken into account for the simulation). It is apparent from the convergence
results in Fig.~\ref{fig:TC2_Results} that optimal higher-order
convergence rates are achieved. It is noteworthy that the Surface
FEM with even element orders converges one order higher than expected
in $\varepsilon_{\mathfrak{e}}$, whereas this is not the case for
the Trace FEM. This can already be traced back to the accuracy in
the numerical integration: Simply integrating the length of the rope
using the Surface or Trace FEM viewpoint shows that an extra order
in the accuracy is achieved with the Surface FEM for even element
orders. For the convergence results in $\varepsilon_{\mathrm{res}}$,
results for linear meshes are omitted because second order derivatives
are needed for this error measure, see Eq.~(\ref{eq:ResidualError}).
In this error norm, Surface and Trace FEM converge optimally with
$p-1$.

\subsection{Membranes with Surface FEM}

\begin{figure}
\centering

\subfigure[Map A, undeformed conf.]{\includegraphics[width=0.35\textwidth]{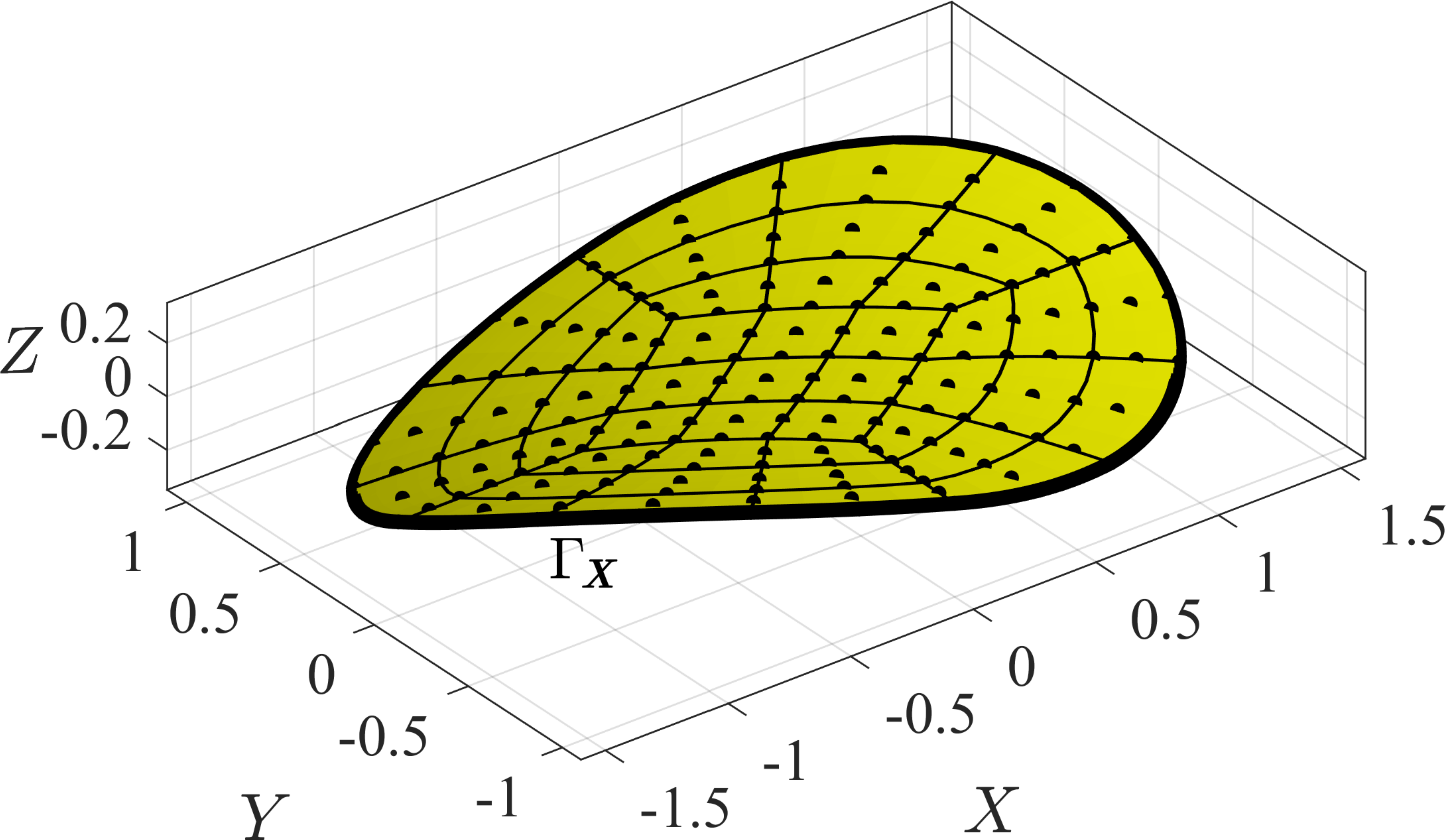}}\hfill\subfigure[Map A, deformed conf.]{\includegraphics[width=0.35\textwidth]{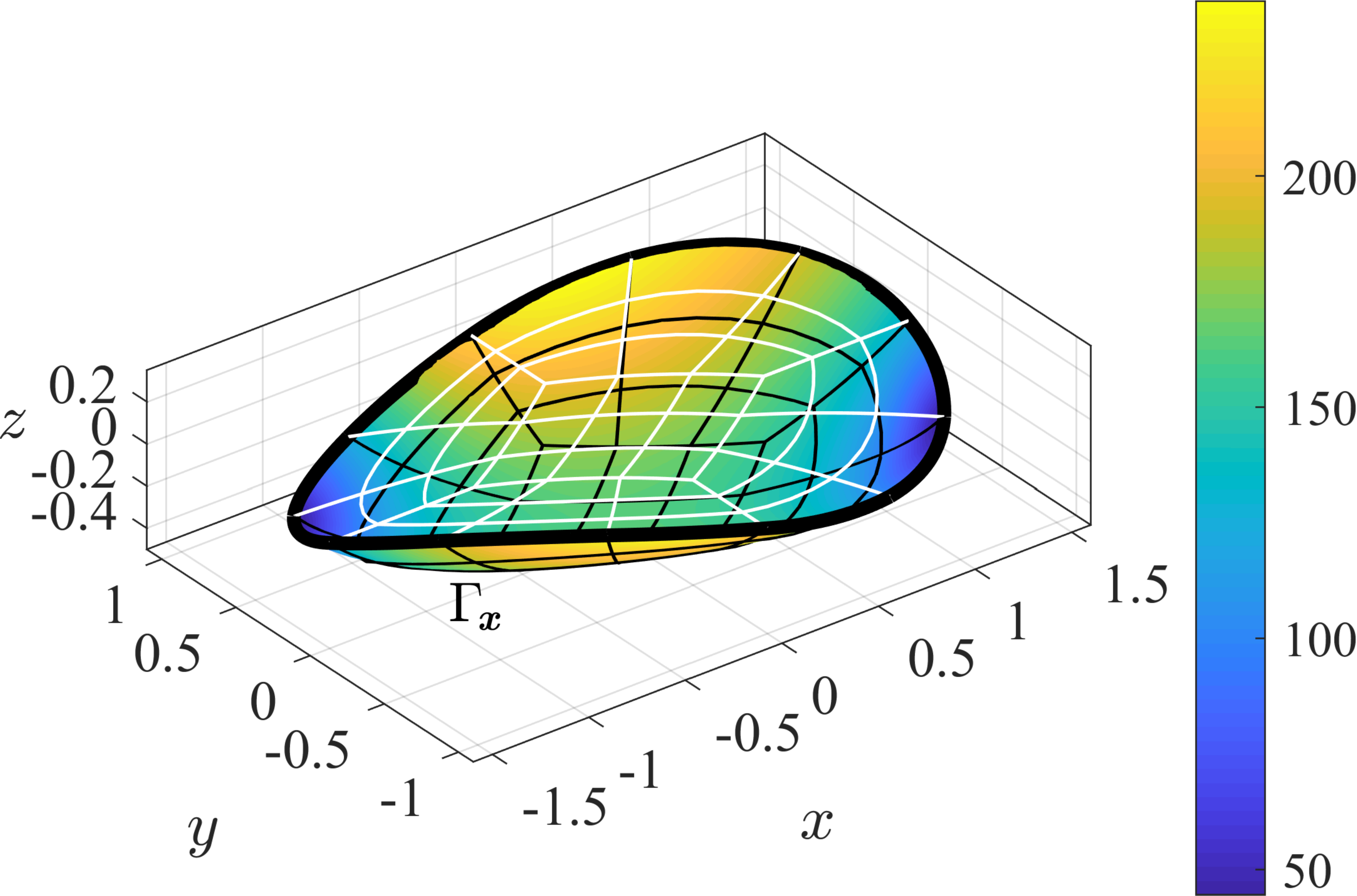}}\hfill\subfigure[Map A, displ.~$w$]{\includegraphics[width=0.25\textwidth]{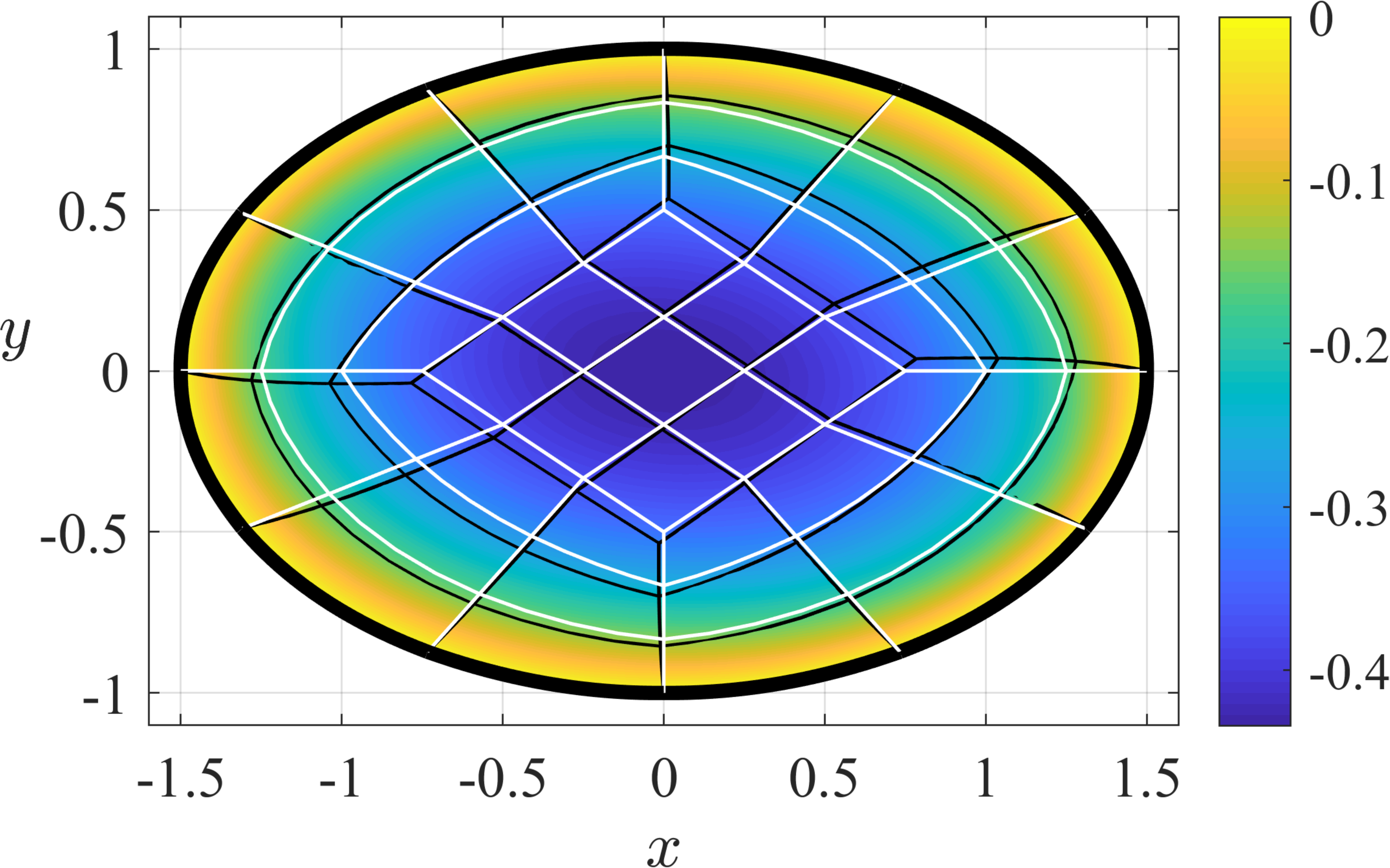}}

\subfigure[Map B, undeformed conf.]{\includegraphics[width=0.35\textwidth]{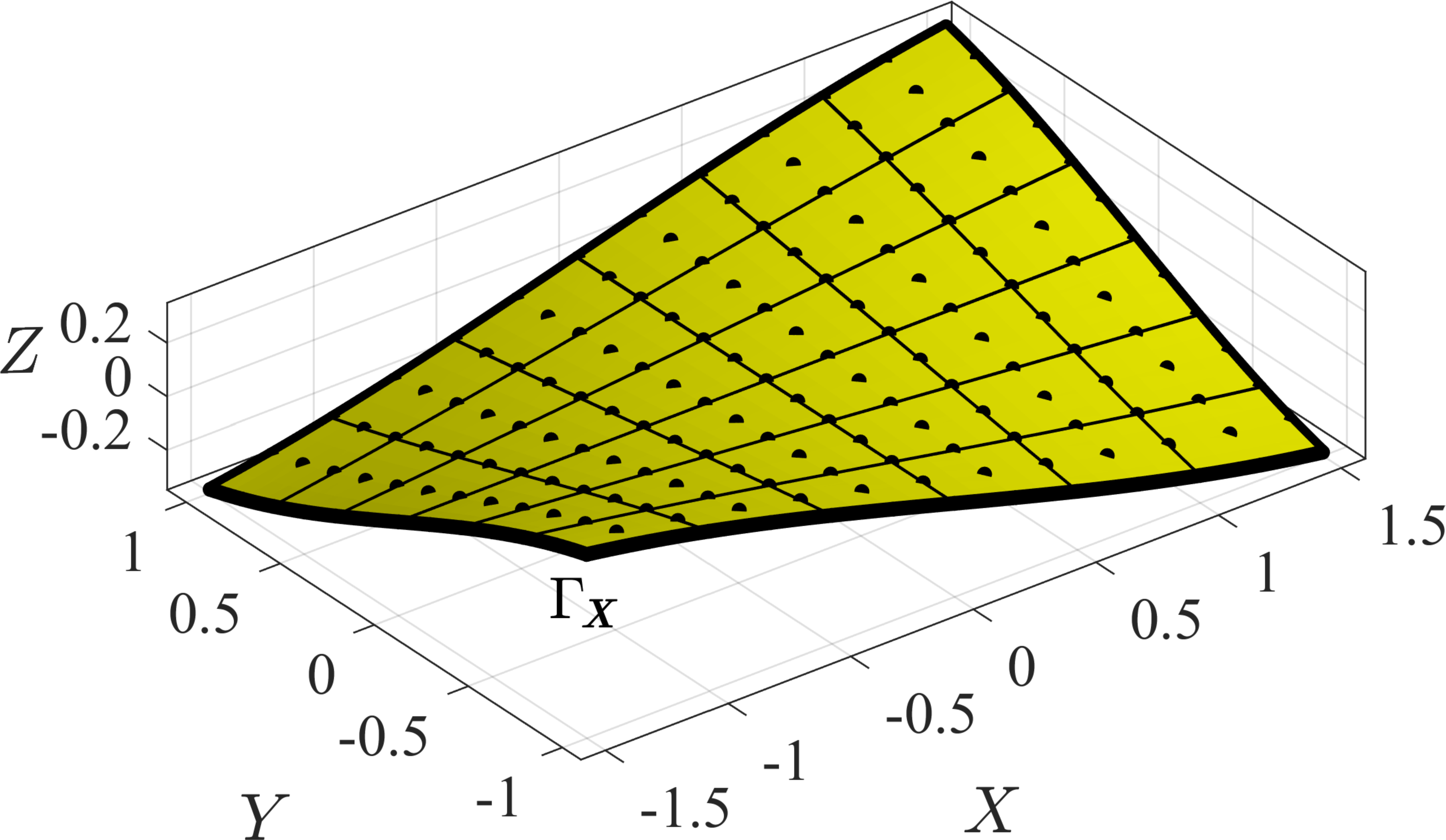}}\hfill\subfigure[Map B, deformed conf.]{\includegraphics[width=0.35\textwidth]{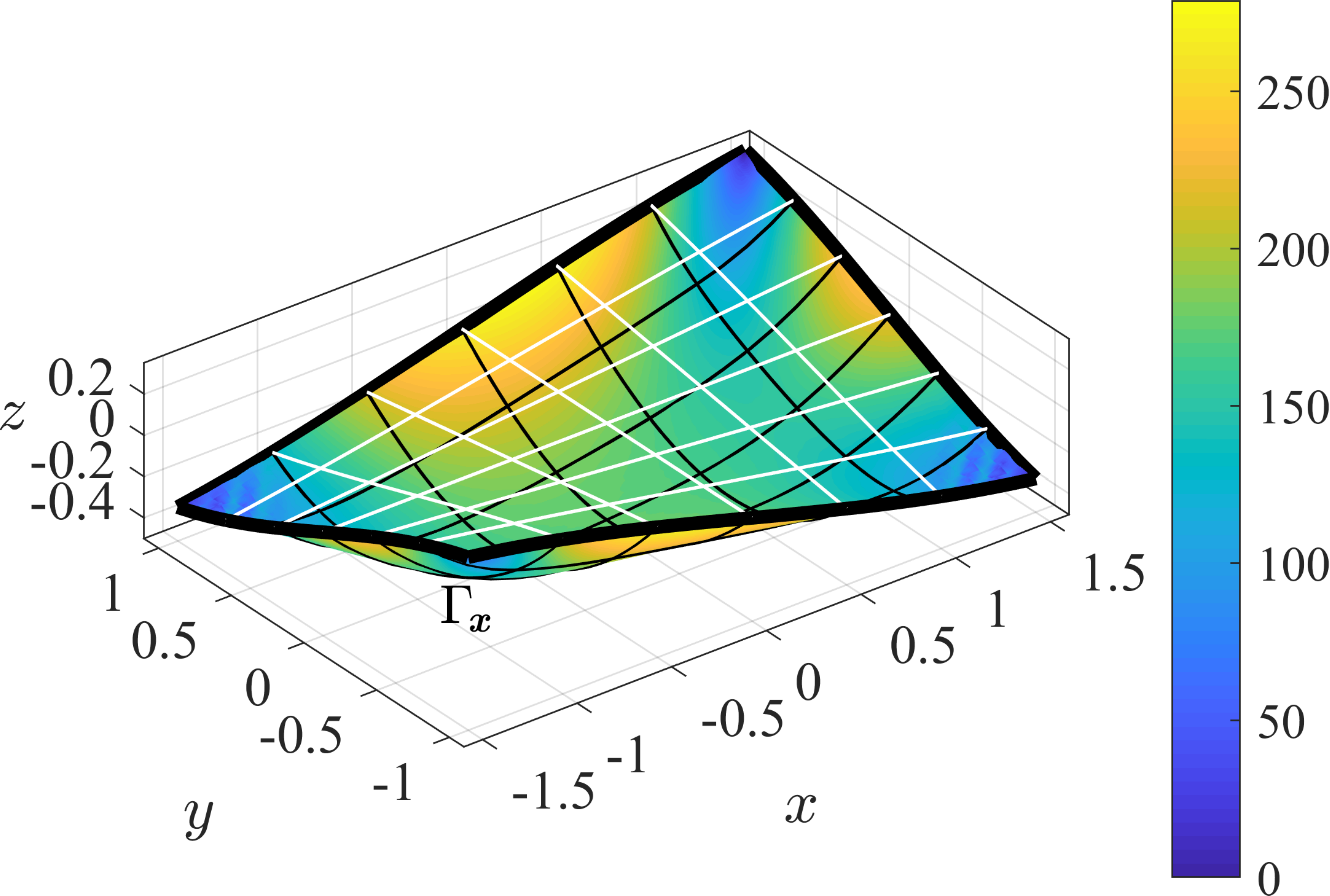}}\hfill\subfigure[Map B, displ.~$w$]{\includegraphics[width=0.25\textwidth]{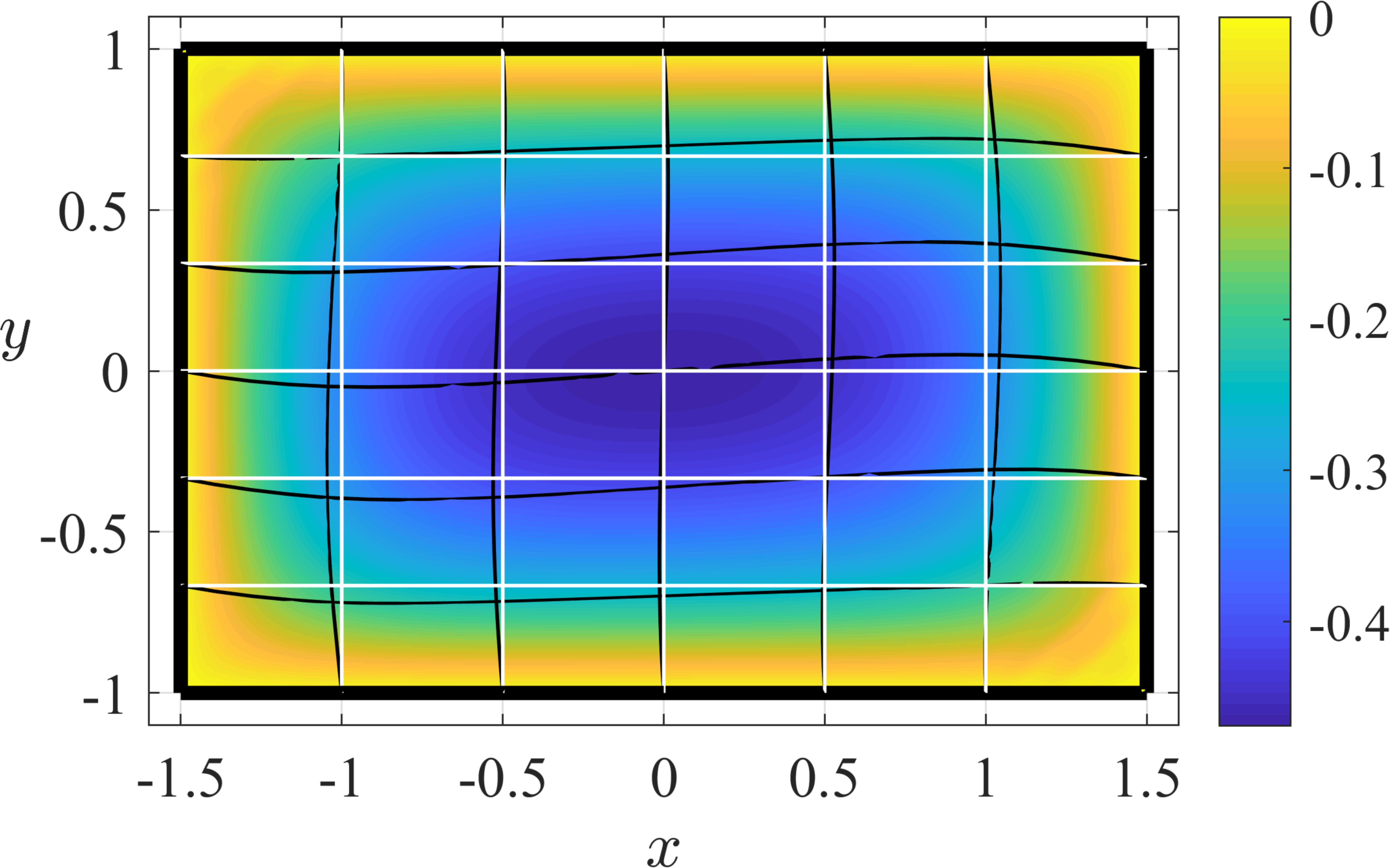}}

\caption{\label{fig:TC3_Situation}Sketch of test case 3 for the two different
maps A and B: (a) and (d) show the undeformed configurations $\Gamma_{\!\vek X}^{h}$
with example meshes composed by quadratic elements, (b) and (e) the
deformed configurations $\Gamma_{\!\vek x}^{h}$ with von-Mises stresses,
(c) and (f) top views of the vertical displacement fields $w_{h}$.
White lines show element edges in $\Gamma_{\vek X}^{h}$, black lines
in $\Gamma_{\vek x}^{h}$.}
\end{figure}

For the next test case, the deformation of membranes is approximated
with the Surface FEM. Two different undeformed configurations are
considered which result from a map of a unit \emph{circle} for map
A and of a unit \emph{square} for map B. The undeformed configurations
$\Gamma_{\!\vek X}$ are given by the parametrizations
\begin{equation}
\vek X\!\left(\vek r\right)=\left[\begin{array}{c}
\nicefrac{3}{2}\cdot r\\
s\\
c\cdot\sin\left(r\cdot s\right)
\end{array}\right]\text{ with }\begin{cases}
\sqrt{r^{2}+s^{2}}\in\left(0,1\right) & \text{for map A,}\\
r,\:s\in\left(-1,1\right) & \text{for map B,}
\end{cases}\label{eq:TC3_Parametrization}
\end{equation}
where $c\in\mathbb{R}$ is a scaling parameter in vertical direction.
The resulting configurations for map A and B are illustrated in Figs.~\ref{fig:TC3_Situation}(a)
and (d) for $c=0.4$, respectively. An important difference is that
the boundary is smooth for map A but involves corners for map B, later
resulting in different convergence behaviors. The thickness of the
membrane is $t=0.01$, Young's modulus is $E=1\,000$ and Poisson
ratio $\nu=0.3$, which is easily converted into the Lam\'e parameters.
The loading is gravity acting on the membrane surface with $\vek F\!\left(\vek X\right)=\left[0,0,-200\cdot t\right]^{\mathrm{T}}$
for all $\vek X\in\Gamma_{\!\vek X}$. The whole boundary is treated
as a Dirichlet boundary with prescribed zero-displacements. The deformed
configurations are displayed in Figs.~\ref{fig:TC3_Situation}(b)
and (e) with computed von-Mises stresses based on the Cauchy stress
tensor. The vertical displacement field is given in Fig.~\ref{fig:TC3_Situation}(c)
and (f) for the two maps in top view, respectively.

\begin{figure}
\centering

\subfigure[map A, $\varepsilon_{\mathfrak{e}}$]{\includegraphics[width=0.45\textwidth]{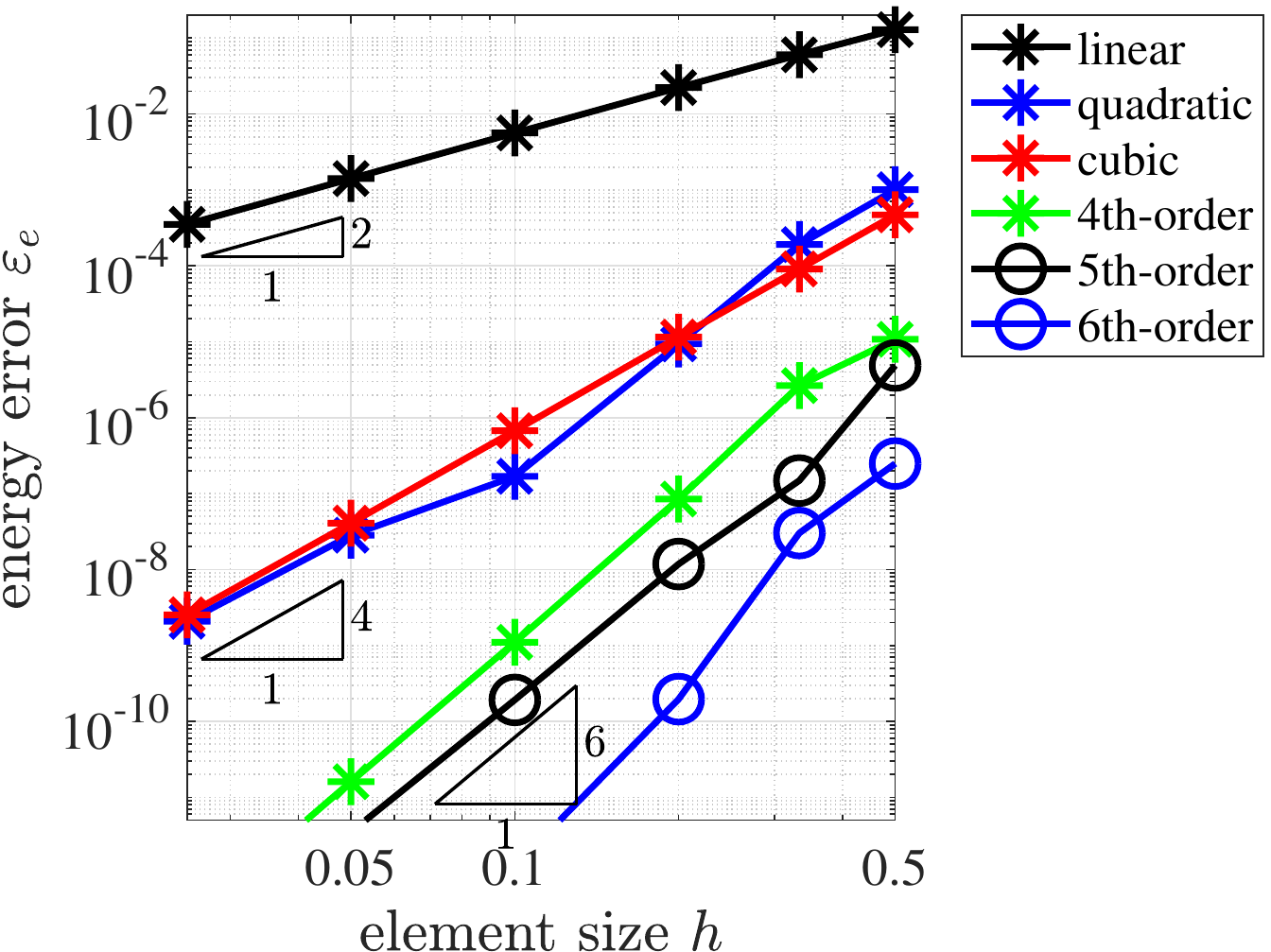}}\hfill\subfigure[map A, $\varepsilon_{\mathrm{res}}$]{\includegraphics[width=0.45\textwidth]{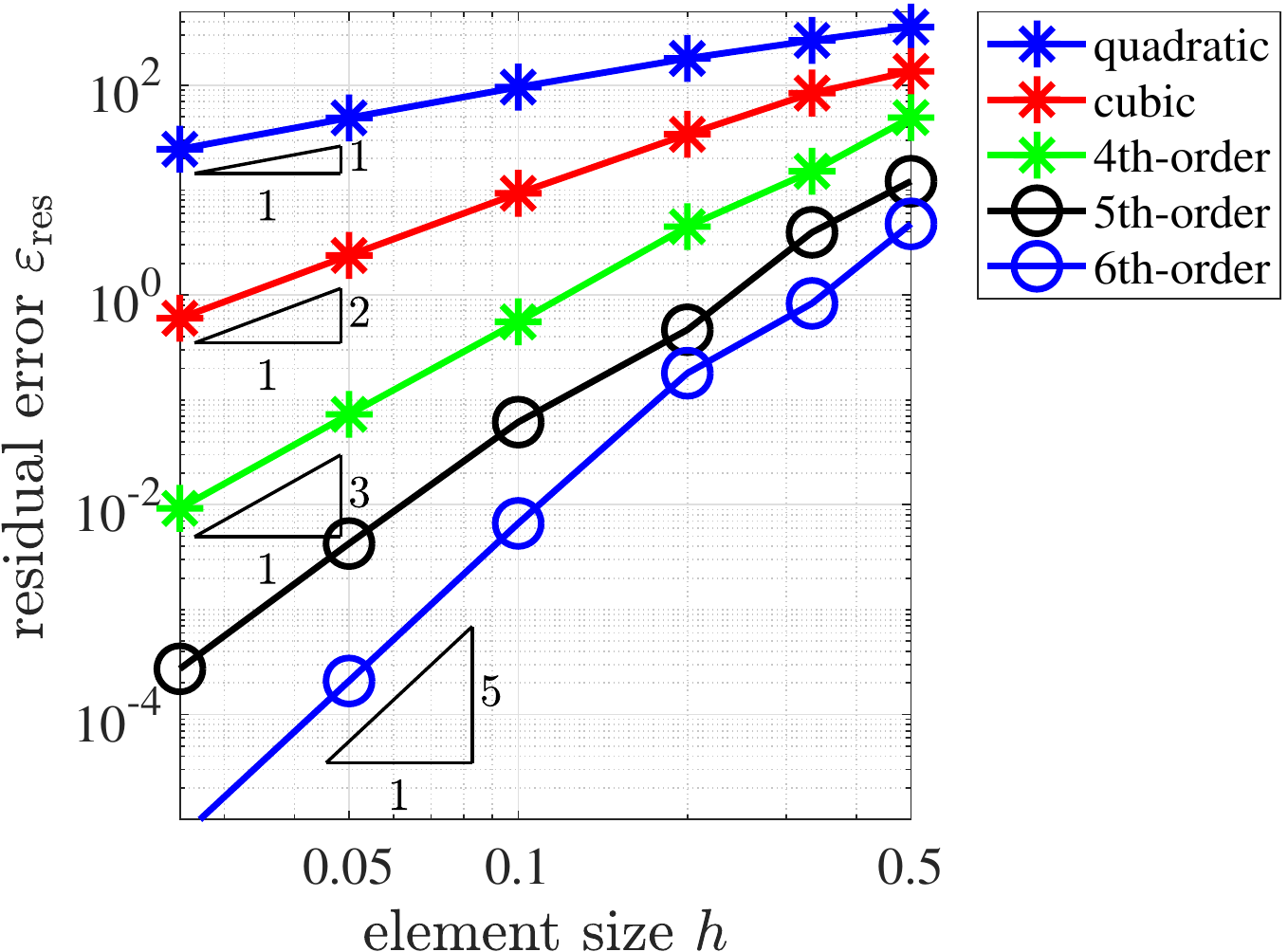}}

\subfigure[map B, $\varepsilon_{\mathfrak{e}}$]{\includegraphics[width=0.45\textwidth]{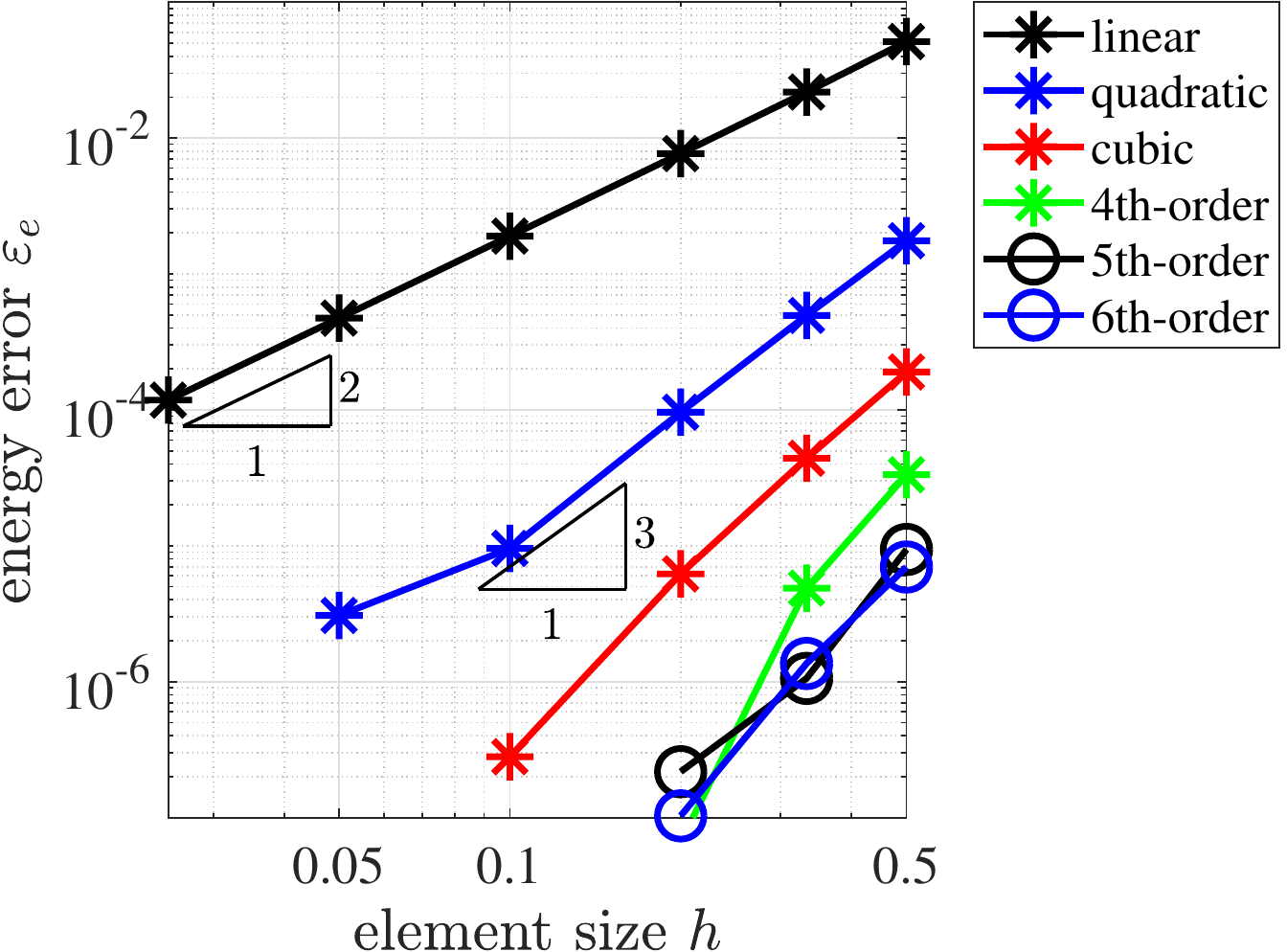}}\hfill\subfigure[map B, $\varepsilon_{\mathrm{res}}$]{\includegraphics[width=0.45\textwidth]{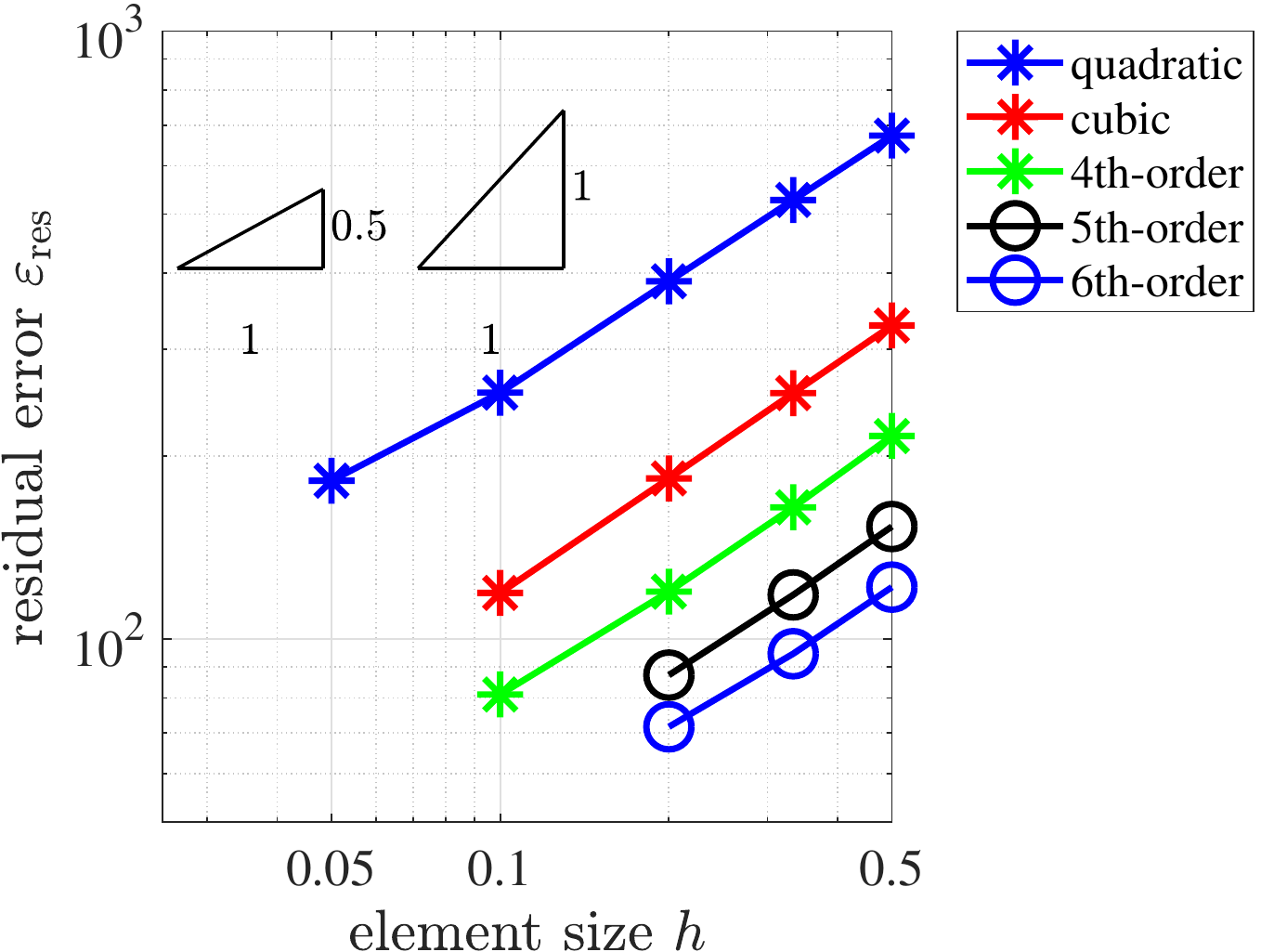}}

\caption{\label{fig:TC3_Results}Convergence results for test case 3: (a) and
(b) show the results for the energy error $\varepsilon_{\mathfrak{e}}$
and residual error $\varepsilon_{\mathrm{res}}$ for map A, (c) and
(d) for map B.}
\end{figure}

Convergence results are given in Fig.~\ref{fig:TC3_Results}. For
map A, where the boundary is smooth, optimal convergence rates are
found in the energy error $\varepsilon_{\mathfrak{e}}$ and residual
error $\varepsilon_{\mathrm{res}}$. For map B, where corners are
present in the membrane geometry, it is seen that the convergence
rates are bounded. Only linear and quadratic elements converge optimally
in $\varepsilon_{\mathfrak{e}}$, higher orders improve the error
level, however, not the convergence rates. It is thus confirmed that
corners in membranes have the potential to reduce the convergence
rates.

\subsection{Coupled ropes and membranes}

\begin{figure}
\centering

\subfigure[inflated ball]{\includegraphics[height=0.2\textheight]{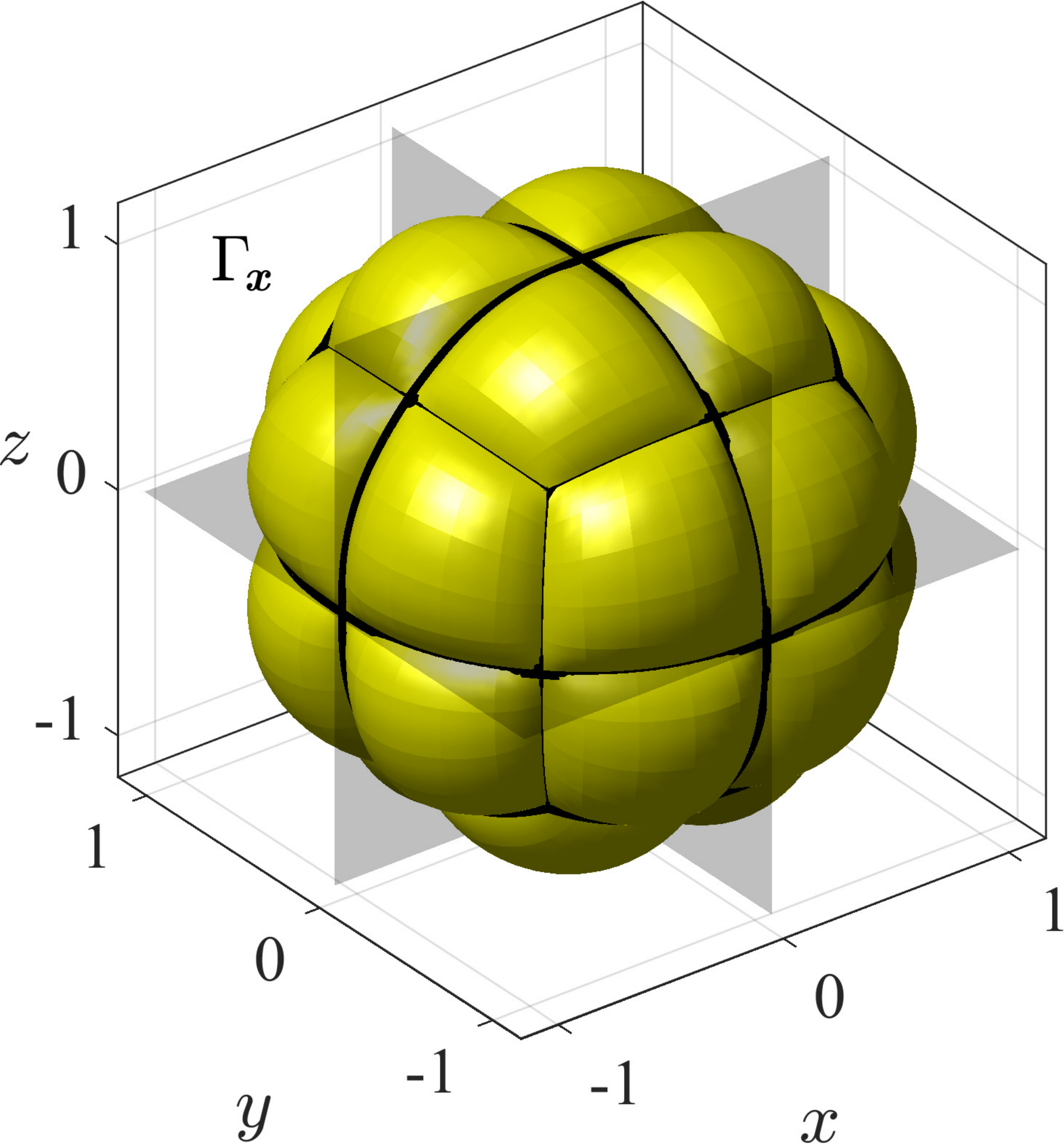}}\hfill\subfigure[domains $\Gamma_{\!\vek X}$ and $\Gamma_{\!\vek x}$]{\includegraphics[height=0.2\textheight]{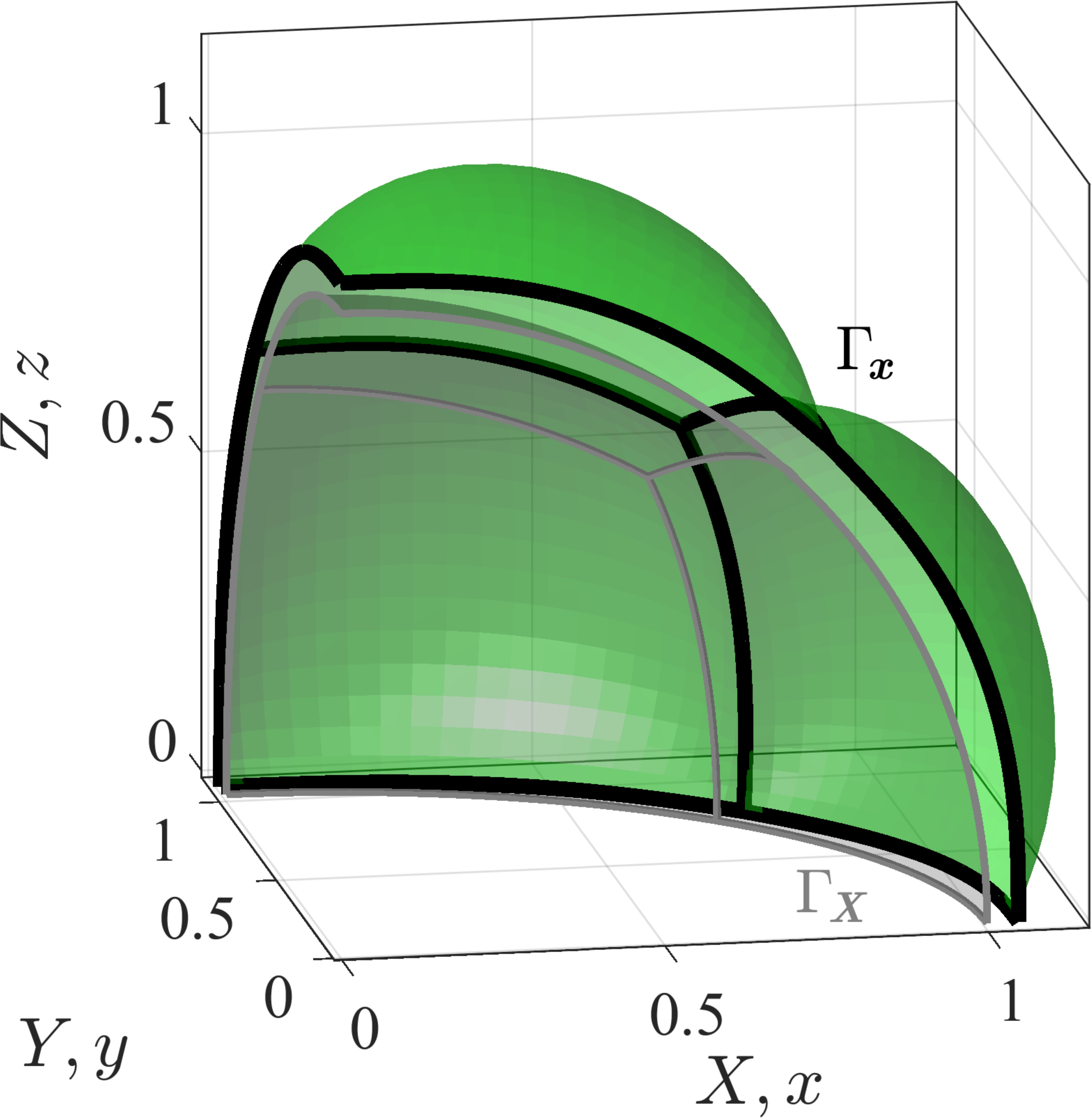}}\hfill\subfigure[von-Mises stress]{\includegraphics[height=0.2\textheight]{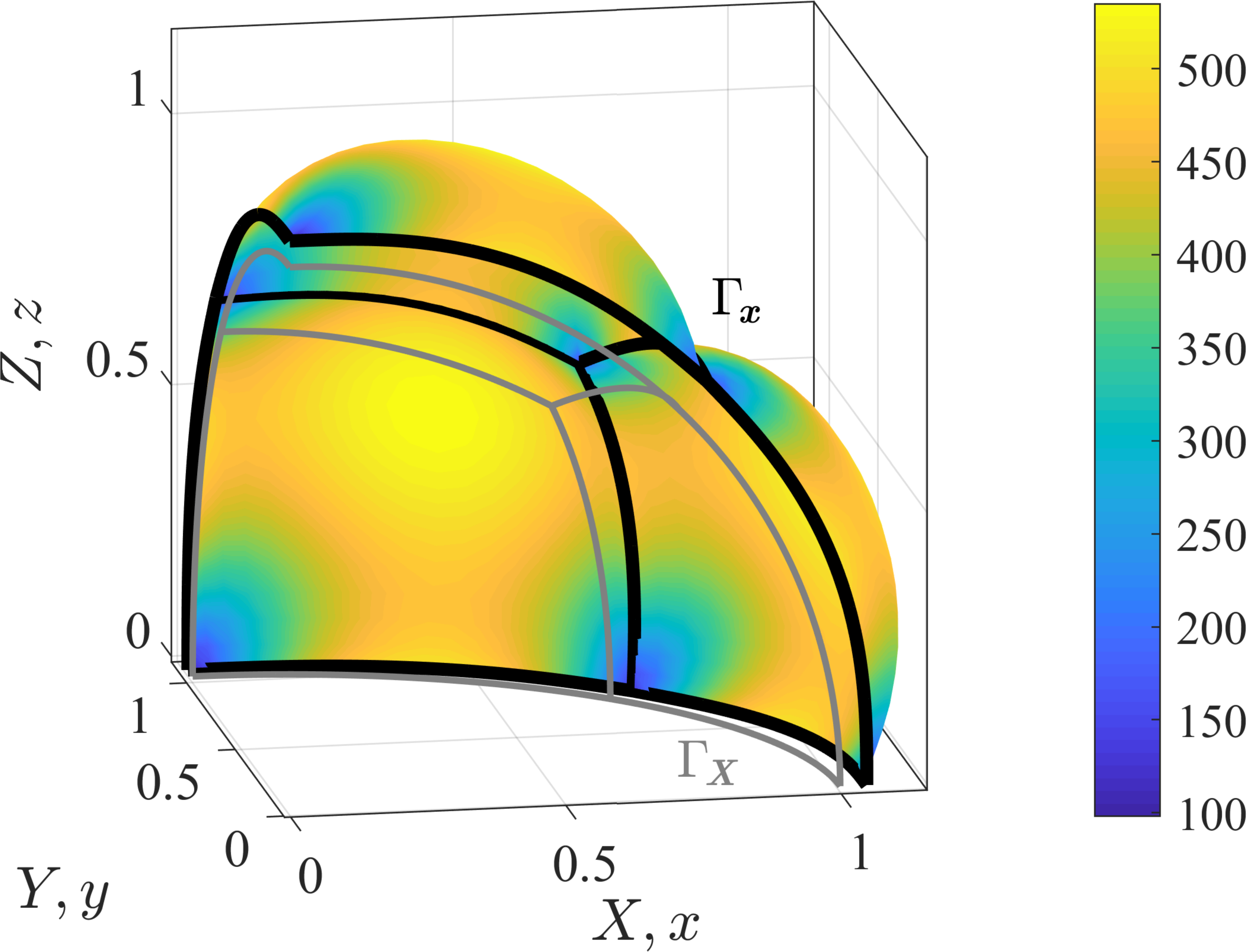}}

\caption{\label{fig:TC4_Situation}Sketch of test case 4: (a) The full inflated
ball reinforced with cables (black lines), (b) deformed and undeformed
domains in the simulations, (c) von-Mises stresses in the membrane.}
\end{figure}

It was mentioned several times that the proposed framework allows
for a unified treatment of ropes and membranes (and even continua).
This shall be confirmed with this last test case where cables and
membranes are coupled. The situation may be described as an inflated
ball with embedded reinforcement cables, see Fig.~\ref{fig:TC4_Situation}(a).
The original radius of the ball is $r=1.0$ and the inner pressure
is $p=20$. This pressure converts to a loading of $\vek f\!\left(\vek x\right)=p\cdot\vek n\left(\vek x\right)$
which depends on the normal vector of the \emph{deformed} configuration,
hence on the displacements $\vek u$, thereby adding a new source
of non-linearity which has to be properly considered in the Newton-Raphson
loop. The membrane is defined by the parameters $t=0.01$, $E=1\,000$
and $\nu=0.3$. The cables which are on the gray planes in Fig.~\ref{fig:TC4_Situation}(a)
feature a Young's modulus of $E=1\,000\,000$ and a cross section
of $A=0.0001$. All other cables have a Young's modulus of $E=500\,000$.
For the modeling, only one eighth of the initial sphere is considered,
see Fig.~\ref{fig:TC4_Situation}(b) for the deformed and undeformed
situations. Symmetry boundary conditions are applied. The resulting
von-Mises stresses in the membrane are shown in Fig.~\ref{fig:TC4_Situation}(c).
The elastic energy stored in the membrane \emph{and} ropes is given
as $\mathfrak{e}\left(\vek u\right)=2.9802127651$.

\begin{figure}
\centering

\subfigure[energy error $\varepsilon_{\mathfrak{e}}$]{\includegraphics[width=0.45\textwidth]{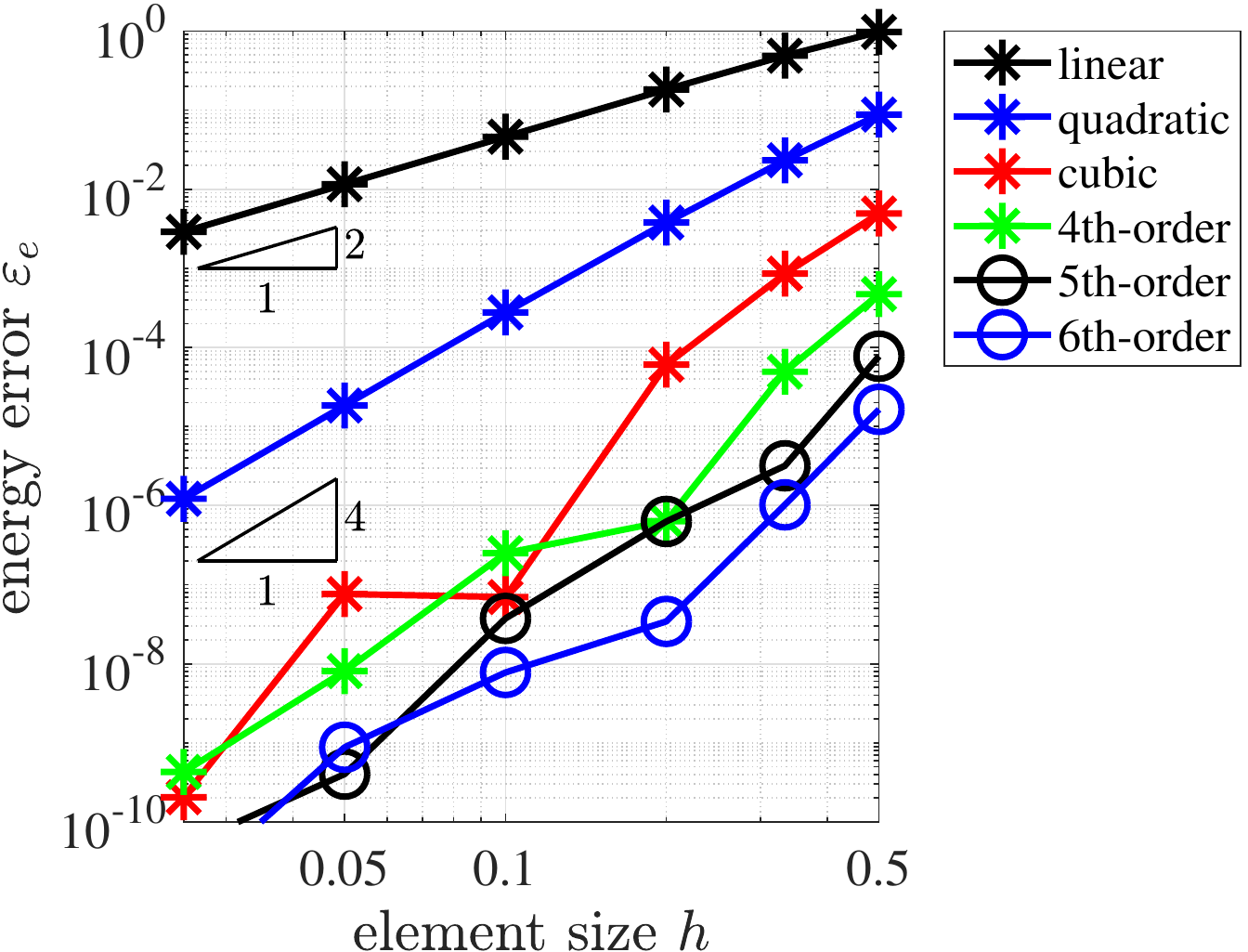}}\hfill\subfigure[residual error $\varepsilon_{\mathrm{res}}$]{\includegraphics[width=0.45\textwidth]{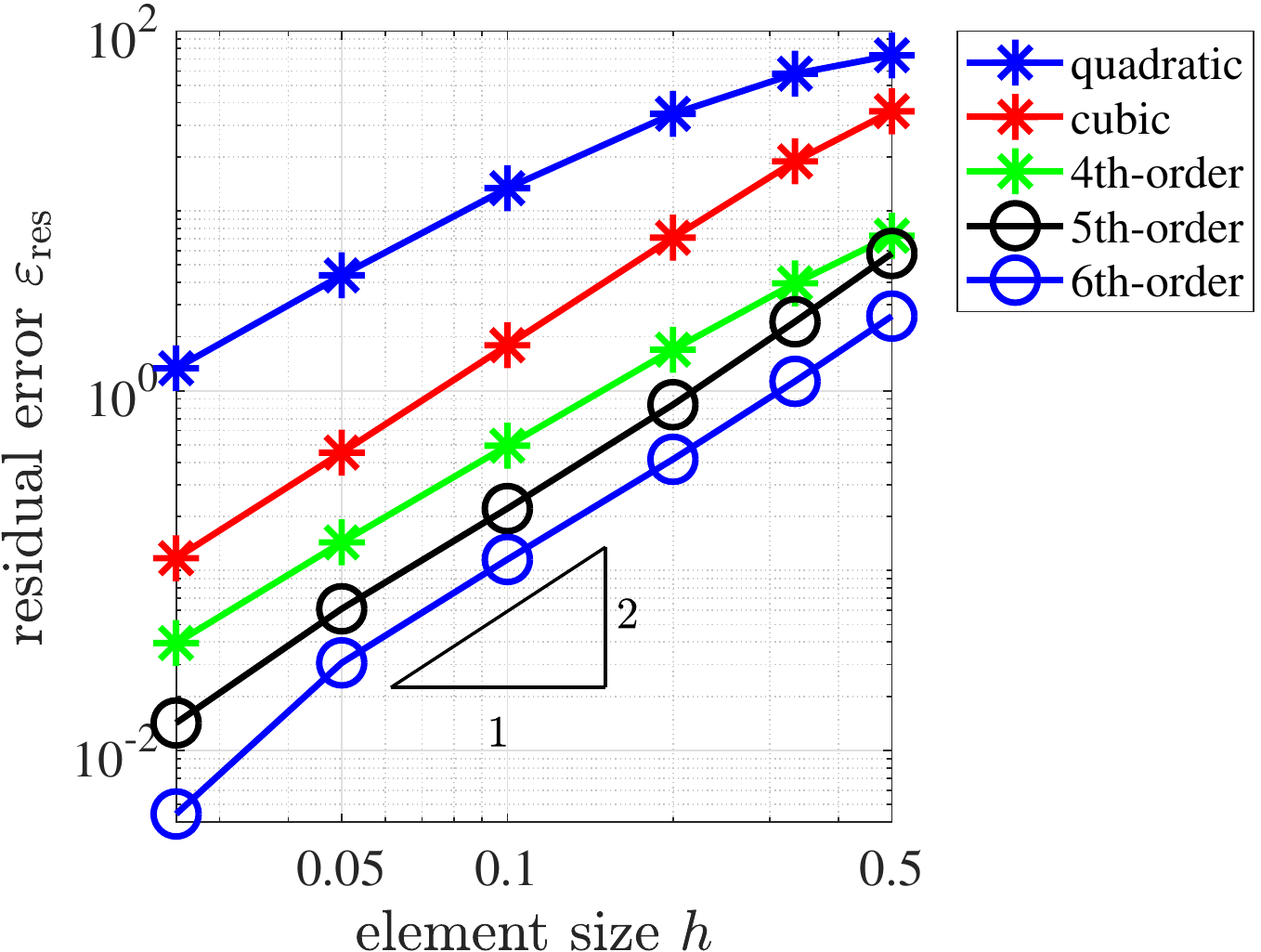}}

\caption{\label{fig:TC4_Results}Convergence results for test case 4: (a) the
energy error $\varepsilon_{\mathfrak{e}}$ and (b) the residual error
$\varepsilon_{\mathrm{res}}$.}
\end{figure}

The convergence results are seen in Fig.~\ref{fig:TC4_Results} for
the energy error $\varepsilon_{\mathfrak{e}}$ and residual error
$\varepsilon_{\mathrm{res}}$. It is again seen that the convergence
rates in $\varepsilon_{\mathfrak{e}}$ are optimal for linear and
quadratic elements. Higher orders achieve better results, however,
do not improve the convergence rates. The reason for the bounded convergence
rates is found in the stress concentrations where the embedded cables
meet, see Fig.~\ref{fig:TC4_Situation}(c).

\section{Conclusions\label{X_Conclusions}}

The modeling of ropes and membranes leads to partial differential
equations on manifolds. Herein, a framework is proposed which unifies
the mechanical modeling of ropes and membranes undergoing large displacements
and, furthermore, applies to parametric and implicit definitions of
manifolds. The fundamental ingredient is the use of Tangential Differential
Calculus (TDC) for the definition of geometric and differential quantities
on the manifolds. The proposed TDC-based formulation is more general
than the classical parametric formulation based on curvilinear coordinates
because it allows for two different numerical approaches, the Surface
and the Trace FEM. The classical approach is restricted to the Surface
FEM and cannot handle manifolds which are defined implicitly (unless
discretized by a surface mesh).

Even in the classical setting using the Surface FEM with parametric
formulations, the new TDC-based formulation leads to a significantly
different implementation, however, ultimately achieving the same results
(as expected). The advantage of the TDC-based formulation is that
it also applies immediately to implicit definitions and the Trace
FEM. More technically speaking, we have one element integration routine
which applies to ropes and membranes (and even continua) no matter
whether we are using the Surface FEM or the Trace FEM. Of course,
in the Trace FEM, additional terms have to be considered for the stabilization.

The numerical results consider various test cases with ropes and membranes
in two and three dimensions. The Surface and Trace FEM with higher-order
elements are used successfully, achieving higher-order convergence
rates. It is confirmed that the smoothness of the geometry and the
coupling of ropes and membranes has an important influence on the
convergence behavior.

\bibliographystyle{schanz}
\addcontentsline{toc}{section}{\refname}\bibliography{FriesRefs}
 
\end{document}